\documentclass[draft]{agujournal2019}
\usepackage{url} 
\usepackage{lineno}
\usepackage{soul}
\usepackage{tikz}
\usetikzlibrary{shapes, arrows}
\usetikzlibrary{cd}
\usetikzlibrary{trees,arrows.meta,shapes.geometric,decorations.pathreplacing,calc}
\usetikzlibrary{positioning, arrows.meta}

\tikzstyle{terminator} = [rectangle, draw, text centered, rounded corners, minimum height=2em]
\tikzstyle{process} = [rectangle, draw, text centered, minimum height=2em]
\tikzstyle{decision} = [diamond, draw, text centered, minimum height=2em]
\tikzstyle{data}=[trapezium, draw, text centered, trapezium left angle=60, trapezium right angle=120, minimum height=2em]
\tikzstyle{connector} = [draw, -latex']
\tikzset{curve/.style={settings={#1},to path={(\tikztostart)
    .. controls ($(\tikztostart)!\pv{pos}!(\tikztotarget)!\pv{height}!270:(\tikztotarget)$)
    and ($(\tikztostart)!1-\pv{pos}!(\tikztotarget)!\pv{height}!270:(\tikztotarget)$)
    .. (\tikztotarget)\tikztonodes}},
    settings/.code={\tikzset{quiver/.cd,#1}
        \def\pv##1{\pgfkeysvalueof{/tikz/quiver/##1}}},
    quiver/.cd,pos/.initial=0.35,height/.initial=0}
    
\newcommand{\ra}[1]{\renewcommand{\arraystretch}{#1}}
\usepackage{amsmath,amssymb,amsfonts}%
\usepackage{amsthm}%
\usepackage{float}
\usepackage{soul}

\journalname{Journal of Advances in Modeling Earth Systems (JAMES)}

\begin{document}

\title{Evaluating near-surface wind speeds simulated by the CRCM6-GEM5 model using AmeriFlux data over North America}

\authors{Tim Whittaker\affil{1}, Alejandro Di Luca\affil{1}, Francois Roberge\affil{1}, Katja Winger\affil{1}}

\affiliation{1}{Centre pour l'étude et la simulation du climat à l'échelle régionale (ESCER), D\'epartement des Sciences de la Terre et de l’Atmosph\`ere, Universit\'e du Qu\'ebec \`a Montr\'eal, Montr\'eal, Qu\'ebec, Canada}

\correspondingauthor{Tim Whittaker}{whittaker.tim@courrier.uqam.ca}

\begin{keypoints} 
\item Introducing a hierarchy of error metrics significantly alters the conclusions regarding the performance and ranking of the simulations. 
\item The inclusion of a turbulent orographic form drag scheme notably improves the simulation of near-surface variables. 
\item Accurately modeling surface winds under very stable atmospheric conditions remains a primary challenge for most configurations. 
\end{keypoints}

\begin{abstract}
    We evaluate the performance of various configurations of the Canadian Regional Climate Model (CRCM6-GEM5) in simulating 10-meter wind speeds using data from 27 AmeriFlux stations across North America. The assessment employs a hierarchy of error metrics, ranging from simple mean bias to advanced metrics that account for the dependence of wind speeds on variables such as friction velocity and stability. The results reveal that (i) the value of roughness length (z0) has a large effect on the simulation of wind speeds, (ii) using a lower limit for the Obhukov length instead of a lower limit for the lowest level wind speed seems to deteriorate the simulation of wind speeds under very stable conditions, (iii) the choice of stability function has a small but noticeable impact on the wind speeds, (iv) the turbulent orographic form drag scheme shows improvement over effective roughness length approach.
\end{abstract}

\section*{Plain Language Summary}
\noindent
Accurate wind speed forecasting is essential for various applications, from predicting health impacts and wildfire risks to assessing wind energy and storm damage. This study examines how well the Canadian Regional Climate Model (CRCM6-GEM5) predicts wind speeds at 10 meters above the ground, using measurements from 27 AmeriFlux stations. The research examines how different model settings, such as surface roughness and atmospheric stability, affect wind speed predictions. By using a variety of error metrics, the study identifies areas where the model performs well and where improvements are needed, particularly in how it handles stable atmospheric conditions and surface roughness.

\section{Introduction}
\label{intro}

The wind speed near the surface is a key variable in a number of applications, including the calculation of heat and windchill indices to quantify impacts in human and animal health \cite{WINDCHILL,Anderson2013}, wildfire risk and fire spread \cite{fwi,https://doi.org/10.1029/2024MS004300}, assessing of the potential for wind-generated electricity \cite{en14123702,Pryor2021} and assessing damages associated with storm hazards \cite{schwierz_modelling_2010,hewson_cyclones_2015,chen_seasonality_2022}. Winds are also important because they transport heat and water from distant locations, thereby influencing local temperature and specific humidity including heat waves, cold waves and precipitation processes. In addition, near-surface winds affect the intensity of turbulence which in turn affects fluxes of energy, water, and momentum between the surface and the atmosphere \cite{Garratt1994}.

Therefore, accurate simulation of winds, especially near the surface, is essential for reliable weather forecasts and climate projections. However, this task is challenging because surface wind speeds are influenced by multiple processes that operate at different spatiotemporal scales while interacting with each other. First, surface winds are influenced by upper-level winds through the vertical transfer of horizontal momentum. This means that upper-level processes, such as Rossby waves, extratropical cyclones, and even mesoscale convective systems, must be accurately simulated \cite{hewson_cyclones_2015,trzeciak_can_2016,schumacher_formation_2020,chen_seasonality_2022}. Second, the vertical transfer of momentum between low levels and the free atmosphere depends on the representation of processes in the planetary boundary layer (PBL) \cite{Oneill_2012,chen_seasonality_2022}, including buoyancy and turbulence, and are strongly parameterized in current numerical weather prediction (NWP) and climate models \cite{beljaars_parametrization_1992,Mahrt1998,suselj_improving_2010,Mctaggart-Cowan2015}. Third, low-level winds are also influenced by the representation of surface-atmosphere interactions within the surface layer, which are typically parameterized using Monin-Obukhov Similarity Theory (MOST) \cite{monin_monin_and_obukhov_1954_1959,Jimenez_2012,Lee_2022}. That is, in NWP and climate models with horizontal grid spacings exceeding a few kilometers and vertical grid spacings larger than a few tens of meters, the evolution of near-surface winds is governed by the resolved dynamics of the equations of motion (e.g., fully compressible Euler equations) and by various subgrid-scale processes. These subgrid-scale processes are parameterized and include key components such as planetary boundary layer dynamics and surface layer interactions.

Few studies have evaluated the impact of PBL processes on near-surface winds \cite{di_luca_atmospheric_2014,dzebre_preliminary_2020}. \citeA{dzebre_preliminary_2020} evaluated the impact of PBL processes on 50 m wind speeds using 11 PBL schemes, generally associated with specific surface layer schemes, available in the Weather Research and Forecasting (WRF) model and data collected at two masts located in Ghana and Angola. Their results show that wind speed differences generally remain within 10 $\%$, although the period of analysis is relatively short. Using a large multi-physics ensemble of WRF simulations, \citeA{di_luca_atmospheric_2014} showed that surface wind speeds over the Mediterranean are similarly sensitive to changes in PBL and surface layer schemes and to changes in deep convective schemes. Given the coupling between the PBL and the surface layer, these errors could propagate downward to 10 m through turbulent processes and vertical momentum transfer.

Most studies looking at the sensitivity of surface wind speeds to models' configuration have focused on how surface-layer processes are represented within the models \cite{suselj_improving_2010,Jimenez_2012,Jimenez_Dudhia_2012,Jimenez_Dudhia_2013}. Most NWP and climate models, such as the National Center for Atmospheric Research WRF model \cite{Jimenez_2012,skamarock_description_2019}, the European Centre for Medium‐Range Forecasts Integrated Forecasting System \cite{ecmwf_ifs_2016,ecmwf_ifs_2023} and the Global Environmental Multiscale (GEM) model \cite{ModernNWP}, to mention just a few, all utilize MOST to represent processes in the surface layer \cite{shin_improved_2019}. The implementation of MOST in numerical models, however, is a complex task. It involves empirically determining certain parameters, such as the roughness length of the surface, and is influenced by the numerical formulation of the model, such as the height of the lowest model level \cite{ModernNWP,shin_improved_2019}. 

For instance, the roughness length parameter (here denoted as $z_0$) indicates the height above the surface at which the wind speed vanishes (i.e., $u(z=z_0)=0$). This parameter can vary by several orders of magnitude over short distances (tens of meters), making it challenging to derive a single value over grid boxes used in current NWP and climate models (generally larger than 5 km$^2$). Over land grid points, $z_0$ is typically estimated following two main approaches \cite{fiedler_geostrophic_1972,TOFD,Jimenez_Dudhia_2012}. The first approach calculates an \textit{effective roughness length} by combining information from both local surface characteristics and subgrid-scale orography \cite{fiedler_geostrophic_1972,Jimenez_Dudhia_2012}. The second approach determines the roughness length based solely on local surface characteristics, while accounting for the influence of orography separately through an additional drag term in the resolved equations \cite{TOFD,Jimenez_Dudhia_2012}.

Wind speeds have been shown to be highly sensitive to the value of $z_0$. For instance, \citeA{Nelli_2020} used station data from a tower in the Arabian Peninsula and found that the $z_0$ value used in the WRF model (with a horizontal grid spacing of 1.33 km) was nearly 10 times larger than the observed $z_0$. This discrepancy resulted in wind speed differences averaging about 0.5 m/s between the default and corrected $z_0$ values. They also demonstrated that the change in $z_0$ caused a nearly constant decrease in wind speeds, mostly independent of the wind speed's magnitude. \citeA{Lee_2022} used station data from towers installed over different land surface types, including early growth soybean, native grassland, and mature soybean. They evaluated MOST under varying conditions, considering the diverse roughness lengths associated with each land surface type. They found that, despite the sites being only 1.7 km apart, the MOST relationships (i.e., variations in the von Kármán constant) differed significantly. This finding suggests that MOST may not hold uniformly even within a single grid cell.

Another parameter that needs to be empirically specified is the functional form of the stability function that is used to correct wind speeds based on the surface layer stability. Several functional forms are proposed in the literature \cite{Beljaars1991,DELAGE1997}, most of them derived from observed data available from a handful of field experiments (e.g., the Kansas experiment, \citeA{kaimal_kansas_1990}). For instance, \citeA{Jimenez_2012} modified the WRF model's surface layer scheme by refining the empirical stability functions for very stable and unstable situations. Their new scheme resulted in sharper afternoon transitions from stable to unstable conditions, aligning well with observations. Additionally, they reported overall improvements in near-surface variables, particularly in terms of diurnal amplitudes.

In addition to the roughness length and the choice of function functional form of the stability function, MOST requires specific assumptions, such as spatial homogeneity and stationarity, to be met. However, these assumptions can be violated during events like thunderstorms. In \citeA{ObservationsofNearSurfaceVerticalWindProfilesand}, near-surface vertical wind profiles and vertical momentum fluxes observations are used to evaluate the performance of MOST on thunderstorm days, when these assumptions do not hold. The study finds that MOST underestimates the nondimensional vertical wind shear near the surface under such conditions, highlighting the inadequacy of MOST, especially during convective events. 

The specific implementation of MOST might also incorporate some ad hoc assumptions to address special conditions. For example, \citeA{Jimenez_2012} reported that earlier implementations of MOST in the WRF model included several limits that were imposed on certain variables in order to avoid undesired effects, such as a lower limit of 0.1 m s$^{-1}$ for the friction velocity to prevent the heat flux from being zero under very stable conditions. Similarly, the current version of the GEM model uses a minimum Obukhov length of 20 m \cite{ModernNWP} while an earlier version used a minimum lowest level wind speed, both limits imposed to maintain land-atmosphere coupling under very stable conditions.

In this work we evaluate the ability of several configurations of the latest version of the Canadian Regional Climate Model (CRCM6-GEM5) to simulate 10-m wind speeds ($u_{10}$) measurements from 27 stations from the AmeriFlux network. We focus on physical processes within the surface layer and test the sensitivity of simulated $u_{10}$ to variations in the representation of very stable regimes, surface roughness, land-surface schemes, and stability function choices.
The wind speed model evaluation involves using a hierarchy of error metrics, ranging from simple metrics such as mean bias, to errors in the diurnal and annual cycles, to more advanced error metrics that evaluate errors in $u_{10}$ based on its dependence on other near-surface variables (e.g., friction velocity and the Obukhov length). 

This paper is organized as follows. In Sect.~\ref{Sec:Data:Sim}, we provide a detailed description of the model and the set of simulations used in the analysis and in Sect.~\ref{Sec:Data:Flux} we describe the AmeriFlux station data, including a height correction that was applied to wind data to facilitate the comparison with model output. Sect.~\ref{Sec:Methods} introduces the hierarchy of error metrics that are used to evaluate the simulations. In Sect.~\ref{Sec:Results} we present the evaluation of the simulations with respect to the hierarchy of metrics. Some implications of our results are discussed in Sect.~\ref{Sec:Discussion}. Finally, the main conclusions are presented in Sect. \ref{Sec:Conclusions}.

\section{Data}
\label{Sec:Data}

\subsection{The CRCM6-GEM5 model and sensitivity experiments}
\label{Sec:Data:Sim}
All simulations were performed using the latest version of the CRCM6-GEM5 \cite{roberge_spatial_2024} which is based on the fifth generation of the GEM model \cite{cote1998,Girard2014,ModernNWP}. GEM5 is developed by the Recherche en Pr\'evision Num\'erique (RPN) group at Environment and Climate Change Canada and is used for numerical weather prediction by the Canadian Meteorological Centre. 

\begin{figure}
    \center
	\includegraphics[width=80mm]{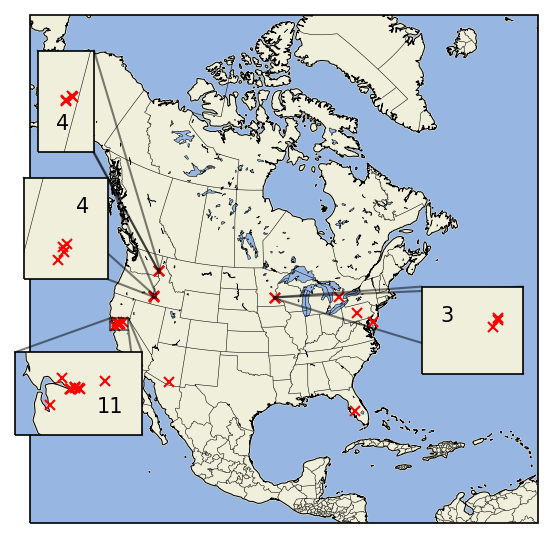}
	\caption{Map of the CORDEX-NA domain used for the simulations with red crosses showing the location of the 27 stations used in the study. Four inlets are shown, zooming in on some areas where there are several stations available.}
	\label{fig:stationmap}
\end{figure}

Simulations were performed using a horizontal grid spacing of 0.11$^\circ$ over the CORDEX North American (CORDEX-NA) domain \cite{Giorgi2015,roberge_spatial_2024}. The CORDEX-NA domain is shown in Fig. \ref{fig:stationmap}. In the vertical, the model uses 71 hybrid levels with the highest prognostic level at 10 hPa and the lowest level at $\sim$10 m for temperature and moisture and at $\sim$20 m for winds. At this horizontal grid spacing, the dynamical core of GEM5 solves the hydrostatic primitive equations using implicit treatment in time and a semi-Lagrangian advection scheme \cite{Girard2014}. It employs the Arakawa-C staggering on a latitude-longitude grid and a terrain-following coordinate based on log-hydrostatic-pressure in the vertical \cite{Girard2014}.

Simulations were initialized and driven at the boundaries and the interior of the domain using data from the ERA5 reanalysis \cite{Hersbach2020} from January 1st 2015 to December 31st 2020. Large-scale nudging using a spectral method was applied within the domain to drive zonal and meridional winds and temperatures \cite{Separovic2012}. This nudging was applied for vertical levels higher than the 0.85 hybrid level which corresponds approximately to 850 hPa and for waves with a wavelength greater than 200 km using a relaxation time scale of 24 hours. The use of spectral nudging ensures a substantial level of agreement between the large-scale patterns from ERA5 and the simulations \cite{Separovic2012}.

All configurations of the CRCM6-GEM5 model use the Bechtold scheme to represent shallow convection \cite{Bechtold2001}, the Kain-Fritsch scheme to represent deep convection \cite{Kain1990,McTaggart-Cowan2019}, an elevated-convection scheme \cite{mctaggart-cowan_convection_2020}, a planetary boundary layer scheme based on a 1.5-order closure with a prognostic equation for the turbulent kinetic energy \cite{Belair1999,Mctaggart-Cowan2015,ModernNWP} and the Predicted Particle Properties (P3) scheme \cite{Morrison2015,Milbrandt2016} to represent cloud and precipitation microphysical processes. In addition, a Subgrid Cloud and Precipitation Fraction scheme has been used to adapt P3 to the model resolution \cite{chosson_adapting_2014,jouan_adaptation_2020}. All simulations also use the Fresh-water Lake model (FLake, \cite{Martynov2012}) to account for changes in surface water temperature and roughness length over lakes. 

The surface layer wind speeds in CRCM6-GEM5 are estimated using MOST, resulting in a log wind profile of the form:
\begin{equation}
	u_z = \frac{u_*}{k}\left[\log\left(\frac{z+z_{0}}{z_{0}}\right) - \Psi_m\right],
	\label{log_profile}
\end{equation}
where $k=0.4$ is the von Kármán constant, $u_*$ the friction velocity, $z_{0}$ the (aerodynamic) roughness length, and $\Psi_m$ an empirical stability function \cite{Beljaars1991,Delage1992,ModernNWP,DELAGE1997} correction with the following general form,
\begin{equation}
	\Psi_m = \begin{cases}
		\Psi_m^{stable} \text{ for } \zeta> 0\\
		0 \text{ for } \zeta = 0\\
		\Psi_m^{unstable} \text{ for } \zeta< 0.
	\end{cases}.
\end{equation}

Under stable conditions, the stability functions used are the following,
\begin{equation}
	\Psi_m^{stable} = \begin{cases}
		\frac{1}{2}\left[a-\frac{z}{h}-\ln \left(1+\frac{b}{2}z+a\right) - \frac{b}{2\ \sqrt[]{c}}\arcsin\left(\frac{b-2cz}{d}\right) \right] \text{ for DEL}\\
		-(a\zeta + b(\zeta-c/d)e^{-d\zeta}+bc/d) \text{ for BEL}
	\end{cases},
\end{equation}

with $a = 1, b = 2/3, c = 5, d = 0.35$. The boundary layer height $h$ for $DEL$ is given by $h = max(h_1 , h_2 , h_3 , h_4)$ where $h_1 = \gamma \cdot (zu + 10 \cdot z_0)$, $\gamma = 1.2$, $zu = 20 $m (height of the model’s lowest prognostic (momentum) level), $h_2 = b_s\ \sqrt[]{\frac{\kappa u_* L}{|f|}}$, $f$ is the Coriolis parameter, $b_s = 1$, $h_3 = \frac{\gamma}{4 a_s \beta} L$, $a_s = 12$, $\beta = 1$ and $h_4 = h_{min}=30 $m.

For unstable conditions, stability functions are given by,
\begin{equation}
	\Psi_m^{unstable} = \begin{cases}
		 2\ln(w+1)+\frac{1}{2}\ln(w^2-w+1)+\frac{3}{2}\ln(w^2+w+1)-\sqrt[]{3}\arctan\frac{(w^2-1)}{\sqrt[]{3} \ w} \text{ for DEL},\\
		\log z - 2 \log(r + 1) - \log(r^2 + 1) + 2 \arctan(r) \text{ for BEL}
	\end{cases},
\end{equation}
with, $w(\zeta) = (1-\beta c_i \zeta)^{1/6}$ and $r=(1 - 16\zeta)^{1/4}$
where $\beta$ and $c_i$ are parameters with default values $\beta=1,c_i=40$.

The stability parameter $\zeta=z/L$ is a dimensionless variable that is inversely proportional to the Obukhov length scale $L$ defined as,
\begin{equation}
	L = - \frac{\bar{\theta}_v u_*^3}{kg(\overline{w'\theta_v'})} = \frac{u_*^2 \bar{\theta}_v}{kg \theta_*},
	\label{L_eq}
\end{equation}
where $\overline{w'\theta_v'}$ is the vertical component of the kinematic buoyancy flux (note that, $\overline{w'\theta_v'} = \overline{w'\theta'}+0.608\bar{\theta}\ \overline{w'q'}$ ; \citeA{Garratt1994}), $\theta_*$ is the potential temperature scale and $\bar{\theta}_v$ is the virtual potential temperature. The Obukhov length scale $L$ is positive when the buoyancy flux is negative (i.e., for stable conditions) which leads to $\Psi_m > 0$ and a negative stability correction. On the other hand, when the buoyancy flux is positive (i.e., for unstable conditions), $L<0$ and the stability correction is positive. In practice, $L$ is calculated in GEM5 using the bulk Richardson number \cite{ModernNWP}.

The friction velocity ($u_*$) is computed as,
\begin{equation}
u_* =    \left(\frac{|\tau_z|}{\rho}\right)^{1/2} =  \frac{k \cdot u(z)}{\log\left(\frac{z + z0}{z_0}\right) - \Psi_m}.
                \label{friction_vel_eq}
\end{equation}

We note that errors in wind speeds can be influenced directly by errors in $u_*$, $\zeta$ and $z_0$. Also note that due to the nonlinear nature of interactions, tracking the source of errors in wind speeds is challenging. For example, while we might know that $u_*$ and $\zeta$ are wrongly simulated, we cannot pinpoint which variable is at the origin of errors since they depend on each other. Additionally, the roughness length value influences the simulation of upper-level winds by primarily affecting the vertical turbulent flux of horizontal momentum (i.e., \(u_*\)). As a result, the roughness length impacts wind speeds both directly, through MOST in Eq.~\ref{log_profile}, and indirectly, by modulating the large-scale flow via \(u_*\).

We performed multiple simulations with the CRCM6-GEM5 model using different configurations. The main characteristics of simulations are included in Table \ref{gem_details}.
Two GEM5 model versions are used, GEM5.0.2 (hereafter GEM50) and GEM5.1.1 (denoted as GEM51). These two versions differ in several respects, but the main difference affecting wind speeds lies in the strategy used to handle very stable regimes that occur with light winds. In both model versions, surface turbulent fluxes and diagnostic fields (e.g., calculation of 2-m temperature and 10-m winds) are calculated by constraining the wind speed at the lowest momentum model level ($z\sim$20) to ensure it does not fall below a specific threshold, which depends on the surface type). This approach maintains a minimum level of mechanical turbulence. In GEM50 (simulation {\bf GEM50-C-DEL}), the minimum wind speed threshold is set to 0.1 m s$^{-1}$ for land surfaces, while for GEM51 ({e.g., \bf GEM51-C-DEL}), it is set to 0.01 m s$^{-1}$. GEM51 also employs an additional condition, limiting the Obukhov length scale $L$ so it can never be lower than a certain threshold value (see Table \ref{gem_details} for details). Again, this prevents the model to produce unrealistically small values of turbulence and wind speeds under strong atmospheric stability.

Two distinct groups of empirical stability functions are tested for both stable/unstable regimes: those from \citeA{DELAGE1997}, here denoted as $DEL$, and those from \citeA{Beljaars1991}, denoted as $BEL$. The comparison between {\bf GEM51-C-DEL} and {\bf GEM51-C-BEL} allows to isolate the effect of changing the stability functions. The analytical form of both empirical functions is given in Sect.~1 of the Supporting Information (SI). 

Two different land-surface models (LSMs) are used to represent surface processes: the Interaction Soil Biosphere Atmosphere \cite{ISBA,belair_operational_2003} and the version 3.6 of the Canadian Land Surface Scheme \cite{CLASS1,verseghy_canadian_2000}. The choice of the LSM can affect wind speeds in multiple ways. A direct impact of the LSM is through the calculation of the roughness length ($z_0$) that is used in Eq. \ref{log_profile}. An indirect impact is related with the calculation and partition of sensible and latent heat fluxes which affect the value of the Obukhov length ($L$) and the stability correction. The effect of the LSM is evaluated by comparing results from the {\bf GEM51-C-BEL} and the {\bf GEM51-I-BEL} simulations. 

The roughness length $z_0$ is calculated using three different methods. Two methods, one for the ISBA and another for the CLASS LSMs, use the concept of ``effective roughness length" by combining an orographic roughness length, $z_{(0,oro)}$, dependent on the subgrid-scale orography, and a vegetation roughness length, $z_{(0,veg)}$, dependent on the surface type (e.g., forest or grass). A third method, only used with the CLASS LSM, uses the turbulent orographic form drag (TOFD) scheme from \citeA{TOFD} to account for the effect of subgrid-scale orography, combined with a pure vegetation roughness length ($z_0 = z_{(0,veg)}$). The determination of $z_{(0,oro)}$ and $z_{(0,veg)}$ is based on high-resolution ($\sim 900$ m) elevation and land cover fields produced by the US Geological Survey \cite{usgs_gtopo30,usgs_land_cover}. While both ISBA and CLASS use the same input data, they employ different methods to determine the vegetation roughness length, as well as the way $z_{(0,veg)}$ is combined with $z_{(0,orog)}$ to obtain the effective roughness length $z_0$. 

Some calculation details and the resulting effective $z_0$ for CLASS, CLASS-TOFD, ISBA and AMF (estimated using the approach described in Section \ref{Sec:Data:Flux}) are provided in the SI (see Fig. S8-S9 and Table S1). We highlight some similarities and differences in the roughness length across different datasets. First, for all four datasets, the $z_0$ varies strongly between stations, with differences attaining about two orders of magnitude between the lowest and largest $z_0$. Second, the average $z_0$ from CLASS and CLASS-TOFD is closer to AMF than ISBA, with the later showing excessively large $z_0$ values. Third, there is no spatial correlation between the AMF $z_0$ and the CLASS, CLASS-TOFD or ISBA $z_0$. Additionally, Fig. S9 reveals that the effective roughness length in the CLASS simulations (both with and without TOFD) displays a pronounced annual cycle at certain stations. This cycle is characterized by lower values during winter, driven by snow cover, and higher values in summer.
Simulations using ISBA z0 do not show any dependence of roughness length with snow. Both ISBA and CLASS simulations show a weak annual cycle can be associated with grid points being covered by either rivers and lakes as freezing influences the effective roughness length since water and ice have different roughness characteristics. The same is true for grid points partially covered with open ocean, where the effective roughness length varies with wind speeds based on the Charnock relationship \cite{charnock_wind_1955}. When using TOFD, a grid point situated in a region with complex topography will be dominated by orographic component leading to a different annual cycle.

To separate the direct and indirect effects introduced by changing the LSM, a simulation was performed using the CLASS LSM with the ISBA roughness length $z_0$ (simulation {\bf GEM51-C-BEL-Iz0}). 
The comparison between {\bf GEM51-C-BEL} and {\bf GEM51-C-BEL-Iz0} allows to isolate the effect of the roughness length. 

In addition, to assess the effects of using an orographic drag in the resolved equations, a simulation was performed using the TOFD scheme. The comparison between {\bf GEM51-C-DEL} and {\bf GEM51-C-DEL-TOFD} allows to isolate the effect of changing the formulation of the roughness length.

\begin{table*}
\resizebox{1.\linewidth}{!}{
\ra{1.8}
\begin{tabular}{@{}llllll@{}}
\hline
& GEM   & Land Surface  &  Stability   &  Very stable  & Roughness   \\ 
Simulation Name    &  version  &  Model &   function  &   regime  &  Length \\ \hline
GEM50-C-DEL & 5.0  & CLASS        & DEL & $u\ge$0.1 m s$^{-1}$& $z_{0}^C$ \\ 
GEM51-C-DEL   & 5.1.1    &   CLASS     & DEL & $u\ge$0.01 m s$^{-1}$, $L \ge $20 m  & $z_{0}^C$ \\ 
GEM51-C-DEL-TOFD   &  5.1.1     &  CLASS     & DEL &  $u\ge$0.01 m s$^{-1}$, $L \ge $20 m  & $z_{(0,veg)}^C$, TOFD\\ 
GEM51-C-BEL   & 5.1.1     &  CLASS    &  BEL & $u\ge$0.01 m s$^{-1}$, $L \ge $20 m  &  $z_{0}^C$ \\ 
GEM51-C-BEL-Iz0   & 5.1.1     &  CLASS    &  BEL & $u\ge$0.01 m s$^{-1}$, $L \ge $20 m  &  $z^I_{0}$\\ 
GEM51-I-BEL   &  5.1.1    &   ISBA    & BEL & $u\ge$0.01 m s$^{-1}$, $L \ge $20 m  & $z_{0}^C$  \\ 
\hline
\end{tabular}
}
\caption{Details of the configuration of simulations, including the GEM model version, the land surface model (LSM), the stability functions, the strategy for very stable regimes and the formulation of the roughness length. For the very stable regime strategy, the threshold values are given for the wind speed of the lowest level ($u$) and the Obukhov length ($L$), both corresponding to those over land grid points. The superscript in $z_0$ values indicates whether the ISBA (I) or the CLASS (C) LSM is being used. In the case that TOFD is used, only the vegetation roughness length is retained.}
\label{gem_details}
\end{table*}

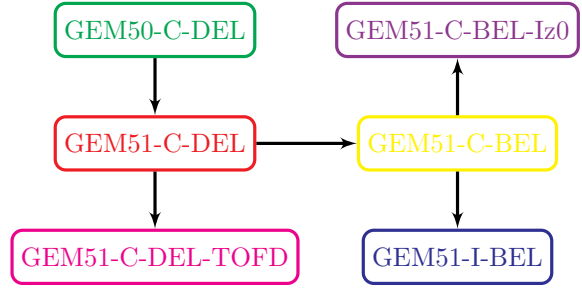
\begin{figure}
\center
\begin{tikzpicture}
\node [terminator, color=Green, fill=white, very thick] at (0,0) (GEM50-C-DEL) {GEM50-C-DEL};
\node [terminator, color=Yellow, fill=white, very thick] at (4,-1.5) (GEM51-C-BEL) {GEM51-C-BEL};
\node [terminator, color=Red, very thick] at (0,-1.5) (GEM51-C-DEL) {GEM51-C-DEL};

\node [terminator, color=Magenta, very thick] at (0,-3) (GEM51-C-DEL-TOFD) {GEM51-C-DEL-TOFD};
\node [terminator, color=Fuchsia, very thick] at (4,0) (GEM51-C-BEL-Iz0) {GEM51-C-BEL-Iz0};
\node [terminator, color=Blue, very thick] at (4,-3) (GEM51-I-BEL) {GEM51-I-BEL};

\path [connector, very thick] (GEM50-C-DEL) -- (GEM51-C-DEL);

\path [connector, very thick] (GEM51-C-DEL) -- (GEM51-C-DEL-TOFD);
\path [connector, very thick] (GEM51-C-DEL) -- (GEM51-C-BEL);
\path [connector, very thick] (GEM51-C-BEL) -- (GEM51-I-BEL);
\path [connector, very thick] (GEM51-C-BEL) -- (GEM51-C-BEL-Iz0);

\end{tikzpicture}
\caption{Flowchart of simulations where each node represents a distinct simulation configuration. Connections between nodes indicate a single degree of change in the configuration from one simulation to the next.}
\label{simulations_chart}
\end{figure}

Figure \ref{simulations_chart} illustrates the the degree of similarity between the configuration of all 6 simulations. Note that each simulation has a corresponding twin simulation that differs by only one configuration change, allowing for a controlled comparison of the effects of individual changes.

\subsection{AmeriFlux data and height adjustment} 
\label{Sec:Data:Flux}

AmeriFlux (denoted as AMF) is a network of sites measuring air-surface fluxes of CO$_2$, water, and energy in North, Central and South America. AMF data contains high-temporal resolution measurements of a range of variables near and at the surface including soil, atmospheric, and biological measurements. In particular, AMF provides direct measurements of atmospheric vertical turbulent fluxes acquired using eddy-covariance flux towers \cite{https://doi.org/10.1046/j.1365-2486.2003.00629.x}. 

We use the AmeriFlux BASE under the CC-BY-4.0 data use policy that includes a total of 311 sites across South, Central, and North America. We select 27 stations within the CORDEX-NA domain (see Fig.~\ref{fig:stationmap} for a map of the stations) which include wind speed ($u$), friction velocity ($u_*$), and the Monin-Obukhov stability parameter ($\zeta$) in the period 2015-2020. While most stations provide measurements every 30 minutes, hourly values were calculated to match the hourly resolution of the simulated data. Hourly values are calculated using the average of the two 30-minute values when both are available or by simply considering a single 30-minute value when only one value is available. Only stations that contain at least one year of hourly measurements over the entire 6-year period are considered. In addition, we excluded stations for which we do not have instrument height information, as it is needed to perform the height adjustment. Considering all stations, a total of 756824 measurements are obtained, which is equivalent to about 86 years of hourly data. More information on the individual stations, including their latitude, longitude, measurement height, name, associated DOI, and total number of measurements, is available in Tab.~S2 of SI.

Since the stations have varying measurement heights, we adjusted the wind speed and stability parameter measurements to the standard height of 10 m, which is commonly used to report near-surface wind speeds and is also the standard height used by the model. We only adjust the wind speeds and the stability parameter as we assume, as usually done within the MOST, that turbulent fluxes are constant within the surface layer \cite{Garratt1994}.

The adjustment of the stability parameter is given by,
\begin{equation}
	\label{eq:zlCorr}
	\zeta_{10} = \frac{10}{h}\zeta_{h},
\end{equation}
where $h$ is the height of the measurement. As for the wind speed, we first estimate the roughness length at each station by using the log-wind relationship in neutral conditions ($\zeta_{10} \sim 0$). We assume that neutral stability conditions are obtained when the modulus of the stability parameter ($|\zeta_{10}|$) is lower than 0.02 \cite{Drawl2014},

\begin{equation}
z_0= \frac{z}{\exp(\frac{u \kappa}{u_*})-1},
\end{equation}
\begin{equation}
	\label{eq:uCorr}
	u_{10} = u_h + \frac{u_*}{\kappa} \left[\log\frac{h}{10}+(\Psi_{10} - \Psi_{h})\right].
\end{equation}
We use BEL stability functions for both the stable and unstable branches. Furthermore, we adjust the stability parameter length ($\zeta$) to a minimum value of -5 and a maximum value of 10 so the stability correction is within a reasonable range \cite{stabilit_mo}. The number of times that the stability parameter is outside the [-5,10] range varies between stations but it is at most 4$\%$ of the time.

An alternative adjustment is performed using the $(z+z_0)$ formulation instead of $(z)$:
\begin{equation}
	\label{eq:uCorr_wz}
	u_{10} = u_h + \frac{u_*}{\kappa} \left[\log\frac{h+z_0}{10}+(\Psi_{10} - \Psi_{h})\right].
\end{equation}

\begin{figure}
    \center
	\includegraphics[width=0.75\textwidth]{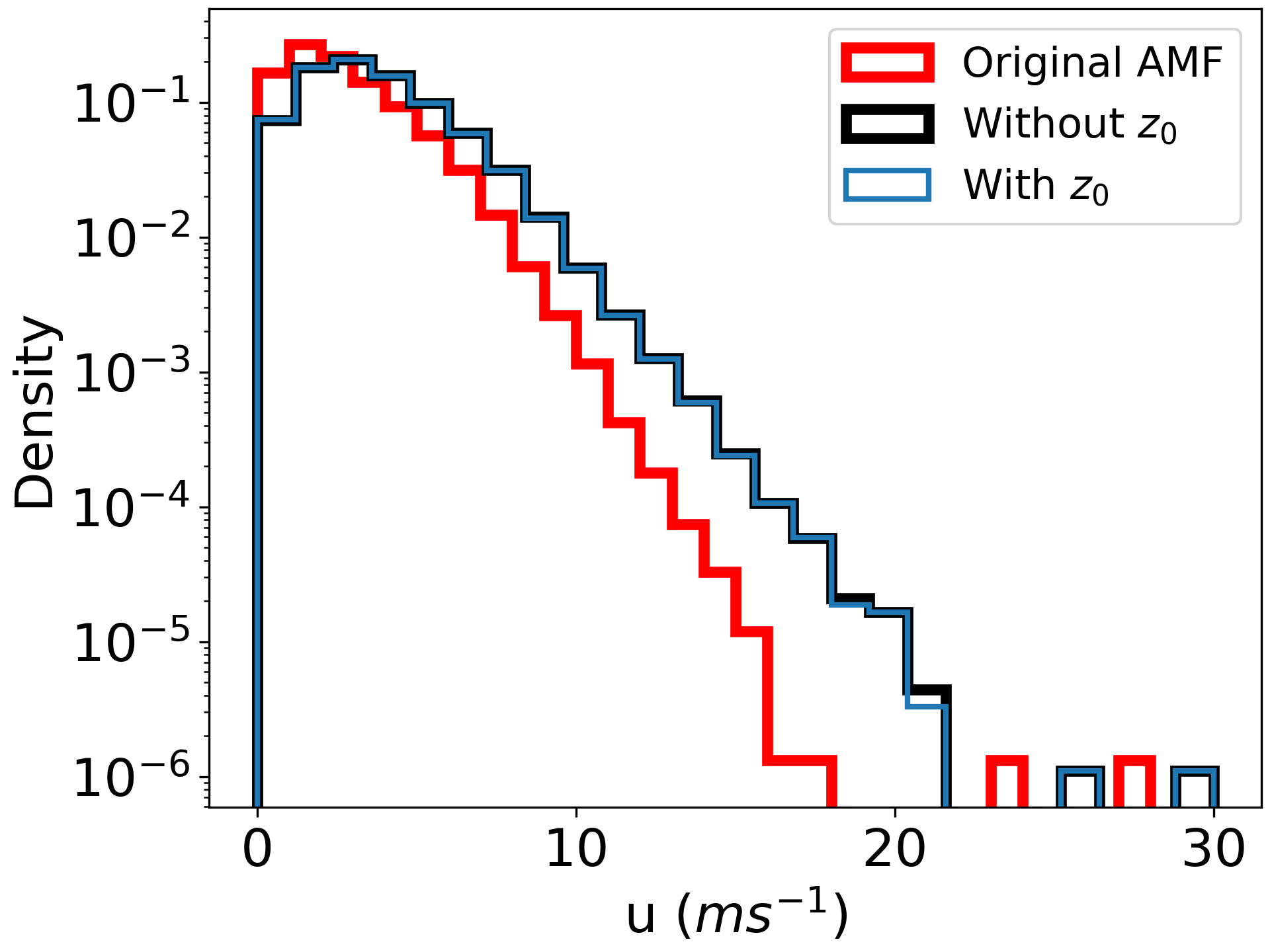}
	\caption{Density histogram of wind speeds for the original AMF stations (in red) and for two adjusted versions of AMF (in black and blue). Adjusted versions differ on the use of $z_0$ in the numerator of the logarithm (see Eqs. \ref{eq:uCorr} and \ref{eq:uCorr_wz}).}
	\label{FIG:uCorrDist}
\end{figure}

In Fig.~\ref{FIG:uCorrDist} we show the distribution on the original AMF wind speeds and two adjusted versions at 10 m (with and without $z_0$). Adjusted 10-m wind speeds distributions are somewhat shifted towards larger wind speeds as the mean height of wind speeds measurements is $\sim 3.5$ $m$ lower than the reference height of 10 m. The shift is substantial and nearly doubled the mean wind speed of the original dataset, showing the key importance of performing the height correction before comparing observations and simulated data. Results for the two proposed methods, with or without $z_0$, show little to no differences. In the rest of the analysis, the corrected data without $z_0$ has been used. 

\section{Methods}
\label{Sec:Methods}

\subsection{Comparing AMF and CRCM6-GEM5 output data}
\label{Sec:Methods1}
The CRCM6-GEM5 model outputs 10-m wind speed $(u_{10})$, friction velocity $(u_*)$, and Obukhov length scale $(L)$ every hour over the entire CORDEX-NA domain and during the whole 2015-2020 period. The comparison between each CRCM6-GEM5 simulation and the AMF adjusted measurements is performed by:
 \begin{enumerate}
        \item selecting the model grid point closest to each AMF station (see Fig.~\ref{fig:stationmap}),
        \item selecting only the modeled hourly time steps for which the station has measurements (see section \ref{Sec:Data:Flux}).
    \end{enumerate}
This procedure guarantees that the number of values available from both simulations and observations is equal and that both datasets sample the same calendar days. As it will be discussed in detail in the next section, comparisons are not made directly between single model grid points and stations. Although comparing individual model grid points directly with stations is sometimes practiced in model evaluation studies \cite{Nelli_2020}, it can be problematic due to differences in observed and simulated roughness lengths.

\subsection{A hierarchy of wind speeds error metrics}
\label{Sec:Methods2}
A hierarchy of error metrics is employed to describe $u_{10}$ errors, progressing from basic error metrics to more advanced and sophisticated measures. All error metrics are derived using different ways of grouping and averaging observed and simulated $u_{10}$. 

For observations and simulations, the $u_{10}$ data can be represented as $u=u_{x,t}$, with the index $x$ representing individual stations and the index $t$ representing hourly time steps at the station $x$. Note that the number of hours available at each station $x$ is generally different. Individual time series can then be mapped into specific hours of the day ($h$) and months ($m$) as $u_{x,t} \rightarrow u_{h,m,n_{h,m}}$, with the index $h$ ranging from 1 to 24, the index $m$ from 1 to 12 and the index $n_{h,m}$ from 1 to $N_{h,m}$, depending on the number of occurrences for the hour $h$ and the month $m$. The average $u_{10}$ across all stations and times can then be calculated as (for simplicity, the subindex $10$ has been removed):
\begin{equation}
	\overline{u} = \frac{1}{{T \cdot X}}\sum_x \sum_t u_{x,t(x)} = \frac{1}{{T \cdot X}}\sum_h \sum_m \sum_{n_{h,m}} u_{h,m,n_{h,m}}= \frac{1}{{T \cdot X}}\sum_h \sum_m N_{h,m} \cdot \overline{u_{h,m}},
	\label{eq::cycles}
\end{equation}
where $T$ is the average number of hourly measurements across all stations, $X$ is the number of stations (i.e., $X=27$) and $T \cdot X$ gives the total number of observations considering all stations ($T \cdot X =756824$). We define:
\begin{equation}
	\overline{u_{h,m}} = \frac{1}{N_{h,m}}\sum_{n_{h,m}} u_{h,m,n_{h,m}}
\end{equation}

Calculating $\overline{u}$ from observed and simulated data allows us to estimate multiple error metrics. The simplest error metric is given by the absolute error of the mean bias:

\begin{equation}
	\epsilon^{s}_{bias}= \left| \overline{u^{s}}  - \overline{u^{o}} \right|
	\label{eq::e_bias}
\end{equation}
where $s$ denotes the sth simulation and $o$ the AMF observations. 

The mean wind speed can also be expressed using the mean diurnal ($\overline{u_h}$) and annual ($\overline{u_m}$) cycle of wind speeds:
\begin{equation}
	\overline{u} = \frac{M}{{T \cdot X}} \sum_h \frac{1}{{M}} \sum_m N_{h,m} \cdot \overline{u_{h,m}} =   \frac{M}{{T \cdot X}} \sum_h \overline{u_h}
	\label{eq::diu_cycle}
\end{equation}

\begin{equation}
	\overline{u} =  \frac{H}{{T \cdot X}} \sum_m \frac{1}{{H}} \sum_h N_{h,m} \cdot \overline{u_{h,m}} =   \frac{H}{{T \cdot X}} \sum_m \overline{u_m}
	\label{eq::ann_cycles}
\end{equation}

The second and third error metrics are given by the sum of absolute errors from the diurnal and annual cycles respectively:
\begin{equation}
	\epsilon^{s}_{diu}=\frac{M}{{T \cdot X}} \sum_h \left | \overline{u^{s}_{h}}  - \overline{u^{o}_{h}} \right|
	\label{eq::e_diu}
\end{equation}
\begin{equation}
	\epsilon^{s}_{ann}=\frac{H}{{T \cdot X}} \sum_m \left | \overline{u^{s}_{m}}  - \overline{u^{o}_{m}} \right|
	\label{eq::e_ann}
\end{equation}

Defined this way, $\epsilon^s_{ann} \ge \epsilon^s_{bias}$ and $\epsilon^s_{diu} \ge \epsilon^s_{bias}$ and the two error metrics quantify errors associated with the forced variability resulting from the diurnal and the annual cycles respectively. Moreover, a metric quantifying simultaneously diurnal and annual cycle errors can be calculated as: 
\begin{equation}
	\epsilon^s_{ann,diu} =\frac{1}{{T \cdot X}}\sum_h \sum_m N_{h,m} \cdot \left |  \overline{u^{s}_{h,m}} - \overline{u^{o}_{h,m}} \right |, 
	\label{eq::e_ann_diu}
\end{equation}
with $\epsilon^s_{ann,diu} \ge \epsilon^s_{ann} \ge \epsilon^s_{bias}$ and $\epsilon^s_{ann,diu} \ge \epsilon^s_{diu} \ge \epsilon^s_{bias}$. 

To quantify errors associated with higher frequency variability (e.g., hourly variability), wind speeds are mapped according to a range of wind speed intensities: $u_{x,t(x)} \rightarrow u_{k,n_{k}}$ where the index $k$ denotes the wind speeds within the bin $B_k = [u_k,u_{k+1}]$. The mean wind speed is given by:
\begin{equation}
	\overline{u} = \frac{1}{T \cdot X} \sum_k \sum_{n_k} u_{k,n_{k}}  = \frac{1}{T \cdot X} \sum_k N_k \overline{u_k}
	\label{eq::cycles_bin}
\end{equation}

with $\overline{u_k}$ the mean wind speed and ${N_k}$ the number of occurrences at the intensity bin $B_k$. An error metric, identified as the frequency-intensity (fi) error metric, can be defined as:
\begin{equation}
	\epsilon^s_{fi}=\frac{1}{T \cdot X} \sum_k  \left | N^{s}_k \overline{u^{s}_k} - N^{o}_k \overline{u^{o}_k} \right | \sim \frac{1}{T \cdot X} \sum_k  \left | N^{s}_k  - N^{o}_k \right | \cdot \overline{u^{o}_k} 
	\label{eq::e_bin}
\end{equation}
because $\overline{u^{s}_k} \sim \overline{u^{o}_k}$. The error metric verifies that $\epsilon^s_{fi} \ge \epsilon^s_{bias}$.

All five error metrics defined above make use of a single variable: $u_{10}$. As discussed in Section \ref{Sec:Data:Sim}, multiple variables affect $u_{10}$, and can be a source of $u_{10}$ errors. Specifically, $u_{10}$ explicitly depends on the friction velocity $u^{*}$ and the stability parameter $\zeta$, two variables that are measured and available in the AMF dataset. To account for the influence of $u_*$ and $\zeta$ in wind speed errors, we first discretize (i.e., bin) $u_*$ and $\zeta$ into different categories (denoted here as “regimes”) with $u_*$ between [$u_{*i}$,$u_{*,i+1}$] and $\zeta$ between [$\zeta_j$,$\zeta_{j+1}$]. Based on hourly values of $u_{*,x,t(x)}$ and $\zeta_{x,t(x)}$, wind speeds ($u_{x,t(x)}$) are mapped into the $(u_*,\zeta)$ regimes phase space so that $u_{x,t(x)} \rightarrow u_{i,j,n_{i,j}}$. Each regime $(i, j)$ thus contains a total number of occurrences $N_{i,j}$ and an average wind speed $\overline{u_{i,j}}$ so the mean wind speed can be estimated as,

\begin{equation}
	\overline{u} =  \frac{1}{T \cdot X}  \sum_i \sum_j \sum_{n_{i,j}} u_{i,j, n_{i}}  =  \frac{1}{T \cdot X}  \sum_i \sum_j N_{i,j} \overline{u_{i,j}}
	\label{eq::cycles_regime}
\end{equation}

A stability/friction velocity ($\zeta,u_*$) regime error metric can be defined as the difference between simulated and observed values,
\begin{eqnarray}
	\epsilon^{s,tot}_{reg}& =&   \frac{1}{T \cdot X}  \sum_i \sum_j \left ( \left ( \Delta N_{i,j} \cdot \overline{u^{o}_{i,j}}\right ) +\left ( \Delta \overline{u_{i,j} \cdot N^{o}_{i,j}} \right ) +\left ( \Delta \overline{u_{i,j}} \cdot \Delta N_{i,j} \right ) \right ) \nonumber \\
        & =&   \sum_i \sum_j \left ( \left ( \Delta p_{i,j} \cdot \overline{u^{o}_{i,j}}\right ) +\left ( \Delta \overline{u_{i,j} \cdot p^{o}_{i,j}} \right ) +\left ( \Delta \overline{u_{i,j}} \cdot \Delta p_{i,j} \right ) \right ) \nonumber \\
        & =&  \epsilon_{reg}^{s,p}+\epsilon_{reg}^{s,u}+\epsilon_{reg}^{s,p \cdot u}
	\label{eq::e_regime}
\end{eqnarray}

That is, the total regime error ($\epsilon^{s,tot}_{reg}$) is expressed as the sum of three terms:
\begin{enumerate}
\item Frequency error: the term $\epsilon_{reg}^{p}$ quantifies the degree to which the simulation produces the right frequency of regimes.
\item Mean wind speed error: the term $\epsilon_{reg}^{u}$ quantifies the degree to which the simulation produces the right mean wind speed at each regime.
\item Interaction error: the term $\epsilon_{reg}^{p \cdot u}$ quantifies the degree to which errors $\Delta N$ and $\Delta \overline{u}$ are correlated (e.g., the error in the frequency of regimes is associated with the error in mean wind speeds).
\end{enumerate}
Using the decomposition above, we can also define an error that penalizes simulations showing important compensating errors as follows:
\begin{equation}
        \epsilon^s_{reg} = |\epsilon_{reg}^{s,p}|+|\epsilon_{reg}^{s,u}|+|\epsilon_{reg}^{s,p \cdot u}|
	\label{eq::e_regime_abs}
\end{equation}

Defined this way, $\epsilon^s_{reg}$ is equal to zero if and only if the simulation produces the right wind speed for different friction velocities and stability regimes, and $\epsilon^s_{reg}$ does not allow for the compensation of errors across regimes. For example, a simulation that would have the right mean wind speed while constantly being in the wrong $(u_*,\zeta)$ regime would be penalized. 

To emphasize errors in the variability of wind speeds, error metrics are also calculated using a simple unbiased version of simulated wind speeds that removes the overall mean wind speed $\overline{u}$, thus replacing $u$ by $u-\overline{u}$. Unbiased errors are denoted using a prime (i.e., $\epsilon^{\prime}$) and are always lower than biased errors (i.e., they verify that $\epsilon^{\prime} \le \epsilon$). 

Finally, to compare all errors in a single plot and assess the performance of simulations ranking, normalized errors are calculated by dividing each error by the maximum across all simulations:
\begin{equation}
	\hat{\epsilon}^{i} = \frac{\epsilon^{i}}{max( \epsilon^{i})}
\end{equation}

\section{Results}
\label{Sec:Results}

\subsection{Hourly and monthly mean wind speed errors}
\label{Sec:Results2}
\begin{figure}
    \centering
    \includegraphics[width=0.95\textwidth]{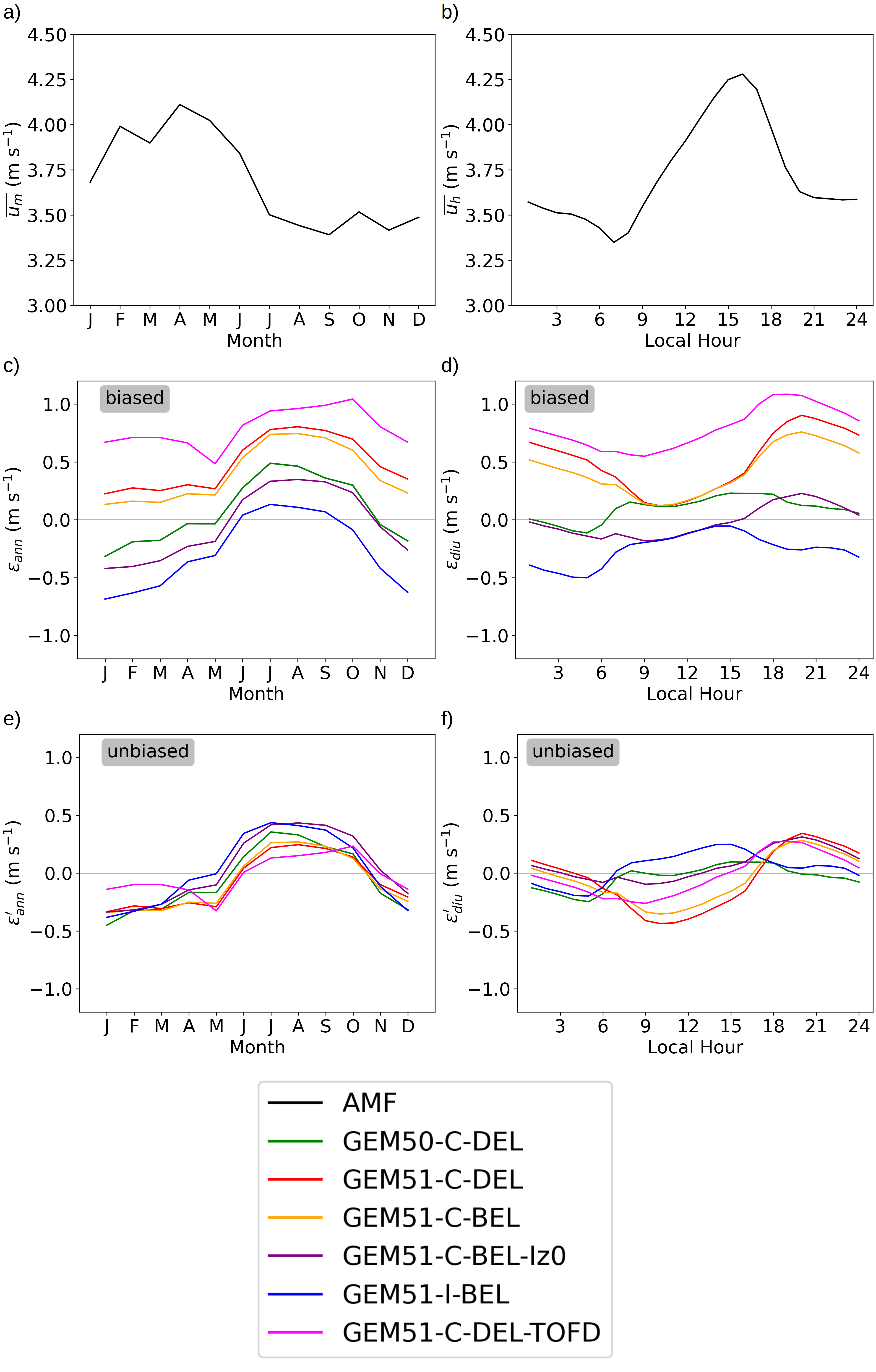}
    \caption{AMF annual (a) and diurnal (b) cycles averaged across all stations. Simulation errors are shown for the annual (left) and diurnal (right) cycles (left) and for the biased (middle panels) and unbiased (bottom panels) errors.}
    \label{fig:res1}
\end{figure}

Top panels in Fig.~\ref{fig:res1} show the annual and diurnal mean cycles of 10-m wind speed from the AMF data. The AMF mean wind speed $\overline{u}$ across all stations and times is 3.7 m s$^{-1}$. The annual cycle is relatively weak with monthly variations within 15$\%$ of $\overline{u^{o}_{m}}$ ($\sim$0.6 m s$^{-1}$) and a maximum value in spring. The diurnal cycle is somewhat stronger than the annual cycle with variations attaining 20$\%$ of $\overline{u^{o}_{h}}$ ($\sim$1.0 m s$^{-1}$), with a minima of 3.3 m s$^{-1}$ at $\sim$07H00 local time and a maxima of 4.3 m s$^{-1}$ at $\sim$ 17H00 local time. 

Simulations mean errors (Fig.~\ref{fig:res1}~c-f) for $\overline{u}$ vary between -0.25 m s$^{-1}$ and 0.80 m s$^{-1}$ for the {\bf GEM51-I-BEL} and {\bf GEM51-C-DEL-TOFD} simulations respectively. Fig.~\ref{fig:res1}~c-f shows that errors in the diurnal and annual cycles (errors in $\overline{u^{o}_{h,m}}$) are dominated by errors in the mean wind speed $\overline{u}$ and there is a strong reduction in the error when considering the unbiased version of errors (bottom panels in Fig.~\ref{fig:res1}). For the annual cycle, most simulations, with maybe the exception of {\bf GEM51-C-DEL-TOFD}, show an overestimation of wind speeds during the summer/fall and an underestimation during winter/spring. For the diurnal cycle, error differences between simulations are substantial, and the simulations are highly sensitive to the choice of model version and of roughness length values. In particular, we note that the mean roughness values (see Tab.~S1) have a large impact on the on the biased annual errors. We note that the roughness length also has an annual cycle (see Fig.~S9) though it is not obvious how the cycle is related to the annual error cycle. For instance, simulations using GEM51 usually show an underestimation in the early afternoon and an overestimation in the late afternoon while the simulations using the ISBA $z_0$ show the opposite behavior.

\begin{figure}[H]
    \centering
    \includegraphics[width=0.95\textwidth]{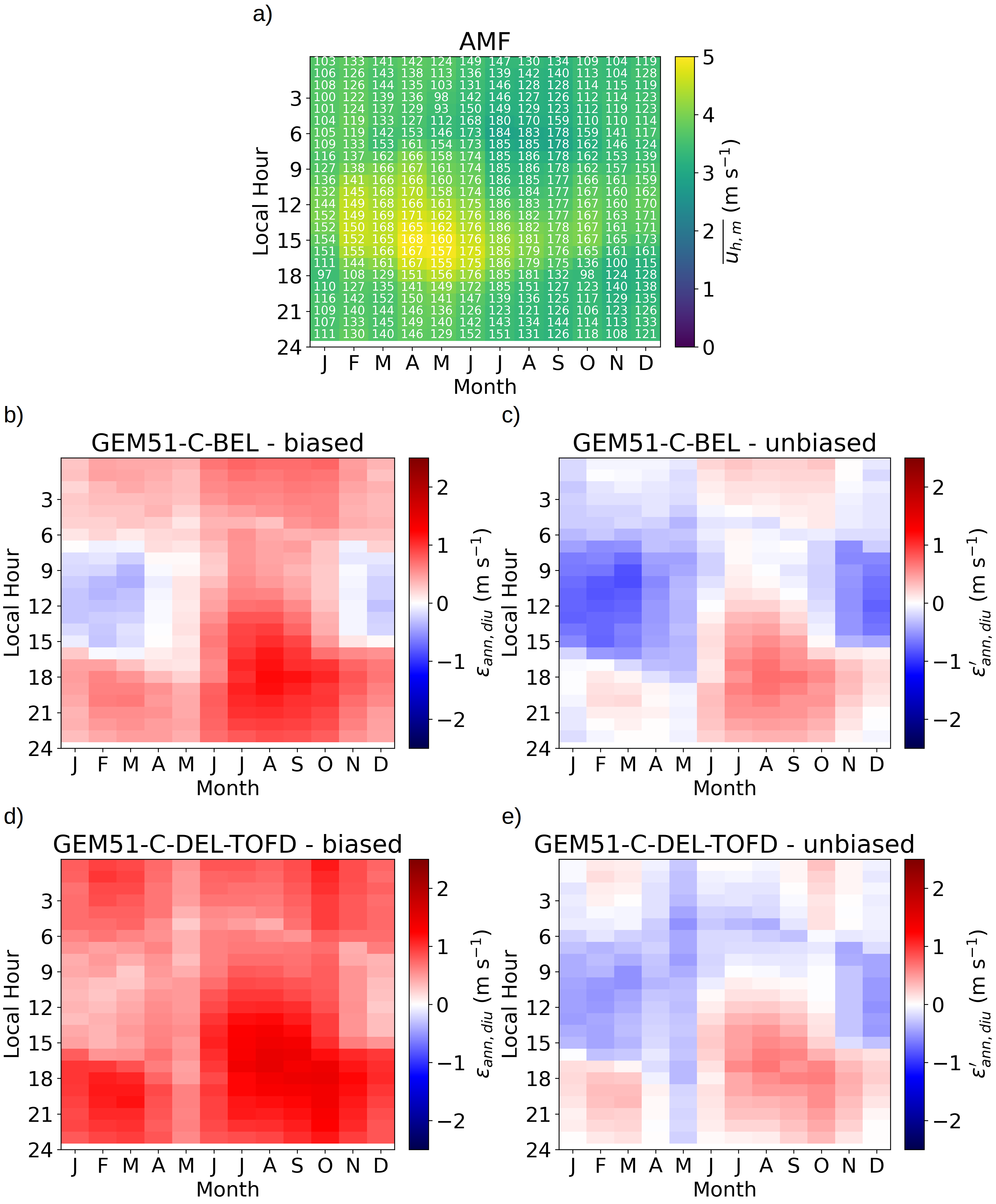}
    \caption{Combined monthly and hourly wind speed cycles from AMF (a), along with the biased (b,d) and unbiased (c,e) errors for GEM50-C-DEL (b,c) and GEM51-C-DEL (d,e).}
    \label{fig:res2}
\end{figure}

Figure~\ref{fig:res2} presents the combined monthly and hourly mean wind speeds of the AMF data (Figure~\ref{fig:res2}a) and biased (Fig.~\ref{fig:res2}b,d) and unbiased errors (Fig.~\ref{fig:res2}c,e) for two selected simulations ({\bf GEM50-C-DEL} and {\bf GEM51-C-DEL}) that display some error patterns observed in other simulations (biased and unbiased errors for all simulations are shown in Fig.~S1, S2 of the SI). Similarly to Sect.~\ref{Sec:Results2}, we observe both the diurnal and annual cycles in the AMF data, with the highest wind speeds of about 5 m s$^{-1}$ occurring in the afternoon during the months of April and May, and the lowest wind speeds of about 2 m s$^{-1}$ occurring in the early morning in August-September. 

Figure~\ref{fig:res2}b, d shows the biased hourly-monthly mean wind speed error for {\bf GEM50-C-DEL} and {\bf GEM51-C-DEL}, respectively. Both simulations show similar error patterns during the months of June to October with a large overestimation of the wind speeds. Between November and May, {\bf GEM50-C-DEL} shows an overall underestimation of wind speeds which is in contrast with {\bf GEM51-C-DEL} which shows a general overestimation except during the daytime when it shows a light underestimation. Once the biases are removed (Fig.~\ref{fig:res2}c,e), the magnitude of errors changes little for {\bf GEM50-C-DEL}, showing a similar error pattern. 
{\bf GEM51-C-DEL}, on the other hand, has a significant reduction in error for the months of June through October, and an increase in underestimation for the other months. 

\subsection{Hourly frequency-intensity wind speed errors}
\label{Sec:Results4}
\begin{figure}
    \centering
    \includegraphics[width=0.95\textwidth]{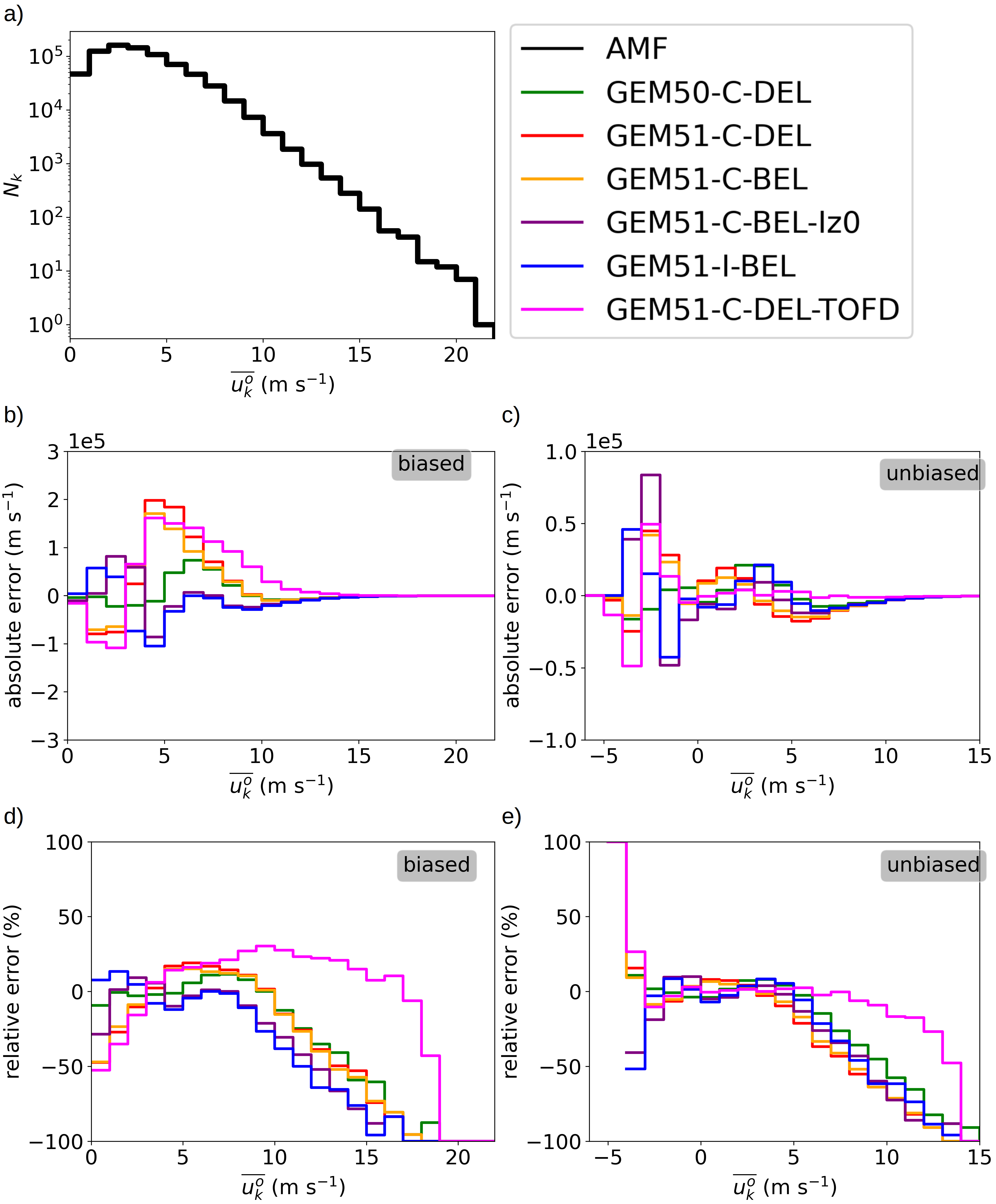}
    \caption{AMF wind speed frequency-intensity distribution (a), the absolute frequency-intensity errors (b,c) and the relative frequency-intensity errors (d,e) for the biased (b,d) and the unbiased (c,e) simulations. Absolute errors are calculated as $(N^{s}_k - N^{o}_k) \cdot \overline{u^{o}_k}$ while relative errors are defined as $(N^{s}_k - N^{o}_k)/(N^{s}_k + N^{o}_k)$.}
    \label{fig:res3}
\end{figure}

Figure~\ref{fig:res3}a shows the frequency-intensity distribution of the AMF wind speeds over all stations. For wind speeds greater than approximately 4 m s\(^{-1}\), the distribution shows an exponential decrease in the frequency of hourly wind speeds as intensity increases, with the highest observed values reaching about 22 m s$^{-1}$.
Figure~\ref{fig:res3}b,c shows the absolute error, which is the frequency error weighted by the mean observed intensity (i.e., $(N^{s}_k - N^{o}_k) \cdot \overline{u^{o}_k}$), for the biased (Fig.~\ref{fig:res3}b) and the unbiased (Fig.~\ref{fig:res3}c) simulations. For both biased and unbiased simulations, total errors are dominated by errors occurring at wind speeds between 0 and about 10 m s$^{-1}$, with a peak at about 5 m s$^{-1}$. Biased simulations with the $z_0$ from CLASS usually underestimate the frequency of wind speeds below 5 m s$^{-1}$ and overestimate the frequency of wind speeds above 5 m s$^{-1}$. Biased simulations with the $z_0$ from ISBA have lower errors and the opposite behavior: an overestimation of low wind speeds and an underestimation of high wind speeds. Removing the mean wind speed from simulations and AMF data (Figure~\ref{fig:res3}c) leads to much lower errors for all simulations (nearly 3 times smaller), especially for the {\bf GEM51-C-DEL-TOFD} that shows very low errors for all wind speed intensities.

Fig.~\ref{fig:res3}d,e shows the relative errors, at each bin, of individual simulations with respect to AMF. For the biased simulations (Fig.~\ref{fig:res3}d), errors are generally within 50\% for wind speeds below 10 m s$^{-1}$. All simulations but {\bf GEM51-C-DEL-TOFD} substantially underestimate the frequency of wind speeds larger than 10 m s$^{-1}$ and they do not produce any value larger than about 17 m s$^{-1}$. For the unbiased simulations (Fig.~\ref{fig:res3}e), we note that relative errors are much smaller and within about 10 \% for wind speeds below 5 m s$^{-1}$ but they increase as we consider higher wind speeds, except for the {\bf GEM51-C-DEL-TOFD} simulation.

\subsection{Wind speed errors conditioned on stability and friction velocity}

\begin{figure}
    \centering
    \includegraphics[width=0.95\textwidth]{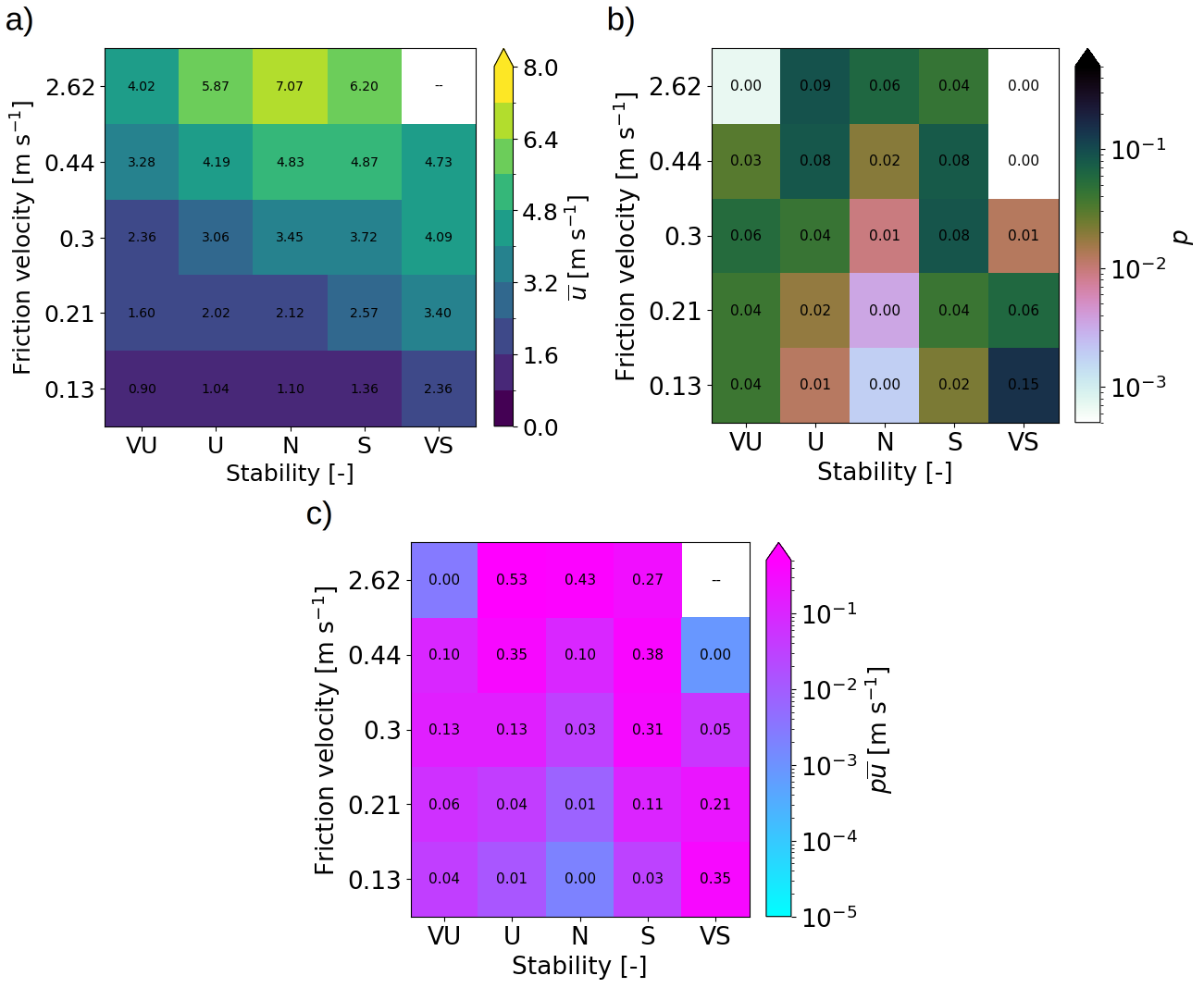}
    \caption{(a) AMF mean wind speed for different regimes of friction velocity and stability and (b) the probability of occurrence of each regime. Panel c) shows the  wind speed per regime, calculated as the product between the mean wind speed and the probability of occurrences. See Eq. \ref{eq::cycles_regime} for more details.}
    \label{fig:res4}
\end{figure}

Figure~\ref{fig:res4} shows the AMF mean wind speed ($\overline{u_{i,j}}$; Fig.~\ref{fig:res4}a) and the probability of occurrences ($p=N_{i,j}/N$, Fig.~\ref{fig:res4}b) for different stability and friction velocity regimes. Results show that the AMF mean wind speed increases as the friction velocity increases, with $\overline{u_{i,j}}$ values below 1 m s$^{-1}$ for $u_*<$0.13 m s$^{-1}$ and $\overline{u_{i,j}}$ values attaining about 7 m s$^{-1}$ for $u_*>$2.6 m s$^{-1}$. Fig.~\ref{fig:res4}a) also shows that mean wind speed usually increases as the surface layer becomes more stable, although this is only observed for weak and moderate $u_*$ values. The number of occurrences for different regimes (Fig.~\ref{fig:res4}b) shows strong variability with very unstable or very stable regimes mostly associated with low to moderate friction velocities, and unstable, neutral, and stable conditions occurring with high friction velocities. Note that, according to AMF data, the surface layer is never in a very stable regime with very high friction velocities. 
The resulting wind speed per regime (Fig.~\ref{fig:res4}c), obtained from the product between the mean wind speed and the occurrence ($\overline{u_{i,j}} \cdot p_{i,j}$), is dominated by the probability of occurrences, and shows the highest values for regimes with large friction velocities under unstable and neutral conditions. Note that the sum of the wind speed per regime across all regimes gives the AMF mean wind speed of 3.7 m s$^{-1}$.

\begin{figure}
    \centering
    \includegraphics[width=0.95\textwidth]{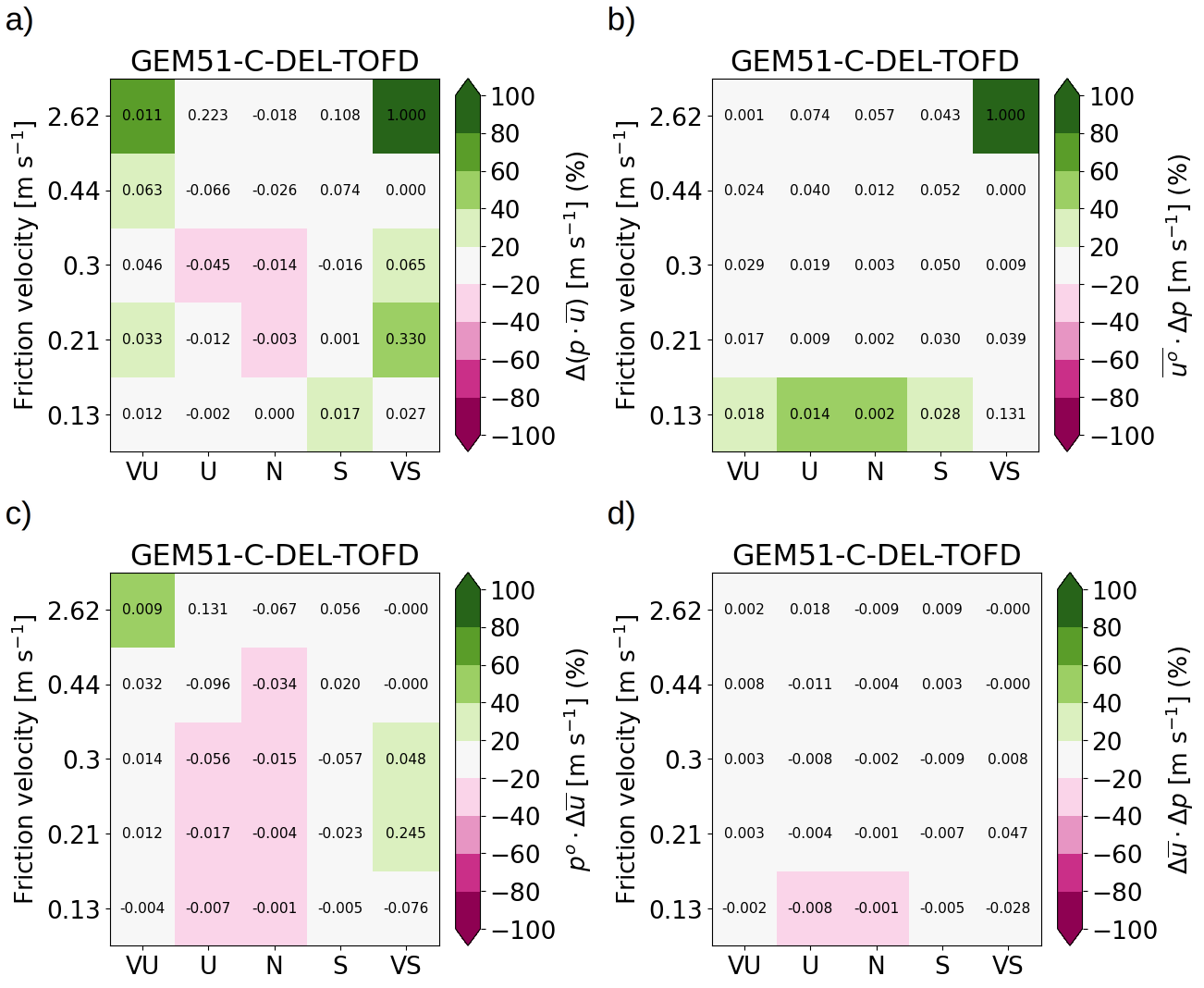}
    \caption{Decomposed regime errors for the {\bf GEM51-C-DEL-TOFD} simulation (see Eq. \ref{eq::e_regime}). (a) shows the total wind speed error ($\epsilon^s_{reg}$) that results from the sum of the decomposition terms: (b) the frequency error ($\epsilon_{reg}^{s,p}$), (c) the intensity error ($\epsilon_{reg}^{s,u}$), and (d) the interaction error ($\epsilon_{reg}^{s,p \cdot u}$). Colors indicate the relative errors (in \%) while absolute errors are annotated for each regime. Errors were calculated using the biased simulation.}
    \label{fig:res5}
\end{figure}

Figure~\ref{fig:res5} shows the different terms of the regime-conditioned error decomposition (see Eq. \ref{eq::e_regime}) for the {\bf GEM51-C-DEL-TOFD} simulation. The other five simulations are shown in Figs. S3-S7. Fig.~\ref{fig:res5}a) shows the total error in each regime ($\epsilon^{s,tot}_{reg}$) that is obtained as the sum of the frequency error ($\epsilon_{reg}^{s,p}$; Fig.~\ref{fig:res5}b), the intensity error ($\epsilon_{reg}^{s,u}$; Fig.~\ref{fig:res5}c) and the interaction error ($\epsilon_{reg}^{s,p \cdot u}$, Fig.~\ref{fig:res5}d) terms. We see that the unstable and neutral cases have an overall negative bias which is dominated by errors in the intensity term (Fig.~\ref{fig:res5}c), indicating that {\bf GEM51-C-DEL-TOFD} underestimates the wind speed intensity in these regimes. 
The very unstable and stable cases, on the other hand, are primarily dominated by frequency error (Fig.~\ref{fig:res5}c) and show a positive bias. Notably, the simulation shows the largest errors for high friction velocity regimes (total error of about 1.3 m s$^{-1}$) and for very stable regimes (total error of about 1.4 m s$^{-1}$). The interaction error (Fig.~\ref{fig:res5}d) is generally small compared to the other terms.

\begin{figure}[H]
    \centering
    \includegraphics[width=0.95\textwidth]{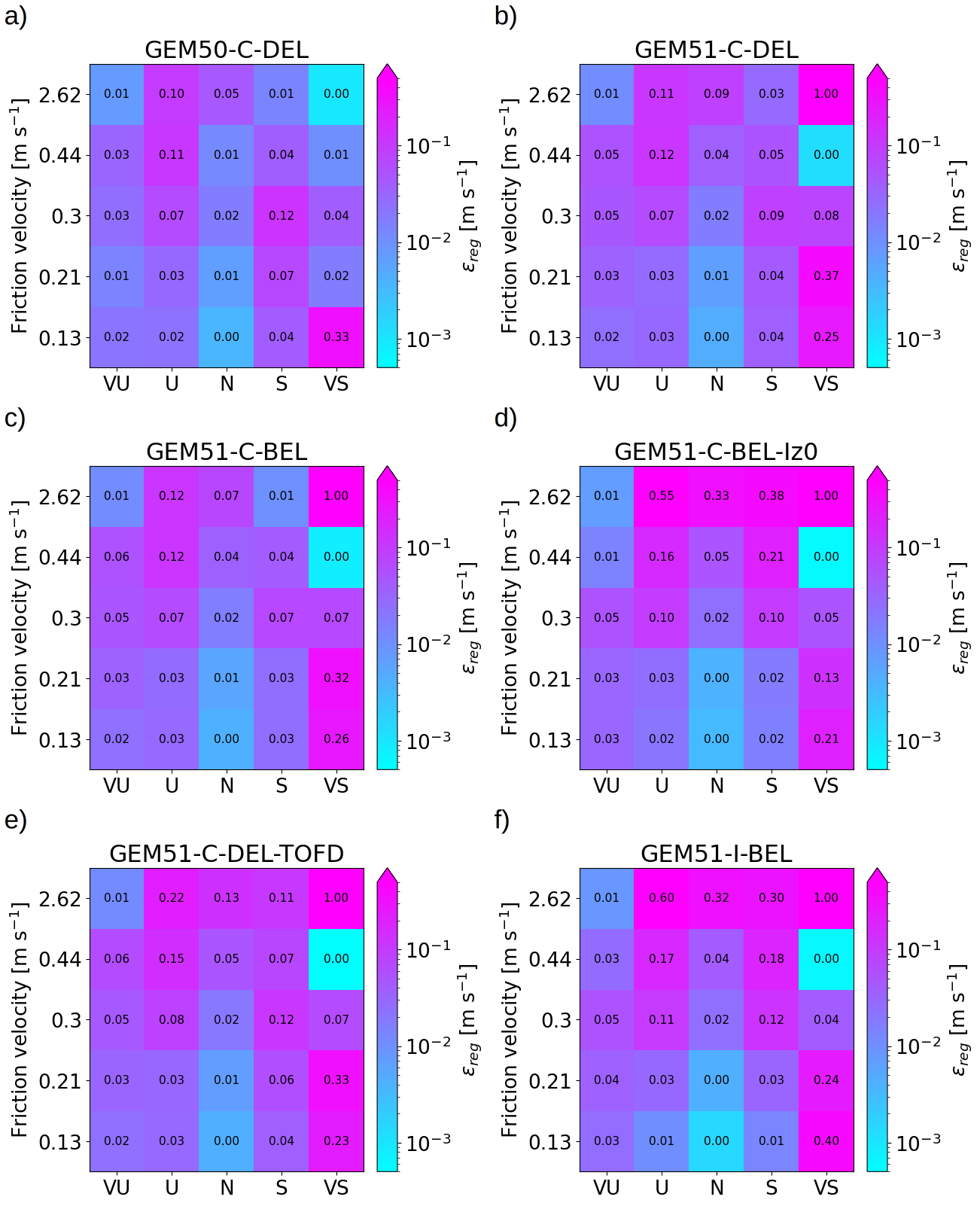}
    \caption{Absolute errors ($\epsilon_{reg}$) within each friction velocity and stability regime for individual simulations. Errors were calculated using Eq. \ref{eq::e_regime} and the biased simulations.}
    \label{fig:res6}
\end{figure}

The regime error ($\epsilon^{s}_{reg}$), as calculated using Eq. \ref{eq::e_regime_abs}, is presented in Fig.~\ref{fig:res6} for all six simulations. With the exception of the {\bf GEM50-C-DEL} simulation, all simulations show the largest errors for the very stable regime. Some simulations, including both simulations using the ISBA $z_0$, also show relatively large errors for high friction velocity regimes. Overall, the simulation {\bf GEM50-C-DEL} shows the lowest errors. All GEM51 simulations show a very large error in the large friction velocity, very stable regime and {\bf GEM50-C-DEL} seems to perform best in that regime.

\subsection{Synthesizing wind speed error metrics}
\begin{figure}[H]
    \centering
    \includegraphics[width=0.95\textwidth]{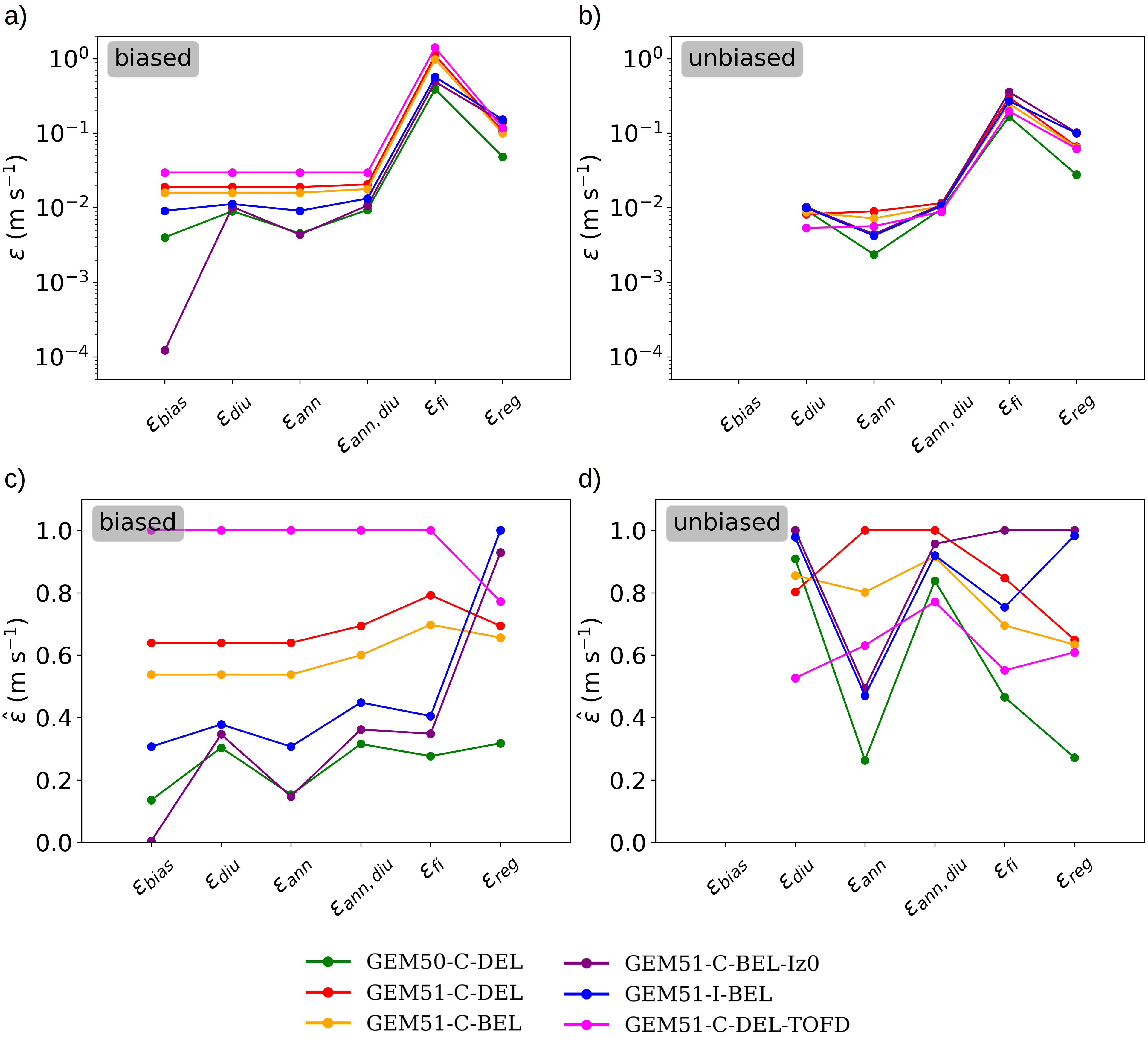}
    \caption{Combined biased errors (a), combined unbiased errors (b), combined normalized biased errors (c), combined normalized unbiased errors (d). }
    \label{fig:res7}
\end{figure}

We combine the results of all error metrics (see Section \ref{Sec:Methods2}) in Fig.~\ref{fig:res7} for both biased and unbiased (left and right panels respectively) and absolute and normalized (top and bottom panels respectively) errors. As expected, the hierarchy of error metrics holds verifying that $\epsilon_{ann,diu} \ge \epsilon_{ann} \ge \epsilon_{bias}$, $\epsilon_{ann,diu} \ge \epsilon_{diu} \ge \epsilon_{bias}$, $\epsilon_{fi} \ge \epsilon_{bias}$ and $\epsilon_{reg} \ge \epsilon_{bias}$.

For biased errors (left panels in Fig.~\ref{fig:res7}), the {\bf GEM50-C-DEL} simulation shows the best overall performance across most metrics, with {\bf GEM51-C-BEL-Iz0} closely following and showing the lowest mean absolute bias. Conversely, {\bf GEM51-C-DEL-TOFD} performs the worst across all error metrics with only the exception of the stability/friction velocity regime-conditioned error. However, once the mean wind speed has been removed from AMF and observations, error metrics are substantially modified (right panels in Fig.~\ref{fig:res7}). For instance, in Fig.~\ref{fig:res7}b), {\bf GEM51-C-DEL-TOFD} exhibits relatively lower errors compared to the other simulations, showing the lowest errors for the diurnal and annual-diurnal cycle metrics and the second lowest errors for the frequency-intensity and regime error metrics. The two simulations differing on the choice of stability function ({\bf GEM51-C-DEL} and {\bf GEM51-C-BEL}), show very similar errors although {\bf GEM51-C-BEL} performs slightly but systematically better than {\bf GEM51-C-DEL}. 

Finally, we note the presence of distinct clusters, particularly in the unbiased stability/friction velocity regime-conditioned error. Three groups seem to emerge: the first group, showing the highest error, includes the two simulations using the ISBA roughness length; the second group comprises the remaining three GEM51 simulations; and the third group, exhibiting the best performance, consists of the GEM50 simulation.

\section{Discussion}
\label{Sec:Discussion}
We have assessed the performance and sensitivity of six configurations of the CRCM6-GEM5 model using data from 27 stations from the AmeriFlux network that cover a variety of North American climates.
We evaluate the performance of simulations at simulating specific aspects of near-surface wind speeds by using a variety of error metrics (presented in Sect.~\ref{Sec:Methods2}) and by establishing a hierarchy of errors based on the extent to which each metric allows for the compensation of errors. The approach we follow is inspired by and has been used to quantify errors in other variables \cite{di_luca_decomposing_2020,di_luca_evaluating_2021,picart_uncertainty_2024}.

At the bottom of this hierarchy, there is the error in mean wind speed which allows for all possible error compensations (e.g., a low $\epsilon_{bias}$ can be obtained by producing the wrong diurnal or annual cycle). Three additional error metrics are defined to avoid the compensation of wind speed errors in the diurnal cycle ($\epsilon_{diu}$), the annual cycle ($\epsilon_{ann}$), and the diurnal/annual cycles combined ($\epsilon_{diu,ann}$). A fifth error metric quantifies differences in the frequency-intensity distribution of hourly wind speeds ($\epsilon_{fi}$). Finally, we define an error metric that quantifies the behavior of wind speed conditioned to the occurrence of multiple stability ($\zeta$) and friction velocity ($u_*$) regimes ($\epsilon_{reg}$). These two variables, $u_*$ and $\zeta$, are selected because they are directly measured in AmeriFlux sites and are two of the three key variables determining wind speeds according to MOST. In addition, we calculate biased and unbiased versions of all errors in order to highlight errors in the variability of wind speeds. 

The analysis of multiple error metrics shows that:
\begin{itemize}
\item With the exception of the regime error metric, all error metrics are dominated by the error in the mean wind speed thus providing the same ranking of simulations. Errors calculated after removing mean wind speeds (i.e., unbiased simulations) provide a completely different view on the relative performance of simulations.
\item The increasing complexity of error metrics reduces the likelihood of error compensation, which in turn highlights whether simulations reproduce observations for the right reasons (e.g., bias being the primary source of error in TOFD simulations).
\end{itemize}

Among all six simulations, {\bf GEM50-C-DEL} performs best with respect to most metrics including the stability/friction velocity regime-conditioned error indicating the least compensation of errors. Similarly, for unbiased simulations, {\bf GEM51-C-DEL-TOFD} ranks as the second-best performer across most metrics.

In terms of the effect of specific changes in the CRCM6-GEM5 model configuration, our results show that:
\begin{itemize}
\item GEM version ({\bf GEM50-C-DEL} $\rightarrow$ {\bf GEM51-C-DEL}): the sensitivity of wind speeds to changes in the model version is very high. The simulation {\bf GEM50-C-DEL} performs better than {\bf GEM51-C-DEL} for all error metrics including the frequency-intensity distribution ($\epsilon_{fi}$) and the physical coherence among variables ($\epsilon_{reg}$). The {\bf GEM51-C-DEL} simulation produces very stable regimes too often, particularly for very low and very high friction velocities. This behavior is likely due to the strategy used to deal with very stable regimes ($L$ cutoff, see Table \ref{gem_details}).
\item Stability functions ({\bf GEM51-C-DEL} $\rightarrow$ {\bf GEM51-C-BEL}): the sensitivity of wind speeds to changes in stability functions is low and these two simulations show relatively minor differences. However, for most metrics, {\bf GEM51-C-BEL} performs slightly but consistently better than {\bf GEM51-C-DEL}. Both simulations show the largest errors for very stable regimes. This suggests that the majority of errors are unlikely due to the choice of stability function but are instead influenced by other factors that affect simulation accuracy in very stable regimes.
\item Roughness length ($z_0$) formulation ({\bf GEM51-C-DEL} $\rightarrow$ {\bf GEM51-C-DEL-TOFD}): the sensitivity of wind speeds to changes in the roughness length formulation is high. The {\bf GEM51-C-DEL-TOFD} simulation, while showing a much larger error in the mean wind speed, outperforms the simulation {\bf GEM51-C-DEL} in all other aspects when the mean bias is removed. Specifically, {\bf GEM51-C-DEL-TOFD} substantially improves the representation of the intensity distribution of hourly wind speeds compared to {\bf GEM51-C-DEL}. 
The TOFD $z_0$ formulation, by avoiding the usage of roughness length in some cases (e.g., when the orographic roughness length is large), seems to outperform the other GEM51 configurations for all error metrics except for the mean bias.
\item Roughness length values ({\bf GEM51-C-BEL} $\rightarrow$ {\bf GEM51-C-BEL-Iz0}): the sensitivity of wind speeds to changes in $z_0$ values is very high. When considering the biased version of errors, the simulation {\bf GEM51-C-BEL-Iz0} outperforms {\bf GEM51-C-BEL} for all error metrics with the exception of the regimes error metric. The reason is that {\bf GEM51-C-BEL-Iz0} shows strong error compensations with some regimes showing an overestimation of the mean wind speed with an underestimation of the regimes' occurrence.
\item Land surface model ({\bf GEM51-C-BEL} $\rightarrow$ {\bf GEM51-I-BEL}): the sensitivity of wind speeds to changes in the LSM seems low to moderate if we exclude the impact of changing the roughness length values. The simulation {\bf GEM51-I-BEL} shows a similar performance to {\bf GEM51-C-BEL-Iz0} suggesting that most differences with {\bf GEM51-C-BEL} are related with changes in $z_0$. The simulation using CLASS shows a slightly better performance than ISBA for all biased metrics and little differences for unbiased error metrics. 
\end{itemize}

Representing the model's roughness length remains an elusive problem. We use two sets of $z_0$ values in our model, both based on two distinct land cover type fields. Additionally, we estimate the roughness length at each station and find little resemblance between the spatial distribution of $z_0$ used in simulations and the one estimated from AMF data (see Figure S1). This discrepancy is expected, as station measurements are often taken over a grass surface (or sometimes bare soil if the station is in a desert area), even if the station is surrounded by a forest. In the best-case scenario, the model will assume a $z_0$ consistent with the surrounding forest, leading to the observed differences. In addition, there is a problem with the spatial representativity of the data. The inability of knowing the roughness length makes it difficult to directly comparing station data with the simulated conditions in the closest grid point. Our approach was to merge together all 27 locations in the analysis and avoid the comparison of single grid points.

The AmeriFlux dataset provides high temporal resolution data for several variables that are rarely measured elsewhere, offering a valuable opportunity to gain deeper insights into the physical processes governing near-surface wind speeds. Nonetheless, its limitations must be acknowledged. First, wind speeds are measured at only a single height, which limits the analysis of vertical wind profiles. Second, these measurements are often taken at non-standard heights, differing from the commonly used 10 m reference height, potentially introducing inconsistencies when applying height adjustment corrections. Third, the dataset lacks information on surface roughness length, which is critical for understanding surface-atmosphere interactions. Additionally, the stations are typically situated in clear, unobstructed areas to minimize interference, which may not accurately represent the spatial heterogeneity of the surrounding landscape. Finally, while the network includes stations that capture diverse conditions across North America, their distribution is uneven and biased toward specific regions.

Finally, numerous studies have highlighted the limitations of MOST \cite{foken_50_2006,kumar_analysis_2012,RandomErrors,stabilit_mo,Sun2020}. Using detailed observations from two field datasets, \citeA{Sun2020} showed that MOST performs reasonably well only under specific cases. First, it holds under neutral conditions associated with weak stratification, where its validity can extend up to approximately 30 m. Second, under neutral conditions with strong stratification, the validity of MOST is limited to heights close to the surface ($z \lessapprox 1-2$ m). \citeA{stabilit_mo} pointed out that the depth of the surface layer where MOST is applicable varies significantly depending on stability conditions. They suggested that under extremely stable conditions, this depth may be as shallow as 1-5 m, which is considerably smaller than the surface layer depth assumed by the current vertical resolution of many models, including ours. \citeA{kumar_analysis_2012} also reported discrepancies between MOST-derived values and observational data, particularly under very stable atmospheric regimes. \cite{RandomErrors} performs an error propagation analysis on MOST. They focus on errors produced on the friction velocity and stability parameter by random errors, as opposed to systematic or instrument errors. They find that errors in the stability parameter to be in the order of $40\%$ in the unstable regime while the friction velocity in the order of $10\%$. The find through statistical analysis that the errors cannot be fully explained by the random errors introduced. They suggest that additional physical process are not being represented, and that this could be remedied by incorporating additional non-dimensional parameters beyond the stability parameter $\zeta = z/L$ in the stability functions.

\section{Conclusions}
\label{Sec:Conclusions}
The evaluation of the ability of several configurations of the version 6 of the Canadian Regional Climate Model (CRCM6-GEM5) to simulate 10-m wind speeds ($u_{10}$) allows us to conclude that the
ranking of simulations depends on the error metric used, and five of the six error metrics used in the study are dominated by errors in the mean wind speed. The only error metric that is only partially affected by errors in mean wind speeds is regime-conditioned error metric, that explores errors in $u_{10}$ based on its dependence on other near-surface variables.

Our results show that:
\begin{itemize}
\item Using a TOFD scheme instead of the simpler effective roughness length approach is a step in the right direction, as it significantly improves the intensity distribution of hourly wind speeds.
\item Using a lower limit for the Obhukov length instead of a lower limit for the lowest-level wind speed seems to deteriorate the simulation of wind speeds under very stable conditions.
\item The value of the roughness length ($z_0$) at different grid points has a massive effect on the simulation of wind speeds. However, determining the true roughness length remains problematic.
\item Using the CLASS LSM instead of the ISBA LSM and the \citeA{Beljaars1991} instead of the \citeA{DELAGE1997} stability functions, results in small but consistent improvements in the simulation of wind speeds.
\end{itemize}

Further work is needed to assess the sensitivity of wind speeds to other changes such as the models' horizontal resolution and the use of other strategies to deal with very stable regimes. Moreover, a more comprehensive evaluation approach should be considered—one that encompasses not only surface-layer processes but also planetary boundary layer dynamics and the representation of wind speeds at higher altitudes. Since surface roughness is a key factor influencing near-surface wind speeds, further research is needed to develop methods for estimating roughness characteristics. This effort should explicitly account for the spatial heterogeneity that governs effective terrain roughness, as current simplifications may obscure important interactions. Another research direction would be to map out errors in wind speeds as a function of ``heterogeneity class'' could help in assessing how MOST performs under non-homogeneous conditions. Additionally, alternative approaches to surface layer processes should be considered, as current methods based on MOST have important limitations \cite{foken_50_2006,kumar_analysis_2012,RandomErrors,stabilit_mo,Sun2020}.

%
%

\section*{Open Research Section}

\begin{itemize}
     \item Data availability: The processed data from the AmeriFlux stations and the CRCM6-GEM5 simulation data are available in the Borealis data repository \cite{SP3/VWMCY0_2024}. 
    \item Code availability: The code to reproduce the study is available at \url{https://github.com/timwhittaker/WindErrorHierarchy}.
\end{itemize}

\acknowledgments
We acknowledge the following AmeriFlux sites for their data records: Tab.~S2. In addition, funding for AmeriFlux data resources was provided by the U.S. Department of Energy’s Office of Science. The authors would like to thank Frédérik Toupin for maintaining a user-friendly local computing facility and for downloading and preparing some of the precipitation datasets. The authors would like to thank Ayrton Zadra for discussion about MOST and its implementation in the CRCM6-GEM5 model, and Danahe Paquin for valuable feedback. In addition, the authors would like to thank the two anonymous reviewers for the valuable input. This research has been conducted as part of the project “Simulation et analyse du climat à haute resolution” thanks to the financial participation of the Government of Québec. Alejandro Di Luca was funded by the Natural Sciences and Engineering Research Council of Canada (NSERC) (grant no. RGPIN-2020-05631). This research was enabled in part by support provided by Calcul Québec (calculquebec.ca) and the Digital Research Alliance of Canada (alliancecan.ca). 

%
%
\newcommand{\includecitedreferences}{
\nocite{osti_1579541}
\nocite{osti_1871134} 
 \nocite{osti_1871135}
\nocite{osti_1832158} 
\nocite{osti_1881573}
\nocite{osti_1881574} 
\nocite{osti_1881575}
\nocite{osti_1832159}
 \nocite{osti_1811363} 
\nocite{osti_1669685}
\nocite{osti_1832163} 
 \nocite{osti_1418682} 
\nocite{osti_1375202}
\nocite{osti_1419507} 
\nocite{osti_1818371}
\nocite{osti_1881590} 
\nocite{osti_1617724}
 \nocite{osti_1375201}
\nocite{osti_1660351} 
\nocite{osti_1871144}
\nocite{osti_1854371}
\nocite{osti_1418685} 
\nocite{osti_1669695}
\nocite{osti_1246147} 
\nocite{osti_1669698} 
\nocite{osti_1246140}
\nocite{osti_1245984} 
}
\includecitedreferences
\bibliography{agusample}

\begin{thebibliography}{}

\bibitem [\protect \citeauthoryear {%
Anderson%
, Bell%
\BCBL {}\ \BBA {} Peng%
}{%
Anderson%
\ \protect \BOthers {.}}{%
{\protect \APACyear {2013}}%
}]{%
Anderson2013}
\APACinsertmetastar {%
Anderson2013}%
\begin{APACrefauthors}%
Anderson, G\BPBI B.%
, Bell, M\BPBI L.%
\BCBL {}\ \BBA {} Peng, R\BPBI D.%
\end{APACrefauthors}%
\unskip\
\newblock
\APACrefYearMonthDay{2013}{Oct}{}.
\newblock
{\BBOQ}\APACrefatitle {Methods to calculate the heat index as an exposure
  metric in environmental health research} {Methods to calculate the heat index
  as an exposure metric in environmental health research}.{\BBCQ}
\newblock
\APACjournalVolNumPages{Environmental health perspectives}{121}{10}{1111-1119}.
\newblock
\begin{APACrefURL} \url{https://doi.org/10.1289/ehp.1206273} \end{APACrefURL}
\newblock
\APACrefnote{23934704[pmid]}
\newblock
\begin{APACrefDOI} \doi{10.1289/ehp.1206273} \end{APACrefDOI}
\PrintBackRefs{\CurrentBib}

\bibitem [\protect \citeauthoryear {%
Baker%
\ \BBA {} Griffis%
}{%
Baker%
\ \BBA {} Griffis%
}{%
{\protect \APACyear {2018}}%
}]{%
osti_1419507}
\APACinsertmetastar {%
osti_1419507}%
\begin{APACrefauthors}%
Baker, J.%
\BCBT {}\ \BBA {} Griffis, T.%
\end{APACrefauthors}%
\unskip\
\newblock
\APACrefYearMonthDay{2018}{}{}.
\newblock
\APACrefbtitle {AmeriFlux US-Ro4 Rosemount Prairie.} {Ameriflux us-ro4
  rosemount prairie.}
\newblock
\begin{APACrefDOI} \doi{10.17190/AMF/1419507} \end{APACrefDOI}
\PrintBackRefs{\CurrentBib}

\bibitem [\protect \citeauthoryear {%
Baker%
\ \BBA {} Griffis%
}{%
Baker%
\ \BBA {} Griffis%
}{%
{\protect \APACyear {2021}}%
}]{%
osti_1818371}
\APACinsertmetastar {%
osti_1818371}%
\begin{APACrefauthors}%
Baker, J.%
\BCBT {}\ \BBA {} Griffis, T.%
\end{APACrefauthors}%
\unskip\
\newblock
\APACrefYearMonthDay{2021}{}{}.
\newblock
\APACrefbtitle {AmeriFlux FLUXNET-1F US-Ro5 Rosemount I18\_South.} {Ameriflux
  fluxnet-1f us-ro5 rosemount i18\_south.}
\newblock
\begin{APACrefDOI} \doi{10.17190/AMF/1818371} \end{APACrefDOI}
\PrintBackRefs{\CurrentBib}

\bibitem [\protect \citeauthoryear {%
Baker%
\ \BBA {} Griffis%
}{%
Baker%
\ \BBA {} Griffis%
}{%
{\protect \APACyear {2022}}%
}]{%
osti_1881590}
\APACinsertmetastar {%
osti_1881590}%
\begin{APACrefauthors}%
Baker, J.%
\BCBT {}\ \BBA {} Griffis, T.%
\end{APACrefauthors}%
\unskip\
\newblock
\APACrefYearMonthDay{2022}{}{}.
\newblock
\APACrefbtitle {AmeriFlux FLUXNET-1F US-Ro6 Rosemount I18\_North.} {Ameriflux
  fluxnet-1f us-ro6 rosemount i18\_north.}
\newblock
\begin{APACrefDOI} \doi{10.17190/AMF/1881590} \end{APACrefDOI}
\PrintBackRefs{\CurrentBib}

\bibitem [\protect \citeauthoryear {%
Baldocchi%
}{%
Baldocchi%
}{%
{\protect \APACyear {2003}}%
}]{%
https://doi.org/10.1046/j.1365-2486.2003.00629.x}
\APACinsertmetastar {%
https://doi.org/10.1046/j.1365-2486.2003.00629.x}%
\begin{APACrefauthors}%
Baldocchi, D.%
\end{APACrefauthors}%
\unskip\
\newblock
\APACrefYearMonthDay{2003}{}{}.
\newblock
{\BBOQ}\APACrefatitle {Assessing the eddy covariance technique for evaluating
  carbon dioxide exchange rates of ecosystems: past, present and future}
  {Assessing the eddy covariance technique for evaluating carbon dioxide
  exchange rates of ecosystems: past, present and future}.{\BBCQ}
\newblock
\APACjournalVolNumPages{Global Change Biology}{9}{4}{479-492}.
\newblock
\begin{APACrefURL}
  \url{https://onlinelibrary.wiley.com/doi/abs/10.1046/j.1365-2486.2003.00629.x}
  \end{APACrefURL}
\newblock
\begin{APACrefDOI} \doi{https://doi.org/10.1046/j.1365-2486.2003.00629.x}
  \end{APACrefDOI}
\PrintBackRefs{\CurrentBib}

\bibitem [\protect \citeauthoryear {%
Bechtold%
, Bazile%
, Guichard%
, Mascart%
\BCBL {}\ \BBA {} Richard%
}{%
Bechtold%
\ \protect \BOthers {.}}{%
{\protect \APACyear {2001}}%
}]{%
Bechtold2001}
\APACinsertmetastar {%
Bechtold2001}%
\begin{APACrefauthors}%
Bechtold, P.%
, Bazile, E.%
, Guichard, F.%
, Mascart, P.%
\BCBL {}\ \BBA {} Richard, E.%
\end{APACrefauthors}%
\unskip\
\newblock
\APACrefYearMonthDay{2001}{apr}{}.
\newblock
{\BBOQ}\APACrefatitle {{A mass-flux convection scheme for regional and global
  models}} {{A mass-flux convection scheme for regional and global
  models}}.{\BBCQ}
\newblock
\APACjournalVolNumPages{Quarterly Journal of the Royal Meteorological
  Society}{127}{573}{869--886}.
\newblock
\begin{APACrefURL}
  \url{https://onlinelibrary.wiley.com/doi/10.1002/qj.49712757309}
  \end{APACrefURL}
\newblock
\begin{APACrefDOI} \doi{10.1002/qj.49712757309} \end{APACrefDOI}
\PrintBackRefs{\CurrentBib}

\bibitem [\protect \citeauthoryear {%
B{\'{e}}lair%
, Mailhot%
, Strapp%
\BCBL {}\ \BBA {} Macpherson%
}{%
B{\'{e}}lair%
\ \protect \BOthers {.}}{%
{\protect \APACyear {1999}}%
}]{%
Belair1999}
\APACinsertmetastar {%
Belair1999}%
\begin{APACrefauthors}%
B{\'{e}}lair, S.%
, Mailhot, J.%
, Strapp, J\BPBI W.%
\BCBL {}\ \BBA {} Macpherson, J\BPBI I.%
\end{APACrefauthors}%
\unskip\
\newblock
\APACrefYearMonthDay{1999}{}{}.
\newblock
{\BBOQ}\APACrefatitle {{An examination of local versus nonlocal aspects of a
  TKE-based boundary layer scheme in clear convective conditions}} {{An
  examination of local versus nonlocal aspects of a TKE-based boundary layer
  scheme in clear convective conditions}}.{\BBCQ}
\newblock
\APACjournalVolNumPages{Journal of Applied Meteorology}{38}{10}{1499--1518}.
\newblock
\begin{APACrefDOI} \doi{10.1175/1520-0450(1999)038<1499:AEOLVN>2.0.CO;2}
  \end{APACrefDOI}
\PrintBackRefs{\CurrentBib}

\bibitem [\protect \citeauthoryear {%
Beljaars%
}{%
Beljaars%
}{%
{\protect \APACyear {1992}}%
}]{%
beljaars_parametrization_1992}
\APACinsertmetastar {%
beljaars_parametrization_1992}%
\begin{APACrefauthors}%
Beljaars, A.%
\end{APACrefauthors}%
\unskip\
\newblock
\APACrefYearMonthDay{1992}{}{}.
\newblock
{\BBOQ}\APACrefatitle {The parametrization of the planetary boundary layer}
  {The parametrization of the planetary boundary layer}.{\BBCQ}
\newblock
\APACjournalVolNumPages{Meteorological Training Course Lecture Series}{}{May
  1992}{1--57}.
\newblock
\begin{APACrefURL}
  \url{https://www.ecmwf.int/sites/default/files/elibrary/2002/16959-parametrization-planetary-boundary-layer.pdf}
  \end{APACrefURL}
\PrintBackRefs{\CurrentBib}

\bibitem [\protect \citeauthoryear {%
Beljaars%
, Brown%
\BCBL {}\ \BBA {} Wood%
}{%
Beljaars%
\ \protect \BOthers {.}}{%
{\protect \APACyear {2004}}%
}]{%
TOFD}
\APACinsertmetastar {%
TOFD}%
\begin{APACrefauthors}%
Beljaars, A.%
, Brown, A\BPBI R.%
\BCBL {}\ \BBA {} Wood, N.%
\end{APACrefauthors}%
\unskip\
\newblock
\APACrefYearMonthDay{2004}{}{}.
\newblock
{\BBOQ}\APACrefatitle {A new parametrization of turbulent orographic form drag}
  {A new parametrization of turbulent orographic form drag}.{\BBCQ}
\newblock
\APACjournalVolNumPages{Quarterly Journal of the Royal Meteorological
  Society}{130}{599}{1327-1347}.
\newblock
\begin{APACrefURL}
  \url{https://rmets.onlinelibrary.wiley.com/doi/abs/10.1256/qj.03.73}
  \end{APACrefURL}
\newblock
\begin{APACrefDOI} \doi{https://doi.org/10.1256/qj.03.73} \end{APACrefDOI}
\PrintBackRefs{\CurrentBib}

\bibitem [\protect \citeauthoryear {%
Beljaars%
\ \BBA {} Holtslag%
}{%
Beljaars%
\ \BBA {} Holtslag%
}{%
{\protect \APACyear {1991}}%
}]{%
Beljaars1991}
\APACinsertmetastar {%
Beljaars1991}%
\begin{APACrefauthors}%
Beljaars, A.%
\BCBT {}\ \BBA {} Holtslag, A\BPBI A\BPBI M.%
\end{APACrefauthors}%
\unskip\
\newblock
\APACrefYearMonthDay{1991}{}{}.
\newblock
{\BBOQ}\APACrefatitle {Flux Parameterization over Land Surfaces for Atmospheric
  Models} {Flux parameterization over land surfaces for atmospheric
  models}.{\BBCQ}
\newblock
\APACjournalVolNumPages{Journal of Applied Meteorology and
  Climatology}{30}{3}{327 - 341}.
\newblock
\begin{APACrefURL}
  \url{https://journals.ametsoc.org/view/journals/apme/30/3/1520-0450_1991_030_0327_fpolsf_2_0_co_2.xml}
  \end{APACrefURL}
\newblock
\begin{APACrefDOI} \doi{10.1175/1520-0450(1991)030<0327:FPOLSF>2.0.CO;2}
  \end{APACrefDOI}
\PrintBackRefs{\CurrentBib}

\bibitem [\protect \citeauthoryear {%
Bergamaschi%
\ \BBA {} Windham-Myers%
}{%
Bergamaschi%
\ \BBA {} Windham-Myers%
}{%
{\protect \APACyear {2018}}%
}]{%
osti_1418685}
\APACinsertmetastar {%
osti_1418685}%
\begin{APACrefauthors}%
Bergamaschi, B.%
\BCBT {}\ \BBA {} Windham-Myers, L.%
\end{APACrefauthors}%
\unskip\
\newblock
\APACrefYearMonthDay{2018}{}{}.
\newblock
\APACrefbtitle {AmeriFlux US-Srr Suisun marsh - Rush Ranch.} {Ameriflux us-srr
  suisun marsh - rush ranch.}
\newblock
\begin{APACrefDOI} \doi{10.17190/AMF/1418685} \end{APACrefDOI}
\PrintBackRefs{\CurrentBib}

\bibitem [\protect \citeauthoryear {%
Bélair%
, Crevier%
, Mailhot%
, Bilodeau%
\BCBL {}\ \BBA {} Delage%
}{%
Bélair%
\ \protect \BOthers {.}}{%
{\protect \APACyear {2003}}%
}]{%
belair_operational_2003}
\APACinsertmetastar {%
belair_operational_2003}%
\begin{APACrefauthors}%
Bélair, S.%
, Crevier, L\BPBI P.%
, Mailhot, J.%
, Bilodeau, B.%
\BCBL {}\ \BBA {} Delage, Y.%
\end{APACrefauthors}%
\unskip\
\newblock
\APACrefYearMonthDay{2003}{}{}.
\newblock
{\BBOQ}\APACrefatitle {Operational implementation of the {ISBA} land surface
  scheme in the {Canadian} regional weather forecast model. {Part} {I}: {Warm}
  season results} {Operational implementation of the {ISBA} land surface scheme
  in the {Canadian} regional weather forecast model. {Part} {I}: {Warm} season
  results}.{\BBCQ}
\newblock
\APACjournalVolNumPages{Journal of Hydrometeorology}{4}{2}{352--370}.
\newblock
\begin{APACrefDOI} \doi{10.1175/1525-7541(2003)4<352:OIOTIL>2.0.CO;2}
  \end{APACrefDOI}
\PrintBackRefs{\CurrentBib}

\bibitem [\protect \citeauthoryear {%
Charnock%
}{%
Charnock%
}{%
{\protect \APACyear {1955}}%
}]{%
charnock_wind_1955}
\APACinsertmetastar {%
charnock_wind_1955}%
\begin{APACrefauthors}%
Charnock, H.%
\end{APACrefauthors}%
\unskip\
\newblock
\APACrefYearMonthDay{1955}{{\APACmonth{10}}}{}.
\newblock
{\BBOQ}\APACrefatitle {Wind stress on a water surface} {Wind stress on a water
  surface}.{\BBCQ}
\newblock
\APACjournalVolNumPages{Quarterly Journal of the Royal Meteorological
  Society}{81}{350}{639--640}.
\newblock
\begin{APACrefURL}
  [{2024-12-04}]\url{https://rmets.onlinelibrary.wiley.com/doi/10.1002/qj.49708135027}
  \end{APACrefURL}
\newblock
\begin{APACrefDOI} \doi{10.1002/qj.49708135027} \end{APACrefDOI}
\PrintBackRefs{\CurrentBib}

\bibitem [\protect \citeauthoryear {%
Chen%
, Di~Luca%
, Winger%
, Laprise%
\BCBL {}\ \BBA {} Thériault%
}{%
Chen%
\ \protect \BOthers {.}}{%
{\protect \APACyear {2022}}%
}]{%
chen_seasonality_2022}
\APACinsertmetastar {%
chen_seasonality_2022}%
\begin{APACrefauthors}%
Chen, T\BHBI C.%
, Di~Luca, A.%
, Winger, K.%
, Laprise, R.%
\BCBL {}\ \BBA {} Thériault, J\BPBI M.%
\end{APACrefauthors}%
\unskip\
\newblock
\APACrefYearMonthDay{2022}{{\APACmonth{08}}}{}.
\newblock
{\BBOQ}\APACrefatitle {Seasonality of {Continental} {Extratropical}‐{Cyclone}
  {Wind} {Speeds} {Over} {Northeastern} {North} {America}} {Seasonality of
  {Continental} {Extratropical}‐{Cyclone} {Wind} {Speeds} {Over}
  {Northeastern} {North} {America}}.{\BBCQ}
\newblock
\APACjournalVolNumPages{Geophysical Research Letters}{49}{15}{}.
\newblock
\begin{APACrefURL}
  \url{https://onlinelibrary.wiley.com/doi/10.1029/2022GL098776
  https://agupubs.onlinelibrary.wiley.com/doi/10.1029/2022GL098776}
  \end{APACrefURL}
\newblock
\begin{APACrefDOI} \doi{10.1029/2022GL098776} \end{APACrefDOI}
\PrintBackRefs{\CurrentBib}

\bibitem [\protect \citeauthoryear {%
Chosson%
, Vaillancourt%
, Milbrandt%
, Yau%
\BCBL {}\ \BBA {} Zadra%
}{%
Chosson%
\ \protect \BOthers {.}}{%
{\protect \APACyear {2014}}%
}]{%
chosson_adapting_2014}
\APACinsertmetastar {%
chosson_adapting_2014}%
\begin{APACrefauthors}%
Chosson, F.%
, Vaillancourt, P\BPBI A.%
, Milbrandt, J\BPBI A.%
, Yau, M\BPBI K.%
\BCBL {}\ \BBA {} Zadra, A.%
\end{APACrefauthors}%
\unskip\
\newblock
\APACrefYearMonthDay{2014}{}{}.
\newblock
{\BBOQ}\APACrefatitle {Adapting two-moment microphysics schemes across model
  resolutions: {Subgrid} cloud and precipitation fraction and microphysical
  sub-time step} {Adapting two-moment microphysics schemes across model
  resolutions: {Subgrid} cloud and precipitation fraction and microphysical
  sub-time step}.{\BBCQ}
\newblock
\APACjournalVolNumPages{Journal of the Atmospheric
  Sciences}{71}{7}{2635--2653}.
\newblock
\begin{APACrefDOI} \doi{10.1175/JAS-D-13-0367.1} \end{APACrefDOI}
\PrintBackRefs{\CurrentBib}

\bibitem [\protect \citeauthoryear {%
C{\^{o}}t{\'{e}}%
\ \protect \BOthers {.}}{%
C{\^{o}}t{\'{e}}%
\ \protect \BOthers {.}}{%
{\protect \APACyear {1998}}%
}]{%
cote1998}
\APACinsertmetastar {%
cote1998}%
\begin{APACrefauthors}%
C{\^{o}}t{\'{e}}, J.%
, Gravel, S.%
, M{\'{e}}thot, A.%
, Patoine, A.%
, Roch, M.%
\BCBL {}\ \BBA {} Staniforth, A.%
\end{APACrefauthors}%
\unskip\
\newblock
\APACrefYearMonthDay{1998}{}{}.
\newblock
{\BBOQ}\APACrefatitle {{The operational CMC-MRB global environmental multiscale
  (GEM) model. Part I: Design considerations and formulation}} {{The
  operational CMC-MRB global environmental multiscale (GEM) model. Part I:
  Design considerations and formulation}}.{\BBCQ}
\newblock
\APACjournalVolNumPages{Monthly Weather Review}{126}{6}{1373--1395}.
\newblock
\begin{APACrefDOI} \doi{10.1175/1520-0493(1998)126<1373:TOCMGE>2.0.CO;2}
  \end{APACrefDOI}
\PrintBackRefs{\CurrentBib}

\bibitem [\protect \citeauthoryear {%
Delage%
}{%
Delage%
}{%
{\protect \APACyear {1997}}%
}]{%
DELAGE1997}
\APACinsertmetastar {%
DELAGE1997}%
\begin{APACrefauthors}%
Delage, Y.%
\end{APACrefauthors}%
\unskip\
\newblock
\APACrefYearMonthDay{1997}{Jan}{01}.
\newblock
{\BBOQ}\APACrefatitle {Parameterising sub-grid scale vertical transport in
  atmospheric models under statically stable conditions} {Parameterising
  sub-grid scale vertical transport in atmospheric models under statically
  stable conditions}.{\BBCQ}
\newblock
\APACjournalVolNumPages{Boundary-Layer Meteorology}{82}{1}{23-48}.
\newblock
\begin{APACrefURL} \url{https://doi.org/10.1023/A:1000132524077}
  \end{APACrefURL}
\newblock
\begin{APACrefDOI} \doi{10.1023/A:1000132524077} \end{APACrefDOI}
\PrintBackRefs{\CurrentBib}

\bibitem [\protect \citeauthoryear {%
Delage%
\ \BBA {} Girard%
}{%
Delage%
\ \BBA {} Girard%
}{%
{\protect \APACyear {1992}}%
}]{%
Delage1992}
\APACinsertmetastar {%
Delage1992}%
\begin{APACrefauthors}%
Delage, Y.%
\BCBT {}\ \BBA {} Girard, C.%
\end{APACrefauthors}%
\unskip\
\newblock
\APACrefYearMonthDay{1992}{Jan}{01}.
\newblock
{\BBOQ}\APACrefatitle {Stability functions correct at the free convection limit
  and consistent for for both the surface and Ekman layers} {Stability
  functions correct at the free convection limit and consistent for for both
  the surface and ekman layers}.{\BBCQ}
\newblock
\APACjournalVolNumPages{Boundary-Layer Meteorology}{58}{1}{19-31}.
\newblock
\begin{APACrefURL} \url{https://doi.org/10.1007/BF00120749} \end{APACrefURL}
\newblock
\begin{APACrefDOI} \doi{10.1007/BF00120749} \end{APACrefDOI}
\PrintBackRefs{\CurrentBib}

\bibitem [\protect \citeauthoryear {%
Di~Luca%
, Argüeso%
, Sherwood%
\BCBL {}\ \BBA {} Evans%
}{%
Di~Luca%
\ \protect \BOthers {.}}{%
{\protect \APACyear {2021}}%
}]{%
di_luca_evaluating_2021}
\APACinsertmetastar {%
di_luca_evaluating_2021}%
\begin{APACrefauthors}%
Di~Luca, A.%
, Argüeso, D.%
, Sherwood, S.%
\BCBL {}\ \BBA {} Evans, J\BPBI P.%
\end{APACrefauthors}%
\unskip\
\newblock
\APACrefYearMonthDay{2021}{{\APACmonth{07}}}{}.
\newblock
{\BBOQ}\APACrefatitle {Evaluating {Precipitation} {Errors} {Using} the
  {Environmentally} {Conditioned} {Intensity}‐{Frequency} {Decomposition}
  {Method}} {Evaluating {Precipitation} {Errors} {Using} the {Environmentally}
  {Conditioned} {Intensity}‐{Frequency} {Decomposition} {Method}}.{\BBCQ}
\newblock
\APACjournalVolNumPages{Journal of Advances in Modeling Earth
  Systems}{13}{7}{1--22}.
\newblock
\begin{APACrefURL}
  \url{https://onlinelibrary.wiley.com/doi/10.1029/2020MS002447}
  \end{APACrefURL}
\newblock
\begin{APACrefDOI} \doi{10.1029/2020MS002447} \end{APACrefDOI}
\PrintBackRefs{\CurrentBib}

\bibitem [\protect \citeauthoryear {%
Di~Luca%
, Flaounas%
, Drobinski%
\BCBL {}\ \BBA {} Brossier%
}{%
Di~Luca%
\ \protect \BOthers {.}}{%
{\protect \APACyear {2014}}%
}]{%
di_luca_atmospheric_2014}
\APACinsertmetastar {%
di_luca_atmospheric_2014}%
\begin{APACrefauthors}%
Di~Luca, A.%
, Flaounas, E.%
, Drobinski, P.%
\BCBL {}\ \BBA {} Brossier, C\BPBI L.%
\end{APACrefauthors}%
\unskip\
\newblock
\APACrefYearMonthDay{2014}{{\APACmonth{11}}}{}.
\newblock
{\BBOQ}\APACrefatitle {The atmospheric component of the {Mediterranean} {Sea}
  water budget in a {WRF} multi-physics ensemble and observations} {The
  atmospheric component of the {Mediterranean} {Sea} water budget in a {WRF}
  multi-physics ensemble and observations}.{\BBCQ}
\newblock
\APACjournalVolNumPages{Climate Dynamics}{43}{9-10}{2349--2375}.
\newblock
\begin{APACrefURL} \url{http://link.springer.com/10.1007/s00382-014-2058-z}
  \end{APACrefURL}
\newblock
\begin{APACrefDOI} \doi{10.1007/s00382-014-2058-z} \end{APACrefDOI}
\PrintBackRefs{\CurrentBib}

\bibitem [\protect \citeauthoryear {%
Di~Luca%
, Pitman%
\BCBL {}\ \BBA {} de Elía%
}{%
Di~Luca%
\ \protect \BOthers {.}}{%
{\protect \APACyear {2020}}%
}]{%
di_luca_decomposing_2020}
\APACinsertmetastar {%
di_luca_decomposing_2020}%
\begin{APACrefauthors}%
Di~Luca, A.%
, Pitman, A\BPBI J.%
\BCBL {}\ \BBA {} de Elía, R.%
\end{APACrefauthors}%
\unskip\
\newblock
\APACrefYearMonthDay{2020}{}{}.
\newblock
{\BBOQ}\APACrefatitle {Decomposing {Temperature} {Extremes} {Errors} in {CMIP5}
  and {CMIP6} {Models}} {Decomposing {Temperature} {Extremes} {Errors} in
  {CMIP5} and {CMIP6} {Models}}.{\BBCQ}
\newblock
\APACjournalVolNumPages{Geophysical Research Letters}{47}{14}{1--10}.
\newblock
\begin{APACrefDOI} \doi{10.1029/2020GL088031} \end{APACrefDOI}
\PrintBackRefs{\CurrentBib}

\bibitem [\protect \citeauthoryear {%
Draxl%
, Hahmann%
, Peña%
\BCBL {}\ \BBA {} Giebel%
}{%
Draxl%
\ \protect \BOthers {.}}{%
{\protect \APACyear {2014}}%
}]{%
Drawl2014}
\APACinsertmetastar {%
Drawl2014}%
\begin{APACrefauthors}%
Draxl, C.%
, Hahmann, A\BPBI N.%
, Peña, A.%
\BCBL {}\ \BBA {} Giebel, G.%
\end{APACrefauthors}%
\unskip\
\newblock
\APACrefYearMonthDay{2014}{}{}.
\newblock
{\BBOQ}\APACrefatitle {Evaluating winds and vertical wind shear from Weather
  Research and Forecasting model forecasts using seven planetary boundary layer
  schemes} {Evaluating winds and vertical wind shear from weather research and
  forecasting model forecasts using seven planetary boundary layer
  schemes}.{\BBCQ}
\newblock
\APACjournalVolNumPages{Wind Energy}{17}{1}{39-55}.
\newblock
\begin{APACrefURL}
  \url{https://onlinelibrary.wiley.com/doi/abs/10.1002/we.1555}
  \end{APACrefURL}
\newblock
\begin{APACrefDOI} \doi{https://doi.org/10.1002/we.1555} \end{APACrefDOI}
\PrintBackRefs{\CurrentBib}

\bibitem [\protect \citeauthoryear {%
Dzebre%
\ \BBA {} Adaramola%
}{%
Dzebre%
\ \BBA {} Adaramola%
}{%
{\protect \APACyear {2020}}%
}]{%
dzebre_preliminary_2020}
\APACinsertmetastar {%
dzebre_preliminary_2020}%
\begin{APACrefauthors}%
Dzebre, D\BPBI E.%
\BCBT {}\ \BBA {} Adaramola, M\BPBI S.%
\end{APACrefauthors}%
\unskip\
\newblock
\APACrefYearMonthDay{2020}{{\APACmonth{02}}}{}.
\newblock
{\BBOQ}\APACrefatitle {A preliminary sensitivity study of {Planetary}
  {Boundary} {Layer} parameterisation schemes in the weather research and
  forecasting model to surface winds in coastal {Ghana}} {A preliminary
  sensitivity study of {Planetary} {Boundary} {Layer} parameterisation schemes
  in the weather research and forecasting model to surface winds in coastal
  {Ghana}}.{\BBCQ}
\newblock
\APACjournalVolNumPages{Renewable Energy}{146}{}{66--86}.
\newblock
\begin{APACrefURL}
  [{2024-11-28}]\url{https://linkinghub.elsevier.com/retrieve/pii/S0960148119309711}
  \end{APACrefURL}
\newblock
\begin{APACrefDOI} \doi{10.1016/j.renene.2019.06.133} \end{APACrefDOI}
\PrintBackRefs{\CurrentBib}

\bibitem [\protect \citeauthoryear {%
{Earth Resources Observation And Science (EROS) Center}%
}{%
{Earth Resources Observation And Science (EROS) Center}%
}{%
{\protect \APACyear {2017}}%
{\protect \APACexlab {{\protect \BCnt {1}}}}}]{%
usgs_gtopo30}
\APACinsertmetastar {%
usgs_gtopo30}%
\begin{APACrefauthors}%
{Earth Resources Observation And Science (EROS) Center}.%
\end{APACrefauthors}%
\unskip\
\newblock
\APACrefYearMonthDay{2017{\protect \BCnt {1}}}{}{}.
\newblock
\APACrefbtitle {Global 30 {Arc}-{Second} {Elevation} ({GTOPO30}).} {Global 30
  {Arc}-{Second} {Elevation} ({GTOPO30}).}
\newblock
\APACaddressPublisher{}{U.S. Geological Survey}.
\newblock
\begin{APACrefURL}
  [{2024-07-11}]\url{https://www.usgs.gov/centers/eros/science/usgs-eros-archive-digital-elevation-global-30-arc-second-elevation-gtopo30?qt-science_center_objects=0#qt-science_center_objects}
  \end{APACrefURL}
\newblock
\begin{APACrefDOI} \doi{10.5066/F7DF6PQS} \end{APACrefDOI}
\PrintBackRefs{\CurrentBib}

\bibitem [\protect \citeauthoryear {%
{Earth Resources Observation And Science (EROS) Center}%
}{%
{Earth Resources Observation And Science (EROS) Center}%
}{%
{\protect \APACyear {2017}}%
{\protect \APACexlab {{\protect \BCnt {2}}}}}]{%
usgs_land_cover}
\APACinsertmetastar {%
usgs_land_cover}%
\begin{APACrefauthors}%
{Earth Resources Observation And Science (EROS) Center}.%
\end{APACrefauthors}%
\unskip\
\newblock
\APACrefYearMonthDay{2017{\protect \BCnt {2}}}{}{}.
\newblock
\APACrefbtitle {Global {Land} {Cover} {Characterization} ({GLCC}).} {Global
  {Land} {Cover} {Characterization} ({GLCC}).}
\newblock
\APACaddressPublisher{}{U.S. Geological Survey}.
\newblock
\begin{APACrefURL}
  [{2024-07-11}]\url{https://www.usgs.gov/centers/eros/science/usgs-eros-archive-land-cover-products-global-land-cover-characterization-glcc}
  \end{APACrefURL}
\newblock
\begin{APACrefDOI} \doi{10.5066/F7GB230D} \end{APACrefDOI}
\PrintBackRefs{\CurrentBib}

\bibitem [\protect \citeauthoryear {%
{ECMWF}%
}{%
{ECMWF}%
}{%
{\protect \APACyear {2016}}%
}]{%
ecmwf_ifs_2016}
\APACinsertmetastar {%
ecmwf_ifs_2016}%
\begin{APACrefauthors}%
{ECMWF}.%
\end{APACrefauthors}%
\unskip\
\newblock
\APACrefYearMonthDay{2016}{}{}.
\newblock
{\BBOQ}\APACrefatitle {{IFS} {Documentation} - {CY41R2}: {Part} {IV}:
  {Physical} {Processes}} {{IFS} {Documentation} - {CY41R2}: {Part} {IV}:
  {Physical} {Processes}}.{\BBCQ}
\newblock
\APACjournalVolNumPages{IFS Documentation CY41R2}{}{March}{213}.
\PrintBackRefs{\CurrentBib}

\bibitem [\protect \citeauthoryear {%
{ECMWF}%
}{%
{ECMWF}%
}{%
{\protect \APACyear {2023}}%
}]{%
ecmwf_ifs_2023}
\APACinsertmetastar {%
ecmwf_ifs_2023}%
\begin{APACrefauthors}%
{ECMWF}.%
\end{APACrefauthors}%
\unskip\
\newblock
\APACrefYearMonthDay{2023}{}{}.
\newblock
\APACrefbtitle {{IFS} {Documentation} {CY48R1} - {Part} {III}: {Dynamics} and
  {Numerical} {Procedures}} {{IFS} {Documentation} {CY48R1} - {Part} {III}:
  {Dynamics} and {Numerical} {Procedures}}\ \APACbVolEdTR{}{\BTR{}}.
\newblock
\begin{APACrefURL}
  [{2024-07-05}]\url{https://www.ecmwf.int/en/elibrary/81369-ifs-documentation-cy48r1-part-iii-dynamics-and-numerical-procedures}
  \end{APACrefURL}
\newblock
\APACrefnote{Publisher: ECMWF}
\newblock
\begin{APACrefDOI} \doi{10.21957/26F0AD3473} \end{APACrefDOI}
\PrintBackRefs{\CurrentBib}

\bibitem [\protect \citeauthoryear {%
Eichelmann%
\ \protect \BOthers {.}}{%
Eichelmann%
\ \protect \BOthers {.}}{%
{\protect \APACyear {2020}}%
}]{%
osti_1669698}
\APACinsertmetastar {%
osti_1669698}%
\begin{APACrefauthors}%
Eichelmann, E.%
, Knox, S.%
, Sanchez, C\BPBI R.%
, Valach, A.%
, Sturtevant, C.%
, Szutu, D.%
\BDBL {}Baldocchi, D.%
\end{APACrefauthors}%
\unskip\
\newblock
\APACrefYearMonthDay{2020}{}{}.
\newblock
\APACrefbtitle {FLUXNET-CH4 US-Tw4 Twitchell East End Wetland.} {Fluxnet-ch4
  us-tw4 twitchell east end wetland.}
\newblock
\begin{APACrefDOI} \doi{10.18140/FLX/1669698} \end{APACrefDOI}
\PrintBackRefs{\CurrentBib}

\bibitem [\protect \citeauthoryear {%
Fiedler%
\ \BBA {} Panofsky%
}{%
Fiedler%
\ \BBA {} Panofsky%
}{%
{\protect \APACyear {1972}}%
}]{%
fiedler_geostrophic_1972}
\APACinsertmetastar {%
fiedler_geostrophic_1972}%
\begin{APACrefauthors}%
Fiedler, F.%
\BCBT {}\ \BBA {} Panofsky, H\BPBI A.%
\end{APACrefauthors}%
\unskip\
\newblock
\APACrefYearMonthDay{1972}{{\APACmonth{01}}}{}.
\newblock
{\BBOQ}\APACrefatitle {The geostrophic drag coefficient and the ‘effective’
  roughness length} {The geostrophic drag coefficient and the ‘effective’
  roughness length}.{\BBCQ}
\newblock
\APACjournalVolNumPages{Quarterly Journal of the Royal Meteorological
  Society}{98}{415}{213--220}.
\newblock
\begin{APACrefURL}
  [{2024-07-04}]\url{https://rmets.onlinelibrary.wiley.com/doi/10.1002/qj.49709841519}
  \end{APACrefURL}
\newblock
\begin{APACrefDOI} \doi{10.1002/qj.49709841519} \end{APACrefDOI}
\PrintBackRefs{\CurrentBib}

\bibitem [\protect \citeauthoryear {%
Flerchinger%
}{%
Flerchinger%
}{%
{\protect \APACyear {2017}}%
{\protect \APACexlab {{\protect \BCnt {1}}}}}]{%
osti_1375202}
\APACinsertmetastar {%
osti_1375202}%
\begin{APACrefauthors}%
Flerchinger, G.%
\end{APACrefauthors}%
\unskip\
\newblock
\APACrefYearMonthDay{2017{\protect \BCnt {1}}}{}{}.
\newblock
\APACrefbtitle {AmeriFlux US-Rms RCEW Mountain Big Sagebrush.} {Ameriflux
  us-rms rcew mountain big sagebrush.}
\newblock
\begin{APACrefDOI} \doi{10.17190/AMF/1375202} \end{APACrefDOI}
\PrintBackRefs{\CurrentBib}

\bibitem [\protect \citeauthoryear {%
Flerchinger%
}{%
Flerchinger%
}{%
{\protect \APACyear {2017}}%
{\protect \APACexlab {{\protect \BCnt {2}}}}}]{%
osti_1375201}
\APACinsertmetastar {%
osti_1375201}%
\begin{APACrefauthors}%
Flerchinger, G.%
\end{APACrefauthors}%
\unskip\
\newblock
\APACrefYearMonthDay{2017{\protect \BCnt {2}}}{}{}.
\newblock
\APACrefbtitle {AmeriFlux US-Rws Reynolds Creek Wyoming big sagebrush.}
  {Ameriflux us-rws reynolds creek wyoming big sagebrush.}
\newblock
\begin{APACrefDOI} \doi{10.17190/AMF/1375201} \end{APACrefDOI}
\PrintBackRefs{\CurrentBib}

\bibitem [\protect \citeauthoryear {%
Flerchinger%
}{%
Flerchinger%
}{%
{\protect \APACyear {2018}}%
}]{%
osti_1418682}
\APACinsertmetastar {%
osti_1418682}%
\begin{APACrefauthors}%
Flerchinger, G.%
\end{APACrefauthors}%
\unskip\
\newblock
\APACrefYearMonthDay{2018}{}{}.
\newblock
\APACrefbtitle {AmeriFlux US-Rls RCEW Low Sagebrush.} {Ameriflux us-rls rcew
  low sagebrush.}
\newblock
\begin{APACrefDOI} \doi{10.17190/AMF/1418682} \end{APACrefDOI}
\PrintBackRefs{\CurrentBib}

\bibitem [\protect \citeauthoryear {%
Flerchinger%
}{%
Flerchinger%
}{%
{\protect \APACyear {2020}}%
}]{%
osti_1617724}
\APACinsertmetastar {%
osti_1617724}%
\begin{APACrefauthors}%
Flerchinger, G.%
\end{APACrefauthors}%
\unskip\
\newblock
\APACrefYearMonthDay{2020}{}{}.
\newblock
\APACrefbtitle {AmeriFlux {US-Rwf} RCEW Upper Sheep Prescibed Fire.} {Ameriflux
  {US-Rwf} rcew upper sheep prescibed fire.}
\newblock
\begin{APACrefDOI} \doi{10.17190/AMF/1617724} \end{APACrefDOI}
\PrintBackRefs{\CurrentBib}

\bibitem [\protect \citeauthoryear {%
Foken%
}{%
Foken%
}{%
{\protect \APACyear {2006}}%
}]{%
foken_50_2006}
\APACinsertmetastar {%
foken_50_2006}%
\begin{APACrefauthors}%
Foken, T.%
\end{APACrefauthors}%
\unskip\
\newblock
\APACrefYearMonthDay{2006}{{\APACmonth{06}}}{}.
\newblock
{\BBOQ}\APACrefatitle {50 {Years} of the {Monin}–{Obukhov} {Similarity}
  {Theory}} {50 {Years} of the {Monin}–{Obukhov} {Similarity}
  {Theory}}.{\BBCQ}
\newblock
\APACjournalVolNumPages{Boundary-Layer Meteorology}{119}{3}{431--447}.
\newblock
\begin{APACrefURL}
  [{2024-11-29}]\url{http://link.springer.com/10.1007/s10546-006-9048-6}
  \end{APACrefURL}
\newblock
\begin{APACrefDOI} \doi{10.1007/s10546-006-9048-6} \end{APACrefDOI}
\PrintBackRefs{\CurrentBib}

\bibitem [\protect \citeauthoryear {%
Garratt%
}{%
Garratt%
}{%
{\protect \APACyear {1994}}%
}]{%
Garratt1994}
\APACinsertmetastar {%
Garratt1994}%
\begin{APACrefauthors}%
Garratt, J\BPBI R.%
\end{APACrefauthors}%
\unskip\
\newblock
\APACrefYear{1994}.
\newblock
\APACrefbtitle {The Atmospheric Boundary Layer} {The atmospheric boundary
  layer}\ (\PrintOrdinal{1st}\ \BEd).
\newblock
\APACaddressPublisher{Cambridge, UK}{Cambridge University Press}.
\PrintBackRefs{\CurrentBib}

\bibitem [\protect \citeauthoryear {%
Giorgi%
\ \BBA {} Gutowski%
}{%
Giorgi%
\ \BBA {} Gutowski%
}{%
{\protect \APACyear {2015}}%
}]{%
Giorgi2015}
\APACinsertmetastar {%
Giorgi2015}%
\begin{APACrefauthors}%
Giorgi, F.%
\BCBT {}\ \BBA {} Gutowski, W\BPBI J.%
\end{APACrefauthors}%
\unskip\
\newblock
\APACrefYearMonthDay{2015}{}{}.
\newblock
{\BBOQ}\APACrefatitle {{Regional Dynamical Downscaling and the CORDEX
  Initiative}} {{Regional Dynamical Downscaling and the CORDEX
  Initiative}}.{\BBCQ}
\newblock
\APACjournalVolNumPages{Annual Review of Environment and
  Resources}{40}{}{467--490}.
\newblock
\begin{APACrefDOI} \doi{10.1146/annurev-environ-102014-021217} \end{APACrefDOI}
\PrintBackRefs{\CurrentBib}

\bibitem [\protect \citeauthoryear {%
Girard%
\ \protect \BOthers {.}}{%
Girard%
\ \protect \BOthers {.}}{%
{\protect \APACyear {2014}}%
}]{%
Girard2014}
\APACinsertmetastar {%
Girard2014}%
\begin{APACrefauthors}%
Girard, C.%
, Plante, A.%
, Desgagné, M.%
, McTaggart-Cowan, R.%
, Côté, J.%
, Charron, M.%
\BDBL {}Zadra, A.%
\end{APACrefauthors}%
\unskip\
\newblock
\APACrefYearMonthDay{2014}{}{}.
\newblock
{\BBOQ}\APACrefatitle {Staggered vertical discretization of the canadian
  environmental multiscale (GEM) model using a coordinate of the
  log-hydrostatic-pressure type} {Staggered vertical discretization of the
  canadian environmental multiscale (gem) model using a coordinate of the
  log-hydrostatic-pressure type}.{\BBCQ}
\newblock
\APACjournalVolNumPages{Monthly Weather Review}{142}{3}{1183--1196}.
\newblock
\begin{APACrefDOI} \doi{10.1175/MWR-D-13-00255.1} \end{APACrefDOI}
\PrintBackRefs{\CurrentBib}

\bibitem [\protect \citeauthoryear {%
Goslee%
}{%
Goslee%
}{%
{\protect \APACyear {2021}}%
}]{%
osti_1811363}
\APACinsertmetastar {%
osti_1811363}%
\begin{APACrefauthors}%
Goslee, S.%
\end{APACrefauthors}%
\unskip\
\newblock
\APACrefYearMonthDay{2021}{}{}.
\newblock
\APACrefbtitle {AmeriFlux US-HWB USDA ARS Pasture Sytems and Watershed
  Management Research Unit- Hawbecker Site.} {Ameriflux us-hwb usda ars pasture
  sytems and watershed management research unit- hawbecker site.}
\newblock
\begin{APACrefURL} \url{https://www.osti.gov/biblio/1811363} \end{APACrefURL}
\newblock
\begin{APACrefDOI} \doi{10.17190/AMF/1811363} \end{APACrefDOI}
\PrintBackRefs{\CurrentBib}

\bibitem [\protect \citeauthoryear {%
Hersbach%
\ \protect \BOthers {.}}{%
Hersbach%
\ \protect \BOthers {.}}{%
{\protect \APACyear {2020}}%
}]{%
Hersbach2020}
\APACinsertmetastar {%
Hersbach2020}%
\begin{APACrefauthors}%
Hersbach, H.%
, Bell, B.%
, Berrisford, P.%
, Hirahara, S.%
, Hor{\'{a}}nyi, A.%
, Mu{\~{n}}oz-Sabater, J.%
\BDBL {}Th{\'{e}}paut, J\BPBI N.%
\end{APACrefauthors}%
\unskip\
\newblock
\APACrefYearMonthDay{2020}{}{}.
\newblock
{\BBOQ}\APACrefatitle {{The ERA5 global reanalysis}} {{The ERA5 global
  reanalysis}}.{\BBCQ}
\newblock
\APACjournalVolNumPages{Quarterly Journal of the Royal Meteorological
  Society}{146}{730}{1999--2049}.
\newblock
\begin{APACrefDOI} \doi{10.1002/qj.3803} \end{APACrefDOI}
\PrintBackRefs{\CurrentBib}

\bibitem [\protect \citeauthoryear {%
Hewson%
\ \BBA {} Neu%
}{%
Hewson%
\ \BBA {} Neu%
}{%
{\protect \APACyear {2015}}%
}]{%
hewson_cyclones_2015}
\APACinsertmetastar {%
hewson_cyclones_2015}%
\begin{APACrefauthors}%
Hewson, T\BPBI D.%
\BCBT {}\ \BBA {} Neu, U.%
\end{APACrefauthors}%
\unskip\
\newblock
\APACrefYearMonthDay{2015}{}{}.
\newblock
{\BBOQ}\APACrefatitle {Cyclones, windstorms and the {IMILAST} project}
  {Cyclones, windstorms and the {IMILAST} project}.{\BBCQ}
\newblock
\APACjournalVolNumPages{Tellus, Series A: Dynamic Meteorology and
  Oceanography}{6}{1}{}.
\newblock
\begin{APACrefDOI} \doi{10.3402/tellusa.v67.27128} \end{APACrefDOI}
\PrintBackRefs{\CurrentBib}

\bibitem [\protect \citeauthoryear {%
Hsu%
\ \BBA {} Yeh%
}{%
Hsu%
\ \BBA {} Yeh%
}{%
{\protect \APACyear {2021}}%
}]{%
en14123702}
\APACinsertmetastar {%
en14123702}%
\begin{APACrefauthors}%
Hsu, W\BHBI K.%
\BCBT {}\ \BBA {} Yeh, C\BHBI K.%
\end{APACrefauthors}%
\unskip\
\newblock
\APACrefYearMonthDay{2021}{}{}.
\newblock
{\BBOQ}\APACrefatitle {Offshore Wind Potential of West Central Taiwan: A Case
  Study} {Offshore wind potential of west central taiwan: A case study}.{\BBCQ}
\newblock
\APACjournalVolNumPages{Energies}{14}{12}{}.
\newblock
\begin{APACrefURL} \url{https://www.mdpi.com/1996-1073/14/12/3702}
  \end{APACrefURL}
\newblock
\begin{APACrefDOI} \doi{10.3390/en14123702} \end{APACrefDOI}
\PrintBackRefs{\CurrentBib}

\bibitem [\protect \citeauthoryear {%
Huggins%
}{%
Huggins%
}{%
{\protect \APACyear {2021}}%
}]{%
osti_1832158}
\APACinsertmetastar {%
osti_1832158}%
\begin{APACrefauthors}%
Huggins, D.%
\end{APACrefauthors}%
\unskip\
\newblock
\APACrefYearMonthDay{2021}{}{}.
\newblock
\APACrefbtitle {AmeriFlux FLUXNET-1F US-CF1 CAF-LTAR Cook East.} {Ameriflux
  fluxnet-1f us-cf1 caf-ltar cook east.}
\newblock
\begin{APACrefDOI} \doi{10.17190/AMF/1832158} \end{APACrefDOI}
\PrintBackRefs{\CurrentBib}

\bibitem [\protect \citeauthoryear {%
Huggins%
}{%
Huggins%
}{%
{\protect \APACyear {2022}}%
{\protect \APACexlab {{\protect \BCnt {1}}}}}]{%
osti_1881573}
\APACinsertmetastar {%
osti_1881573}%
\begin{APACrefauthors}%
Huggins, D.%
\end{APACrefauthors}%
\unskip\
\newblock
\APACrefYearMonthDay{2022{\protect \BCnt {1}}}{}{}.
\newblock
\APACrefbtitle {AmeriFlux FLUXNET-1F US-CF2 CAF-LTAR Cook West.} {Ameriflux
  fluxnet-1f us-cf2 caf-ltar cook west.}
\newblock
\begin{APACrefDOI} \doi{10.17190/AMF/1881573} \end{APACrefDOI}
\PrintBackRefs{\CurrentBib}

\bibitem [\protect \citeauthoryear {%
Huggins%
}{%
Huggins%
}{%
{\protect \APACyear {2022}}%
{\protect \APACexlab {{\protect \BCnt {2}}}}}]{%
osti_1881574}
\APACinsertmetastar {%
osti_1881574}%
\begin{APACrefauthors}%
Huggins, D.%
\end{APACrefauthors}%
\unskip\
\newblock
\APACrefYearMonthDay{2022{\protect \BCnt {2}}}{}{}.
\newblock
\APACrefbtitle {AmeriFlux FLUXNET-1F US-CF3 CAF-LTAR Boyd North.} {Ameriflux
  fluxnet-1f us-cf3 caf-ltar boyd north.}
\newblock
\begin{APACrefDOI} \doi{10.17190/AMF/1881574} \end{APACrefDOI}
\PrintBackRefs{\CurrentBib}

\bibitem [\protect \citeauthoryear {%
Huggins%
}{%
Huggins%
}{%
{\protect \APACyear {2022}}%
{\protect \APACexlab {{\protect \BCnt {3}}}}}]{%
osti_1881575}
\APACinsertmetastar {%
osti_1881575}%
\begin{APACrefauthors}%
Huggins, D.%
\end{APACrefauthors}%
\unskip\
\newblock
\APACrefYearMonthDay{2022{\protect \BCnt {3}}}{}{}.
\newblock
\APACrefbtitle {AmeriFlux FLUXNET-1F US-CF4 CAF-LTAR Boyd South.} {Ameriflux
  fluxnet-1f us-cf4 caf-ltar boyd south.}
\newblock
\begin{APACrefDOI} \doi{10.17190/AMF/1881575} \end{APACrefDOI}
\PrintBackRefs{\CurrentBib}

\bibitem [\protect \citeauthoryear {%
Hung%
\ \protect \BOthers {.}}{%
Hung%
\ \protect \BOthers {.}}{%
{\protect \APACyear {2024}}%
}]{%
https://doi.org/10.1029/2024MS004300}
\APACinsertmetastar {%
https://doi.org/10.1029/2024MS004300}%
\begin{APACrefauthors}%
Hung, W\BHBI T.%
, Campbell, P\BPBI C.%
, Moon, Z.%
, Saylor, R.%
, Kochendorfer, J.%
, Lee, T\BPBI R.%
\BCBL {}\ \BBA {} Massman, W.%
\end{APACrefauthors}%
\unskip\
\newblock
\APACrefYearMonthDay{2024}{}{}.
\newblock
{\BBOQ}\APACrefatitle {Evaluation of an In-Canopy Wind and Wind Adjustment
  Factor Model for Wildfire Spread Applications Across Scales} {Evaluation of
  an in-canopy wind and wind adjustment factor model for wildfire spread
  applications across scales}.{\BBCQ}
\newblock
\APACjournalVolNumPages{Journal of Advances in Modeling Earth
  Systems}{16}{7}{e2024MS004300}.
\newblock
\begin{APACrefURL}
  \url{https://agupubs.onlinelibrary.wiley.com/doi/abs/10.1029/2024MS004300}
  \end{APACrefURL}
\newblock
\APACrefnote{e2024MS004300 2024MS004300}
\newblock
\begin{APACrefDOI} \doi{https://doi.org/10.1029/2024MS004300} \end{APACrefDOI}
\PrintBackRefs{\CurrentBib}

\bibitem [\protect \citeauthoryear {%
Jiménez%
\ \BBA {} Dudhia%
}{%
Jiménez%
\ \BBA {} Dudhia%
}{%
{\protect \APACyear {2012}}%
}]{%
Jimenez_Dudhia_2012}
\APACinsertmetastar {%
Jimenez_Dudhia_2012}%
\begin{APACrefauthors}%
Jiménez, P\BPBI A.%
\BCBT {}\ \BBA {} Dudhia, J.%
\end{APACrefauthors}%
\unskip\
\newblock
\APACrefYearMonthDay{2012}{}{}.
\newblock
{\BBOQ}\APACrefatitle {Improving the Representation of Resolved and Unresolved
  Topographic Effects on Surface Wind in the WRF Model} {Improving the
  representation of resolved and unresolved topographic effects on surface wind
  in the wrf model}.{\BBCQ}
\newblock
\APACjournalVolNumPages{Journal of Applied Meteorology and
  Climatology}{51}{2}{300 - 316}.
\newblock
\begin{APACrefURL}
  \url{https://journals.ametsoc.org/view/journals/apme/51/2/jamc-d-11-084.1.xml}
  \end{APACrefURL}
\newblock
\begin{APACrefDOI} \doi{10.1175/JAMC-D-11-084.1} \end{APACrefDOI}
\PrintBackRefs{\CurrentBib}

\bibitem [\protect \citeauthoryear {%
Jiménez%
\ \BBA {} Dudhia%
}{%
Jiménez%
\ \BBA {} Dudhia%
}{%
{\protect \APACyear {2013}}%
}]{%
Jimenez_Dudhia_2013}
\APACinsertmetastar {%
Jimenez_Dudhia_2013}%
\begin{APACrefauthors}%
Jiménez, P\BPBI A.%
\BCBT {}\ \BBA {} Dudhia, J.%
\end{APACrefauthors}%
\unskip\
\newblock
\APACrefYearMonthDay{2013}{}{}.
\newblock
{\BBOQ}\APACrefatitle {On the Ability of the {WRF} Model to Reproduce the
  Surface Wind Direction over Complex Terrain} {On the ability of the {WRF}
  model to reproduce the surface wind direction over complex terrain}.{\BBCQ}
\newblock
\APACjournalVolNumPages{Journal of Applied Meteorology and
  Climatology}{52}{7}{1610 - 1617}.
\newblock
\begin{APACrefURL}
  \url{https://journals.ametsoc.org/view/journals/apme/52/7/jamc-d-12-0266.1.xml}
  \end{APACrefURL}
\newblock
\begin{APACrefDOI} \doi{10.1175/JAMC-D-12-0266.1} \end{APACrefDOI}
\PrintBackRefs{\CurrentBib}

\bibitem [\protect \citeauthoryear {%
Jiménez%
\ \protect \BOthers {.}}{%
Jiménez%
\ \protect \BOthers {.}}{%
{\protect \APACyear {2012}}%
}]{%
Jimenez_2012}
\APACinsertmetastar {%
Jimenez_2012}%
\begin{APACrefauthors}%
Jiménez, P\BPBI A.%
, Dudhia, J.%
, González-Rouco, J\BPBI F.%
, Navarro, J.%
, Montávez, J\BPBI P.%
\BCBL {}\ \BBA {} García-Bustamante, E.%
\end{APACrefauthors}%
\unskip\
\newblock
\APACrefYearMonthDay{2012}{}{}.
\newblock
{\BBOQ}\APACrefatitle {A Revised Scheme for the {WRF} Surface Layer
  Formulation} {A revised scheme for the {WRF} surface layer
  formulation}.{\BBCQ}
\newblock
\APACjournalVolNumPages{Monthly Weather Review}{140}{3}{898 - 918}.
\newblock
\begin{APACrefURL}
  \url{https://journals.ametsoc.org/view/journals/mwre/140/3/mwr-d-11-00056.1.xml}
  \end{APACrefURL}
\newblock
\begin{APACrefDOI} \doi{10.1175/MWR-D-11-00056.1} \end{APACrefDOI}
\PrintBackRefs{\CurrentBib}

\bibitem [\protect \citeauthoryear {%
Jouan%
, Milbrandt%
, Vaillancourt%
, Chosson%
\BCBL {}\ \BBA {} Morrison%
}{%
Jouan%
\ \protect \BOthers {.}}{%
{\protect \APACyear {2020}}%
}]{%
jouan_adaptation_2020}
\APACinsertmetastar {%
jouan_adaptation_2020}%
\begin{APACrefauthors}%
Jouan, C.%
, Milbrandt, J\BPBI A.%
, Vaillancourt, P\BPBI A.%
, Chosson, F.%
\BCBL {}\ \BBA {} Morrison, H.%
\end{APACrefauthors}%
\unskip\
\newblock
\APACrefYearMonthDay{2020}{}{}.
\newblock
{\BBOQ}\APACrefatitle {Adaptation of the predicted particles properties ({P3})
  microphysics scheme for large-scale numerical weather prediction} {Adaptation
  of the predicted particles properties ({P3}) microphysics scheme for
  large-scale numerical weather prediction}.{\BBCQ}
\newblock
\APACjournalVolNumPages{Weather and Forecasting}{35}{6}{2541--2565}.
\newblock
\begin{APACrefDOI} \doi{10.1175/WAF-D-20-0111.1} \end{APACrefDOI}
\PrintBackRefs{\CurrentBib}

\bibitem [\protect \citeauthoryear {%
Kaimal%
\ \BBA {} Wyngaard%
}{%
Kaimal%
\ \BBA {} Wyngaard%
}{%
{\protect \APACyear {1990}}%
}]{%
kaimal_kansas_1990}
\APACinsertmetastar {%
kaimal_kansas_1990}%
\begin{APACrefauthors}%
Kaimal, J\BPBI C.%
\BCBT {}\ \BBA {} Wyngaard, J\BPBI C.%
\end{APACrefauthors}%
\unskip\
\newblock
\APACrefYearMonthDay{1990}{{\APACmonth{03}}}{}.
\newblock
{\BBOQ}\APACrefatitle {The {Kansas} and {Minnesota} experiments} {The {Kansas}
  and {Minnesota} experiments}.{\BBCQ}
\newblock
\APACjournalVolNumPages{Boundary-Layer Meteorology}{50}{1-4}{31--47}.
\newblock
\begin{APACrefURL}
  [{2024-07-05}]\url{http://link.springer.com/10.1007/BF00120517}
  \end{APACrefURL}
\newblock
\begin{APACrefDOI} \doi{10.1007/BF00120517} \end{APACrefDOI}
\PrintBackRefs{\CurrentBib}

\bibitem [\protect \citeauthoryear {%
Kain%
\ \BBA {} Fritsch%
}{%
Kain%
\ \BBA {} Fritsch%
}{%
{\protect \APACyear {1990}}%
}]{%
Kain1990}
\APACinsertmetastar {%
Kain1990}%
\begin{APACrefauthors}%
Kain, J\BPBI S.%
\BCBT {}\ \BBA {} Fritsch, J\BPBI M.%
\end{APACrefauthors}%
\unskip\
\newblock
\APACrefYearMonthDay{1990}{dec}{}.
\newblock
{\BBOQ}\APACrefatitle {{A One-Dimensional Entraining/Detraining Plume Model and
  Its Application in Convective Parameterization}} {{A One-Dimensional
  Entraining/Detraining Plume Model and Its Application in Convective
  Parameterization}}.{\BBCQ}
\newblock
\APACjournalVolNumPages{Journal of the Atmospheric
  Sciences}{47}{23}{2784--2802}.
\newblock
\begin{APACrefURL}
  \url{http://journals.ametsoc.org/doi/10.1175/1520-0469(1990)047%3C2784:AODEPM%3E2.0.CO;2}
  \end{APACrefURL}
\newblock
\begin{APACrefDOI} \doi{10.1175/1520-0469(1990)047<2784:AODEPM>2.0.CO;2}
  \end{APACrefDOI}
\PrintBackRefs{\CurrentBib}

\bibitem [\protect \citeauthoryear {%
Knox%
, Matthes%
, Verfaillie%
\BCBL {}\ \BBA {} Baldocchi%
}{%
Knox%
\ \protect \BOthers {.}}{%
{\protect \APACyear {2016}}%
}]{%
osti_1246140}
\APACinsertmetastar {%
osti_1246140}%
\begin{APACrefauthors}%
Knox, S.%
, Matthes, J\BPBI H.%
, Verfaillie, J.%
\BCBL {}\ \BBA {} Baldocchi, D.%
\end{APACrefauthors}%
\unskip\
\newblock
\APACrefYearMonthDay{2016}{}{}.
\newblock
\APACrefbtitle {AmeriFlux US-Twt Twitchell Island.} {Ameriflux us-twt twitchell
  island.}
\newblock
\begin{APACrefURL} \url{https://www.osti.gov/biblio/1246140} \end{APACrefURL}
\newblock
\begin{APACrefDOI} \doi{10.17190/AMF/1246140} \end{APACrefDOI}
\PrintBackRefs{\CurrentBib}

\bibitem [\protect \citeauthoryear {%
Kumar%
\ \BBA {} Sharan%
}{%
Kumar%
\ \BBA {} Sharan%
}{%
{\protect \APACyear {2012}}%
}]{%
kumar_analysis_2012}
\APACinsertmetastar {%
kumar_analysis_2012}%
\begin{APACrefauthors}%
Kumar, P.%
\BCBT {}\ \BBA {} Sharan, M.%
\end{APACrefauthors}%
\unskip\
\newblock
\APACrefYearMonthDay{2012}{{\APACmonth{06}}}{}.
\newblock
{\BBOQ}\APACrefatitle {An {Analysis} for the {Applicability} of
  {Monin}–{Obukhov} {Similarity} {Theory} in {Stable} {Conditions}} {An
  {Analysis} for the {Applicability} of {Monin}–{Obukhov} {Similarity}
  {Theory} in {Stable} {Conditions}}.{\BBCQ}
\newblock
\APACjournalVolNumPages{Journal of the Atmospheric
  Sciences}{69}{6}{1910--1915}.
\newblock
\begin{APACrefURL}
  [{2024-11-27}]\url{https://journals.ametsoc.org/doi/10.1175/JAS-D-11-0250.1}
  \end{APACrefURL}
\newblock
\begin{APACrefDOI} \doi{10.1175/JAS-D-11-0250.1} \end{APACrefDOI}
\PrintBackRefs{\CurrentBib}

\bibitem [\protect \citeauthoryear {%
Kusak%
, Sanchez%
, Szutu%
\BCBL {}\ \BBA {} Baldocchi%
}{%
Kusak%
\ \protect \BOthers {.}}{%
{\protect \APACyear {2022}}%
}]{%
osti_1854371}
\APACinsertmetastar {%
osti_1854371}%
\begin{APACrefauthors}%
Kusak, K.%
, Sanchez, C\BPBI R.%
, Szutu, D.%
\BCBL {}\ \BBA {} Baldocchi, D.%
\end{APACrefauthors}%
\unskip\
\newblock
\APACrefYearMonthDay{2022}{}{}.
\newblock
\APACrefbtitle {AmeriFlux FLUXNET-1F US-Snf Sherman Barn.} {Ameriflux
  fluxnet-1f us-snf sherman barn.}
\newblock
\begin{APACrefDOI} \doi{10.17190/AMF/1854371} \end{APACrefDOI}
\PrintBackRefs{\CurrentBib}

\bibitem [\protect \citeauthoryear {%
Lee%
\ \BBA {} Buban%
}{%
Lee%
\ \BBA {} Buban%
}{%
{\protect \APACyear {2020}}%
}]{%
Lee_2022}
\APACinsertmetastar {%
Lee_2022}%
\begin{APACrefauthors}%
Lee, T\BPBI R.%
\BCBT {}\ \BBA {} Buban, M.%
\end{APACrefauthors}%
\unskip\
\newblock
\APACrefYearMonthDay{2020}{}{}.
\newblock
{\BBOQ}\APACrefatitle {Evaluation of {Monin–Obukhov} and Bulk {Richardson}
  Parameterizations for Surface–Atmosphere Exchange} {Evaluation of
  {Monin–Obukhov} and bulk {Richardson} parameterizations for
  surface–atmosphere exchange}.{\BBCQ}
\newblock
\APACjournalVolNumPages{Journal of Applied Meteorology and
  Climatology}{59}{6}{1091 - 1107}.
\newblock
\begin{APACrefURL}
  \url{https://journals.ametsoc.org/view/journals/apme/59/6/jamcD190057.xml}
  \end{APACrefURL}
\newblock
\begin{APACrefDOI} \doi{10.1175/JAMC-D-19-0057.1} \end{APACrefDOI}
\PrintBackRefs{\CurrentBib}

\bibitem [\protect \citeauthoryear {%
Ma%
, Xu%
, Verfaillie%
\BCBL {}\ \BBA {} Baldocchi%
}{%
Ma%
\ \protect \BOthers {.}}{%
{\protect \APACyear {2016}}%
}]{%
osti_1245984}
\APACinsertmetastar {%
osti_1245984}%
\begin{APACrefauthors}%
Ma, S.%
, Xu, L.%
, Verfaillie, J.%
\BCBL {}\ \BBA {} Baldocchi, D.%
\end{APACrefauthors}%
\unskip\
\newblock
\APACrefYearMonthDay{2016}{}{}.
\newblock
\APACrefbtitle {AmeriFlux US-Var Vaira Ranch- Ione.} {Ameriflux us-var vaira
  ranch- ione.}
\newblock
\begin{APACrefDOI} \doi{10.17190/AMF/1245984} \end{APACrefDOI}
\PrintBackRefs{\CurrentBib}

\bibitem [\protect \citeauthoryear {%
Mahrt%
}{%
Mahrt%
}{%
{\protect \APACyear {1998}}%
}]{%
Mahrt1998}
\APACinsertmetastar {%
Mahrt1998}%
\begin{APACrefauthors}%
Mahrt, L.%
\end{APACrefauthors}%
\unskip\
\newblock
\APACrefYearMonthDay{1998}{Jun}{01}.
\newblock
{\BBOQ}\APACrefatitle {Stratified Atmospheric Boundary Layers and Breakdown of
  Models} {Stratified atmospheric boundary layers and breakdown of
  models}.{\BBCQ}
\newblock
\APACjournalVolNumPages{Theoretical and Computational Fluid
  Dynamics}{11}{3}{263-279}.
\newblock
\begin{APACrefURL} \url{https://doi.org/10.1007/s001620050093} \end{APACrefURL}
\newblock
\begin{APACrefDOI} \doi{10.1007/s001620050093} \end{APACrefDOI}
\PrintBackRefs{\CurrentBib}

\bibitem [\protect \citeauthoryear {%
Markowski%
, Lis%
, Turner%
, Lee%
\BCBL {}\ \BBA {} Buban%
}{%
Markowski%
\ \protect \BOthers {.}}{%
{\protect \APACyear {2019}}%
}]{%
ObservationsofNearSurfaceVerticalWindProfilesand}
\APACinsertmetastar {%
ObservationsofNearSurfaceVerticalWindProfilesand}%
\begin{APACrefauthors}%
Markowski, P\BPBI M.%
, Lis, N\BPBI T.%
, Turner, D\BPBI D.%
, Lee, T\BPBI R.%
\BCBL {}\ \BBA {} Buban, M\BPBI S.%
\end{APACrefauthors}%
\unskip\
\newblock
\APACrefYearMonthDay{2019}{}{}.
\newblock
{\BBOQ}\APACrefatitle {Observations of Near-Surface Vertical Wind Profiles and
  Vertical Momentum Fluxes from VORTEX-SE 2017: Comparisons to Monin–Obukhov
  Similarity Theory} {Observations of near-surface vertical wind profiles and
  vertical momentum fluxes from vortex-se 2017: Comparisons to monin–obukhov
  similarity theory}.{\BBCQ}
\newblock
\APACjournalVolNumPages{Monthly Weather Review}{147}{10}{3811 - 3824}.
\newblock
\begin{APACrefURL}
  \url{https://journals.ametsoc.org/view/journals/mwre/147/10/mwr-d-19-0091.1.xml}
  \end{APACrefURL}
\newblock
\begin{APACrefDOI} \doi{10.1175/MWR-D-19-0091.1} \end{APACrefDOI}
\PrintBackRefs{\CurrentBib}

\bibitem [\protect \citeauthoryear {%
Martynov%
, Sushama%
, Laprise%
, Winger%
\BCBL {}\ \BBA {} Dugas%
}{%
Martynov%
\ \protect \BOthers {.}}{%
{\protect \APACyear {2012}}%
}]{%
Martynov2012}
\APACinsertmetastar {%
Martynov2012}%
\begin{APACrefauthors}%
Martynov, A.%
, Sushama, L.%
, Laprise, R.%
, Winger, K.%
\BCBL {}\ \BBA {} Dugas, B.%
\end{APACrefauthors}%
\unskip\
\newblock
\APACrefYearMonthDay{2012}{}{}.
\newblock
{\BBOQ}\APACrefatitle {{Interactive lakes in the Canadian Regional Climate
  Model, version 5: The role of lakes in the regional climate of North
  America}} {{Interactive lakes in the Canadian Regional Climate Model, version
  5: The role of lakes in the regional climate of North America}}.{\BBCQ}
\newblock
\APACjournalVolNumPages{Tellus, Series A: Dynamic Meteorology and
  Oceanography}{64}{1}{0--22}.
\newblock
\begin{APACrefDOI} \doi{10.3402/tellusa.v64i0.16226} \end{APACrefDOI}
\PrintBackRefs{\CurrentBib}

\bibitem [\protect \citeauthoryear {%
Matthes%
\ \protect \BOthers {.}}{%
Matthes%
\ \protect \BOthers {.}}{%
{\protect \APACyear {2020}}%
}]{%
osti_1669685}
\APACinsertmetastar {%
osti_1669685}%
\begin{APACrefauthors}%
Matthes, J\BPBI H.%
, Sturtevant, C.%
, Oikawa, P.%
, Chamberlain, S\BPBI D.%
, Szutu, D.%
, Ortiz, A\BPBI A.%
\BDBL {}Baldocchi, D.%
\end{APACrefauthors}%
\unskip\
\newblock
\APACrefYearMonthDay{2020}{}{}.
\newblock
\APACrefbtitle {FLUXNET-CH4 US-Myb Mayberry Wetland.} {Fluxnet-ch4 us-myb
  mayberry wetland.}
\newblock
\begin{APACrefDOI} \doi{10.18140/FLX/1669685} \end{APACrefDOI}
\PrintBackRefs{\CurrentBib}

\bibitem [\protect \citeauthoryear {%
McTaggart-Cowan%
, Vaillancourt%
, Separovic%
, Corvec%
\BCBL {}\ \BBA {} Zadra%
}{%
McTaggart-Cowan%
\ \protect \BOthers {.}}{%
{\protect \APACyear {2020}}%
}]{%
mctaggart-cowan_convection_2020}
\APACinsertmetastar {%
mctaggart-cowan_convection_2020}%
\begin{APACrefauthors}%
McTaggart-Cowan, R.%
, Vaillancourt, P\BPBI A.%
, Separovic, L.%
, Corvec, S.%
\BCBL {}\ \BBA {} Zadra, A.%
\end{APACrefauthors}%
\unskip\
\newblock
\APACrefYearMonthDay{2020}{{\APACmonth{12}}}{}.
\newblock
{\BBOQ}\APACrefatitle {A {Convection} {Parameterization} for {Low}-{CAPE}
  {Environments}} {A {Convection} {Parameterization} for {Low}-{CAPE}
  {Environments}}.{\BBCQ}
\newblock
\APACjournalVolNumPages{Monthly Weather Review}{148}{12}{4917--4941}.
\newblock
\begin{APACrefURL}
  [{2024-07-11}]\url{https://journals.ametsoc.org/view/journals/mwre/148/12/mwr-d-20-0020.1.xml}
  \end{APACrefURL}
\newblock
\begin{APACrefDOI} \doi{10.1175/MWR-D-20-0020.1} \end{APACrefDOI}
\PrintBackRefs{\CurrentBib}

\bibitem [\protect \citeauthoryear {%
McTaggart-Cowan%
, Vaillancourt%
, Zadra%
, Chamberland%
\BCBL {}\ \protect \BOthers {.}}{%
McTaggart-Cowan%
, Vaillancourt%
, Zadra%
, Chamberland%
\BCBL {}\ \protect \BOthers {.}}{%
{\protect \APACyear {2019}}%
}]{%
ModernNWP}
\APACinsertmetastar {%
ModernNWP}%
\begin{APACrefauthors}%
McTaggart-Cowan, R.%
, Vaillancourt, P\BPBI A.%
, Zadra, A.%
, Chamberland, S.%
, Charron, M.%
, Corvec, S.%
\BDBL {}Yang, J.%
\end{APACrefauthors}%
\unskip\
\newblock
\APACrefYearMonthDay{2019}{}{}.
\newblock
{\BBOQ}\APACrefatitle {Modernization of Atmospheric Physics Parameterization in
  Canadian NWP} {Modernization of atmospheric physics parameterization in
  canadian nwp}.{\BBCQ}
\newblock
\APACjournalVolNumPages{Journal of Advances in Modeling Earth
  Systems}{11}{11}{3593-3635}.
\newblock
\begin{APACrefURL}
  \url{https://agupubs.onlinelibrary.wiley.com/doi/abs/10.1029/2019MS001781}
  \end{APACrefURL}
\newblock
\begin{APACrefDOI} \doi{https://doi.org/10.1029/2019MS001781} \end{APACrefDOI}
\PrintBackRefs{\CurrentBib}

\bibitem [\protect \citeauthoryear {%
McTaggart-Cowan%
, Vaillancourt%
, Zadra%
, Separovic%
\BCBL {}\ \protect \BOthers {.}}{%
McTaggart-Cowan%
, Vaillancourt%
, Zadra%
, Separovic%
\BCBL {}\ \protect \BOthers {.}}{%
{\protect \APACyear {2019}}%
}]{%
McTaggart-Cowan2019}
\APACinsertmetastar {%
McTaggart-Cowan2019}%
\begin{APACrefauthors}%
McTaggart-Cowan, R.%
, Vaillancourt, P\BPBI A.%
, Zadra, A.%
, Separovic, L.%
, Corvec, S.%
\BCBL {}\ \BBA {} Kirshbaum, D.%
\end{APACrefauthors}%
\unskip\
\newblock
\APACrefYearMonthDay{2019}{}{}.
\newblock
{\BBOQ}\APACrefatitle {{A lagrangian perspective on parameterizing deep
  convection}} {{A lagrangian perspective on parameterizing deep
  convection}}.{\BBCQ}
\newblock
\APACjournalVolNumPages{Monthly Weather Review}{147}{11}{4127--4149}.
\newblock
\begin{APACrefDOI} \doi{10.1175/MWR-D-19-0164.1} \end{APACrefDOI}
\PrintBackRefs{\CurrentBib}

\bibitem [\protect \citeauthoryear {%
Mctaggart-Cowan%
\ \BBA {} Zadra%
}{%
Mctaggart-Cowan%
\ \BBA {} Zadra%
}{%
{\protect \APACyear {2015}}%
}]{%
Mctaggart-Cowan2015}
\APACinsertmetastar {%
Mctaggart-Cowan2015}%
\begin{APACrefauthors}%
Mctaggart-Cowan, R.%
\BCBT {}\ \BBA {} Zadra, A.%
\end{APACrefauthors}%
\unskip\
\newblock
\APACrefYearMonthDay{2015}{}{}.
\newblock
{\BBOQ}\APACrefatitle {{Representing richardson number hysteresis in the NWP
  boundary layer}} {{Representing richardson number hysteresis in the NWP
  boundary layer}}.{\BBCQ}
\newblock
\APACjournalVolNumPages{Monthly Weather Review}{143}{4}{1232--1258}.
\newblock
\begin{APACrefDOI} \doi{10.1175/MWR-D-14-00179.1} \end{APACrefDOI}
\PrintBackRefs{\CurrentBib}

\bibitem [\protect \citeauthoryear {%
Milbrandt%
\ \BBA {} Morrison%
}{%
Milbrandt%
\ \BBA {} Morrison%
}{%
{\protect \APACyear {2016}}%
}]{%
Milbrandt2016}
\APACinsertmetastar {%
Milbrandt2016}%
\begin{APACrefauthors}%
Milbrandt, J\BPBI A.%
\BCBT {}\ \BBA {} Morrison, H.%
\end{APACrefauthors}%
\unskip\
\newblock
\APACrefYearMonthDay{2016}{}{}.
\newblock
{\BBOQ}\APACrefatitle {{Parameterization of cloud microphysics based on the
  prediction of bulk ice particle properties. Part III: Introduction of
  multiple free categories}} {{Parameterization of cloud microphysics based on
  the prediction of bulk ice particle properties. Part III: Introduction of
  multiple free categories}}.{\BBCQ}
\newblock
\APACjournalVolNumPages{Journal of the Atmospheric Sciences}{73}{3}{975--995}.
\newblock
\begin{APACrefDOI} \doi{10.1175/JAS-D-15-0204.1} \end{APACrefDOI}
\PrintBackRefs{\CurrentBib}

\bibitem [\protect \citeauthoryear {%
Monin%
\ \BBA {} Obukhov%
}{%
Monin%
\ \BBA {} Obukhov%
}{%
{\protect \APACyear {1954}}%
}]{%
monin_monin_and_obukhov_1954_1959}
\APACinsertmetastar {%
monin_monin_and_obukhov_1954_1959}%
\begin{APACrefauthors}%
Monin, A\BPBI S.%
\BCBT {}\ \BBA {} Obukhov, A\BPBI M.%
\end{APACrefauthors}%
\unskip\
\newblock
\APACrefYearMonthDay{1954}{}{}.
\newblock
{\BBOQ}\APACrefatitle {{Osnovne zakonomernosti turbulentnogo peremeshivaniya v
  prizemnom sloe atmosfery} (Basic Laws of Turbulent Mixing in the Ground Layer
  of the Atmosphere)} {{Osnovne zakonomernosti turbulentnogo peremeshivaniya v
  prizemnom sloe atmosfery} (basic laws of turbulent mixing in the ground layer
  of the atmosphere)}.{\BBCQ}
\newblock
\APACjournalVolNumPages{Trudy Geofizicheskogo Instituta, Akademiya Nauk
  SSSR}{24}{151}{163--187}.
\PrintBackRefs{\CurrentBib}

\bibitem [\protect \citeauthoryear {%
Morrison%
\ \BBA {} Milbrandt%
}{%
Morrison%
\ \BBA {} Milbrandt%
}{%
{\protect \APACyear {2015}}%
}]{%
Morrison2015}
\APACinsertmetastar {%
Morrison2015}%
\begin{APACrefauthors}%
Morrison, H.%
\BCBT {}\ \BBA {} Milbrandt, J\BPBI A.%
\end{APACrefauthors}%
\unskip\
\newblock
\APACrefYearMonthDay{2015}{}{}.
\newblock
{\BBOQ}\APACrefatitle {{Parameterization of cloud microphysics based on the
  prediction of bulk ice particle properties. Part I: Scheme description and
  idealized tests}} {{Parameterization of cloud microphysics based on the
  prediction of bulk ice particle properties. Part I: Scheme description and
  idealized tests}}.{\BBCQ}
\newblock
\APACjournalVolNumPages{Journal of the Atmospheric Sciences}{72}{1}{287--311}.
\newblock
\begin{APACrefDOI} \doi{10.1175/JAS-D-14-0065.1} \end{APACrefDOI}
\PrintBackRefs{\CurrentBib}

\bibitem [\protect \citeauthoryear {%
Nelli%
\ \protect \BOthers {.}}{%
Nelli%
\ \protect \BOthers {.}}{%
{\protect \APACyear {2020}}%
}]{%
Nelli_2020}
\APACinsertmetastar {%
Nelli_2020}%
\begin{APACrefauthors}%
Nelli, N\BPBI R.%
, Temimi, M.%
, Fonseca, R\BPBI M.%
, Weston, M\BPBI J.%
, Thota, M\BPBI S.%
, Valappil, V\BPBI K.%
\BDBL {}Al~Naqbi, H.%
\end{APACrefauthors}%
\unskip\
\newblock
\APACrefYearMonthDay{2020}{}{}.
\newblock
{\BBOQ}\APACrefatitle {Impact of Roughness Length on WRF Simulated
  Land-Atmosphere Interactions Over a Hyper-Arid Region} {Impact of roughness
  length on wrf simulated land-atmosphere interactions over a hyper-arid
  region}.{\BBCQ}
\newblock
\APACjournalVolNumPages{Earth and Space Science}{7}{6}{e2020EA001165}.
\newblock
\begin{APACrefURL}
  \url{https://agupubs.onlinelibrary.wiley.com/doi/abs/10.1029/2020EA001165}
  \end{APACrefURL}
\newblock
\APACrefnote{e2020EA001165 2020EA001165}
\newblock
\begin{APACrefDOI} \doi{https://doi.org/10.1029/2020EA001165} \end{APACrefDOI}
\PrintBackRefs{\CurrentBib}

\bibitem [\protect \citeauthoryear {%
Noilhan%
\ \BBA {} Mahfouf%
}{%
Noilhan%
\ \BBA {} Mahfouf%
}{%
{\protect \APACyear {1996}}%
}]{%
ISBA}
\APACinsertmetastar {%
ISBA}%
\begin{APACrefauthors}%
Noilhan, J.%
\BCBT {}\ \BBA {} Mahfouf, J\BHBI F.%
\end{APACrefauthors}%
\unskip\
\newblock
\APACrefYearMonthDay{1996}{}{}.
\newblock
{\BBOQ}\APACrefatitle {The ISBA land surface parameterisation scheme} {The isba
  land surface parameterisation scheme}.{\BBCQ}
\newblock
\APACjournalVolNumPages{Global and Planetary Change}{13}{1}{145-159}.
\newblock
\begin{APACrefURL}
  \url{https://www.sciencedirect.com/science/article/pii/0921818195000437}
  \end{APACrefURL}
\newblock
\APACrefnote{Soil Moisture Simulation}
\newblock
\begin{APACrefDOI} \doi{https://doi.org/10.1016/0921-8181(95)00043-7}
  \end{APACrefDOI}
\PrintBackRefs{\CurrentBib}

\bibitem [\protect \citeauthoryear {%
Oikawa%
}{%
Oikawa%
}{%
{\protect \APACyear {2021}}%
}]{%
osti_1832159}
\APACinsertmetastar {%
osti_1832159}%
\begin{APACrefauthors}%
Oikawa, P.%
\end{APACrefauthors}%
\unskip\
\newblock
\APACrefYearMonthDay{2021}{}{}.
\newblock
\APACrefbtitle {AmeriFlux FLUXNET-1F US-EDN Eden Landing Ecological Reserve.}
  {Ameriflux fluxnet-1f us-edn eden landing ecological reserve.}
\newblock
\begin{APACrefDOI} \doi{10.17190/AMF/1832159} \end{APACrefDOI}
\PrintBackRefs{\CurrentBib}

\bibitem [\protect \citeauthoryear {%
Optis%
, Monahan%
\BCBL {}\ \BBA {} Bosveld%
}{%
Optis%
\ \protect \BOthers {.}}{%
{\protect \APACyear {2016}}%
}]{%
stabilit_mo}
\APACinsertmetastar {%
stabilit_mo}%
\begin{APACrefauthors}%
Optis, M.%
, Monahan, A.%
\BCBL {}\ \BBA {} Bosveld, F\BPBI C.%
\end{APACrefauthors}%
\unskip\
\newblock
\APACrefYearMonthDay{2016}{}{}.
\newblock
{\BBOQ}\APACrefatitle {Limitations and breakdown of Monin–Obukhov similarity
  theory for wind profile extrapolation under stable stratification}
  {Limitations and breakdown of monin–obukhov similarity theory for wind
  profile extrapolation under stable stratification}.{\BBCQ}
\newblock
\APACjournalVolNumPages{Wind Energy}{19}{6}{1053-1072}.
\newblock
\begin{APACrefURL}
  \url{https://onlinelibrary.wiley.com/doi/abs/10.1002/we.1883}
  \end{APACrefURL}
\newblock
\begin{APACrefDOI} \doi{https://doi.org/10.1002/we.1883} \end{APACrefDOI}
\PrintBackRefs{\CurrentBib}

\bibitem [\protect \citeauthoryear {%
Osczevski%
\ \BBA {} Bluestein%
}{%
Osczevski%
\ \BBA {} Bluestein%
}{%
{\protect \APACyear {2005}}%
}]{%
WINDCHILL}
\APACinsertmetastar {%
WINDCHILL}%
\begin{APACrefauthors}%
Osczevski, R.%
\BCBT {}\ \BBA {} Bluestein, M.%
\end{APACrefauthors}%
\unskip\
\newblock
\APACrefYearMonthDay{2005}{}{}.
\newblock
{\BBOQ}\APACrefatitle {THE NEW WIND CHILL EQUIVALENT TEMPERATURE CHART} {The
  new wind chill equivalent temperature chart}.{\BBCQ}
\newblock
\APACjournalVolNumPages{Bulletin of the American Meteorological
  Society}{86}{10}{1453 - 1458}.
\newblock
\begin{APACrefURL}
  \url{https://journals.ametsoc.org/view/journals/bams/86/10/bams-86-10-1453.xml}
  \end{APACrefURL}
\newblock
\begin{APACrefDOI} \doi{10.1175/BAMS-86-10-1453} \end{APACrefDOI}
\PrintBackRefs{\CurrentBib}

\bibitem [\protect \citeauthoryear {%
O’Neill%
}{%
O’Neill%
}{%
{\protect \APACyear {2012}}%
}]{%
Oneill_2012}
\APACinsertmetastar {%
Oneill_2012}%
\begin{APACrefauthors}%
O’Neill, L\BPBI W.%
\end{APACrefauthors}%
\unskip\
\newblock
\APACrefYearMonthDay{2012}{}{}.
\newblock
{\BBOQ}\APACrefatitle {Wind Speed and Stability Effects on Coupling between
  Surface Wind Stress and SST Observed from Buoys and Satellite} {Wind speed
  and stability effects on coupling between surface wind stress and sst
  observed from buoys and satellite}.{\BBCQ}
\newblock
\APACjournalVolNumPages{Journal of Climate}{25}{5}{1544 - 1569}.
\newblock
\begin{APACrefURL}
  \url{https://journals.ametsoc.org/view/journals/clim/25/5/jcli-d-11-00121.1.xml}
  \end{APACrefURL}
\newblock
\begin{APACrefDOI} \doi{10.1175/JCLI-D-11-00121.1} \end{APACrefDOI}
\PrintBackRefs{\CurrentBib}

\bibitem [\protect \citeauthoryear {%
Picart%
, Di~Luca%
\BCBL {}\ \BBA {} Laprise%
}{%
Picart%
\ \protect \BOthers {.}}{%
{\protect \APACyear {2024}}%
}]{%
picart_uncertainty_2024}
\APACinsertmetastar {%
picart_uncertainty_2024}%
\begin{APACrefauthors}%
Picart, T.%
, Di~Luca, A.%
\BCBL {}\ \BBA {} Laprise, R.%
\end{APACrefauthors}%
\unskip\
\newblock
\APACrefYearMonthDay{2024}{{\APACmonth{02}}}{}.
\newblock
{\BBOQ}\APACrefatitle {Uncertainty and outliers in high‐resolution gridded
  precipitation products over eastern {North} {America}} {Uncertainty and
  outliers in high‐resolution gridded precipitation products over eastern
  {North} {America}}.{\BBCQ}
\newblock
\APACjournalVolNumPages{International Journal of Climatology}{}{June
  2023}{1--22}.
\newblock
\begin{APACrefURL}
  \url{https://rmets.onlinelibrary.wiley.com/doi/10.1002/joc.8369}
  \end{APACrefURL}
\newblock
\begin{APACrefDOI} \doi{10.1002/joc.8369} \end{APACrefDOI}
\PrintBackRefs{\CurrentBib}

\bibitem [\protect \citeauthoryear {%
Pryor%
\ \BBA {} Barthelmie%
}{%
Pryor%
\ \BBA {} Barthelmie%
}{%
{\protect \APACyear {2021}}%
}]{%
Pryor2021}
\APACinsertmetastar {%
Pryor2021}%
\begin{APACrefauthors}%
Pryor, S\BPBI C.%
\BCBT {}\ \BBA {} Barthelmie, R\BPBI J.%
\end{APACrefauthors}%
\unskip\
\newblock
\APACrefYearMonthDay{2021}{Mar}{01}.
\newblock
{\BBOQ}\APACrefatitle {A global assessment of extreme wind speeds for wind
  energy applications} {A global assessment of extreme wind speeds for wind
  energy applications}.{\BBCQ}
\newblock
\APACjournalVolNumPages{Nature Energy}{6}{3}{268-276}.
\newblock
\begin{APACrefURL} \url{https://doi.org/10.1038/s41560-020-00773-7}
  \end{APACrefURL}
\newblock
\begin{APACrefDOI} \doi{10.1038/s41560-020-00773-7} \end{APACrefDOI}
\PrintBackRefs{\CurrentBib}

\bibitem [\protect \citeauthoryear {%
Rey-Sanchez%
, Wang%
, Szutu%
, Hemes%
\BCBL {}\ \protect \BOthers {.}}{%
Rey-Sanchez%
, Wang%
, Szutu%
, Hemes%
\BCBL {}\ \protect \BOthers {.}}{%
{\protect \APACyear {2022}}%
}]{%
osti_1871135}
\APACinsertmetastar {%
osti_1871135}%
\begin{APACrefauthors}%
Rey-Sanchez, C.%
, Wang, C\BPBI T.%
, Szutu, D.%
, Hemes, K.%
, Verfaillie, J.%
\BCBL {}\ \BBA {} Baldocchi, D.%
\end{APACrefauthors}%
\unskip\
\newblock
\APACrefYearMonthDay{2022}{}{}.
\newblock
\APACrefbtitle {AmeriFlux FLUXNET-1F US-Bi2 Bouldin Island corn.} {Ameriflux
  fluxnet-1f us-bi2 bouldin island corn.}
\newblock
\begin{APACrefDOI} \doi{10.17190/AMF/1871135} \end{APACrefDOI}
\PrintBackRefs{\CurrentBib}

\bibitem [\protect \citeauthoryear {%
Rey-Sanchez%
, Wang%
, Szutu%
, Shortt%
\BCBL {}\ \protect \BOthers {.}}{%
Rey-Sanchez%
, Wang%
, Szutu%
, Shortt%
\BCBL {}\ \protect \BOthers {.}}{%
{\protect \APACyear {2022}}%
}]{%
osti_1871134}
\APACinsertmetastar {%
osti_1871134}%
\begin{APACrefauthors}%
Rey-Sanchez, C.%
, Wang, C\BPBI T.%
, Szutu, D.%
, Shortt, R.%
, Chamberlain, S\BPBI D.%
, Verfaillie, J.%
\BCBL {}\ \BBA {} Baldocchi, D.%
\end{APACrefauthors}%
\unskip\
\newblock
\APACrefYearMonthDay{2022}{}{}.
\newblock
\APACrefbtitle {AmeriFlux FLUXNET-1F US-Bi1 Bouldin Island Alfalfa.} {Ameriflux
  fluxnet-1f us-bi1 bouldin island alfalfa.}
\newblock
\begin{APACrefDOI} \doi{10.17190/AMF/1871134} \end{APACrefDOI}
\PrintBackRefs{\CurrentBib}

\bibitem [\protect \citeauthoryear {%
Roberge%
, Di~Luca%
, Laprise%
, Lucas-Picher%
\BCBL {}\ \BBA {} Thériault%
}{%
Roberge%
\ \protect \BOthers {.}}{%
{\protect \APACyear {2024}}%
}]{%
roberge_spatial_2024}
\APACinsertmetastar {%
roberge_spatial_2024}%
\begin{APACrefauthors}%
Roberge, F.%
, Di~Luca, A.%
, Laprise, R.%
, Lucas-Picher, P.%
\BCBL {}\ \BBA {} Thériault, J.%
\end{APACrefauthors}%
\unskip\
\newblock
\APACrefYearMonthDay{2024}{{\APACmonth{02}}}{}.
\newblock
{\BBOQ}\APACrefatitle {Spatial spin-up of precipitation in limited-area
  convection-permitting simulations over {North} {America} using the
  {CRCM6}/{GEM5}.0 model} {Spatial spin-up of precipitation in limited-area
  convection-permitting simulations over {North} {America} using the
  {CRCM6}/{GEM5}.0 model}.{\BBCQ}
\newblock
\APACjournalVolNumPages{Geoscientific Model Development}{17}{4}{1497--1510}.
\newblock
\begin{APACrefURL} \url{https://gmd.copernicus.org/articles/17/1497/2024/}
  \end{APACrefURL}
\newblock
\begin{APACrefDOI} \doi{10.5194/gmd-17-1497-2024} \end{APACrefDOI}
\PrintBackRefs{\CurrentBib}

\bibitem [\protect \citeauthoryear {%
Salesky%
\ \BBA {} Chamecki%
}{%
Salesky%
\ \BBA {} Chamecki%
}{%
{\protect \APACyear {2012}}%
}]{%
RandomErrors}
\APACinsertmetastar {%
RandomErrors}%
\begin{APACrefauthors}%
Salesky, S\BPBI T.%
\BCBT {}\ \BBA {} Chamecki, M.%
\end{APACrefauthors}%
\unskip\
\newblock
\APACrefYearMonthDay{2012}{}{}.
\newblock
{\BBOQ}\APACrefatitle {Random Errors in Turbulence Measurements in the
  Atmospheric Surface Layer: Implications for Monin–Obukhov Similarity
  Theory} {Random errors in turbulence measurements in the atmospheric surface
  layer: Implications for monin–obukhov similarity theory}.{\BBCQ}
\newblock
\APACjournalVolNumPages{Journal of the Atmospheric Sciences}{69}{12}{3700 -
  3714}.
\newblock
\begin{APACrefURL}
  \url{https://journals.ametsoc.org/view/journals/atsc/69/12/jas-d-12-096.1.xml}
  \end{APACrefURL}
\newblock
\begin{APACrefDOI} \doi{10.1175/JAS-D-12-096.1} \end{APACrefDOI}
\PrintBackRefs{\CurrentBib}

\bibitem [\protect \citeauthoryear {%
Schumacher%
\ \BBA {} Rasmussen%
}{%
Schumacher%
\ \BBA {} Rasmussen%
}{%
{\protect \APACyear {2020}}%
}]{%
schumacher_formation_2020}
\APACinsertmetastar {%
schumacher_formation_2020}%
\begin{APACrefauthors}%
Schumacher, R\BPBI S.%
\BCBT {}\ \BBA {} Rasmussen, K\BPBI L.%
\end{APACrefauthors}%
\unskip\
\newblock
\APACrefYearMonthDay{2020}{{\APACmonth{06}}}{}.
\newblock
{\BBOQ}\APACrefatitle {The formation, character and changing nature of
  mesoscale convective systems} {The formation, character and changing nature
  of mesoscale convective systems}.{\BBCQ}
\newblock
\APACjournalVolNumPages{Nature Reviews Earth \& Environment}{1}{6}{300--314}.
\newblock
\begin{APACrefURL}
  [{2024-07-05}]\url{https://www.nature.com/articles/s43017-020-0057-7}
  \end{APACrefURL}
\newblock
\begin{APACrefDOI} \doi{10.1038/s43017-020-0057-7} \end{APACrefDOI}
\PrintBackRefs{\CurrentBib}

\bibitem [\protect \citeauthoryear {%
Schwierz%
\ \protect \BOthers {.}}{%
Schwierz%
\ \protect \BOthers {.}}{%
{\protect \APACyear {2010}}%
}]{%
schwierz_modelling_2010}
\APACinsertmetastar {%
schwierz_modelling_2010}%
\begin{APACrefauthors}%
Schwierz, C.%
, Köllner-Heck, P.%
, Zenklusen~Mutter, E.%
, Bresch, D\BPBI N.%
, Vidale, P\BHBI L.%
, Wild, M.%
\BCBL {}\ \BBA {} Schär, C.%
\end{APACrefauthors}%
\unskip\
\newblock
\APACrefYearMonthDay{2010}{{\APACmonth{08}}}{}.
\newblock
{\BBOQ}\APACrefatitle {Modelling {European} winter wind storm losses in current
  and future climate} {Modelling {European} winter wind storm losses in current
  and future climate}.{\BBCQ}
\newblock
\APACjournalVolNumPages{Climatic Change}{101}{3}{485--514}.
\newblock
\begin{APACrefURL} \url{https://doi.org/10.1007/s10584-009-9712-1}
  \end{APACrefURL}
\newblock
\begin{APACrefDOI} \doi{10.1007/s10584-009-9712-1} \end{APACrefDOI}
\PrintBackRefs{\CurrentBib}

\bibitem [\protect \citeauthoryear {%
Separovic%
, de El{\'{i}}a%
\BCBL {}\ \BBA {} Laprise%
}{%
Separovic%
\ \protect \BOthers {.}}{%
{\protect \APACyear {2012}}%
}]{%
Separovic2012}
\APACinsertmetastar {%
Separovic2012}%
\begin{APACrefauthors}%
Separovic, L.%
, de El{\'{i}}a, R.%
\BCBL {}\ \BBA {} Laprise, R.%
\end{APACrefauthors}%
\unskip\
\newblock
\APACrefYearMonthDay{2012}{}{}.
\newblock
{\BBOQ}\APACrefatitle {{Impact of spectral nudging and domain size in studies
  of RCM response to parameter modification}} {{Impact of spectral nudging and
  domain size in studies of RCM response to parameter modification}}.{\BBCQ}
\newblock
\APACjournalVolNumPages{Climate Dynamics}{38}{7-8}{1325--1343}.
\newblock
\begin{APACrefDOI} \doi{10.1007/s00382-011-1072-7} \end{APACrefDOI}
\PrintBackRefs{\CurrentBib}

\bibitem [\protect \citeauthoryear {%
Shin%
, Ming%
, Zhao%
, Chen%
\BCBL {}\ \BBA {} Lin%
}{%
Shin%
\ \protect \BOthers {.}}{%
{\protect \APACyear {2019}}%
}]{%
shin_improved_2019}
\APACinsertmetastar {%
shin_improved_2019}%
\begin{APACrefauthors}%
Shin, H\BPBI H.%
, Ming, Y.%
, Zhao, M.%
, Chen, X.%
\BCBL {}\ \BBA {} Lin, S.%
\end{APACrefauthors}%
\unskip\
\newblock
\APACrefYearMonthDay{2019}{{\APACmonth{04}}}{}.
\newblock
{\BBOQ}\APACrefatitle {Improved {Surface} {Layer} {Simulation} {Using}
  {Refined} {Vertical} {Resolution} in the {GFDL} {Atmospheric} {General}
  {Circulation} {Model}} {Improved {Surface} {Layer} {Simulation} {Using}
  {Refined} {Vertical} {Resolution} in the {GFDL} {Atmospheric} {General}
  {Circulation} {Model}}.{\BBCQ}
\newblock
\APACjournalVolNumPages{Journal of Advances in Modeling Earth
  Systems}{11}{4}{905--917}.
\newblock
\begin{APACrefURL}
  [{2024-07-04}]\url{https://agupubs.onlinelibrary.wiley.com/doi/10.1029/2018MS001437}
  \end{APACrefURL}
\newblock
\begin{APACrefDOI} \doi{10.1029/2018MS001437} \end{APACrefDOI}
\PrintBackRefs{\CurrentBib}

\bibitem [\protect \citeauthoryear {%
Shortt%
, Hemes%
, Szutu%
, Verfaillie%
\BCBL {}\ \BBA {} Baldocchi%
}{%
Shortt%
\ \protect \BOthers {.}}{%
{\protect \APACyear {2022}}%
}]{%
osti_1871144}
\APACinsertmetastar {%
osti_1871144}%
\begin{APACrefauthors}%
Shortt, R.%
, Hemes, K.%
, Szutu, D.%
, Verfaillie, J.%
\BCBL {}\ \BBA {} Baldocchi, D.%
\end{APACrefauthors}%
\unskip\
\newblock
\APACrefYearMonthDay{2022}{}{}.
\newblock
\APACrefbtitle {AmeriFlux FLUXNET-1F US-Sne Sherman Island Restored Wetland.}
  {Ameriflux fluxnet-1f us-sne sherman island restored wetland.}
\newblock
\begin{APACrefDOI} \doi{10.17190/AMF/1871144} \end{APACrefDOI}
\PrintBackRefs{\CurrentBib}

\bibitem [\protect \citeauthoryear {%
Silveira%
}{%
Silveira%
}{%
{\protect \APACyear {2021}}%
}]{%
osti_1832163}
\APACinsertmetastar {%
osti_1832163}%
\begin{APACrefauthors}%
Silveira, M.%
\end{APACrefauthors}%
\unskip\
\newblock
\APACrefYearMonthDay{2021}{}{}.
\newblock
\APACrefbtitle {AmeriFlux FLUXNET-1F US-ONA Florida pine flatwoods.} {Ameriflux
  fluxnet-1f us-ona florida pine flatwoods.}
\newblock
\begin{APACrefDOI} \doi{10.17190/AMF/1832163} \end{APACrefDOI}
\PrintBackRefs{\CurrentBib}

\bibitem [\protect \citeauthoryear {%
Skamarock%
\ \protect \BOthers {.}}{%
Skamarock%
\ \protect \BOthers {.}}{%
{\protect \APACyear {2019}}%
}]{%
skamarock_description_2019}
\APACinsertmetastar {%
skamarock_description_2019}%
\begin{APACrefauthors}%
Skamarock, W\BPBI C.%
, Klemp, J\BPBI B.%
, Dudhia, J.%
, Gill, D\BPBI O.%
, Liu, Z.%
, Berner, J.%
\BDBL {}Huang, X\BHBI Y.%
\end{APACrefauthors}%
\unskip\
\newblock
\APACrefYearMonthDay{2019}{{\APACmonth{03}}}{}.
\newblock
\APACrefbtitle {A {Description} of the {Advanced} {Research} {WRF} {Model}
  {Version} 4} {A {Description} of the {Advanced} {Research} {WRF} {Model}
  {Version} 4}\ \APACbVolEdTR{}{\BTR{}}.
\newblock
\APACaddressInstitution{}{UCAR/NCAR}.
\newblock
\begin{APACrefURL}
  [{2024-07-05}]\url{https://opensky.ucar.edu/islandora/object/opensky:2898}
  \end{APACrefURL}
\newblock
\begin{APACrefDOI} \doi{10.5065/1DFH-6P97} \end{APACrefDOI}
\PrintBackRefs{\CurrentBib}

\bibitem [\protect \citeauthoryear {%
Stocks%
\ \protect \BOthers {.}}{%
Stocks%
\ \protect \BOthers {.}}{%
{\protect \APACyear {1989}}%
}]{%
fwi}
\APACinsertmetastar {%
fwi}%
\begin{APACrefauthors}%
Stocks, B\BPBI J.%
, Lawson, B\BPBI D.%
, Alexander, M\BPBI E.%
, Wagner, C\BPBI E\BPBI V.%
, McAlpine, R\BPBI S.%
, Lynham, T\BPBI J.%
\BCBL {}\ \BBA {} Dubé, D\BPBI E.%
\end{APACrefauthors}%
\unskip\
\newblock
\APACrefYearMonthDay{1989}{}{}.
\newblock
{\BBOQ}\APACrefatitle {The Canadian Forest Fire Danger Rating System: An
  Overview} {The canadian forest fire danger rating system: An
  overview}.{\BBCQ}
\newblock
\APACjournalVolNumPages{The Forestry Chronicle}{65}{6}{450-457}.
\newblock
\begin{APACrefURL} \url{https://doi.org/10.5558/tfc65450-6} \end{APACrefURL}
\newblock
\begin{APACrefDOI} \doi{10.5558/tfc65450-6} \end{APACrefDOI}
\PrintBackRefs{\CurrentBib}

\bibitem [\protect \citeauthoryear {%
Sun%
, Takle%
\BCBL {}\ \BBA {} Acevedo%
}{%
Sun%
\ \protect \BOthers {.}}{%
{\protect \APACyear {2020}}%
}]{%
Sun2020}
\APACinsertmetastar {%
Sun2020}%
\begin{APACrefauthors}%
Sun, J.%
, Takle, E\BPBI S.%
\BCBL {}\ \BBA {} Acevedo, O\BPBI C.%
\end{APACrefauthors}%
\unskip\
\newblock
\APACrefYearMonthDay{2020}{Oct}{01}.
\newblock
{\BBOQ}\APACrefatitle {Understanding Physical Processes Represented by the
  Monin--Obukhov Bulk Formula for Momentum Transfer} {Understanding physical
  processes represented by the monin--obukhov bulk formula for momentum
  transfer}.{\BBCQ}
\newblock
\APACjournalVolNumPages{Boundary-Layer Meteorology}{177}{1}{69-95}.
\newblock
\begin{APACrefURL} \url{https://doi.org/10.1007/s10546-020-00546-5}
  \end{APACrefURL}
\newblock
\begin{APACrefDOI} \doi{10.1007/s10546-020-00546-5} \end{APACrefDOI}
\PrintBackRefs{\CurrentBib}

\bibitem [\protect \citeauthoryear {%
Sušelj%
\ \BBA {} Sood%
}{%
Sušelj%
\ \BBA {} Sood%
}{%
{\protect \APACyear {2010}}%
}]{%
suselj_improving_2010}
\APACinsertmetastar {%
suselj_improving_2010}%
\begin{APACrefauthors}%
Sušelj, K.%
\BCBT {}\ \BBA {} Sood, A.%
\end{APACrefauthors}%
\unskip\
\newblock
\APACrefYearMonthDay{2010}{}{}.
\newblock
{\BBOQ}\APACrefatitle {Improving the {Mellor}-{Yamada}-{Janjić}
  {Parameterization} for wind conditions in the marine planetary boundary
  layer} {Improving the {Mellor}-{Yamada}-{Janjić} {Parameterization} for wind
  conditions in the marine planetary boundary layer}.{\BBCQ}
\newblock
\APACjournalVolNumPages{Boundary-Layer Meteorology}{136}{2}{301--324}.
\newblock
\begin{APACrefDOI} \doi{10.1007/s10546-010-9502-3} \end{APACrefDOI}
\PrintBackRefs{\CurrentBib}

\bibitem [\protect \citeauthoryear {%
Trzeciak%
, Knippertz%
, Pirret%
\BCBL {}\ \BBA {} Williams%
}{%
Trzeciak%
\ \protect \BOthers {.}}{%
{\protect \APACyear {2016}}%
}]{%
trzeciak_can_2016}
\APACinsertmetastar {%
trzeciak_can_2016}%
\begin{APACrefauthors}%
Trzeciak, T\BPBI M.%
, Knippertz, P.%
, Pirret, J\BPBI S\BPBI R.%
\BCBL {}\ \BBA {} Williams, K\BPBI D.%
\end{APACrefauthors}%
\unskip\
\newblock
\APACrefYearMonthDay{2016}{{\APACmonth{06}}}{}.
\newblock
{\BBOQ}\APACrefatitle {Can we trust climate models to realistically represent
  severe {European} windstorms?} {Can we trust climate models to realistically
  represent severe {European} windstorms?}{\BBCQ}
\newblock
\APACjournalVolNumPages{Climate Dynamics}{46}{11}{3431--3451}.
\newblock
\begin{APACrefURL} \url{https://doi.org/10.1007/s00382-015-2777-9}
  \end{APACrefURL}
\newblock
\begin{APACrefDOI} \doi{10.1007/s00382-015-2777-9} \end{APACrefDOI}
\PrintBackRefs{\CurrentBib}

\bibitem [\protect \citeauthoryear {%
Valach%
\ \protect \BOthers {.}}{%
Valach%
\ \protect \BOthers {.}}{%
{\protect \APACyear {2016}}%
}]{%
osti_1246147}
\APACinsertmetastar {%
osti_1246147}%
\begin{APACrefauthors}%
Valach, A.%
, Shortt, R.%
, Szutu, D.%
, Eichelmann, E.%
, Knox, S.%
, Hemes, K.%
\BDBL {}Baldocchi, D.%
\end{APACrefauthors}%
\unskip\
\newblock
\APACrefYearMonthDay{2016}{}{}.
\newblock
\APACrefbtitle {AmeriFlux US-Tw1 Twitchell Wetland West Pond.} {Ameriflux
  us-tw1 twitchell wetland west pond.}
\newblock
\begin{APACrefDOI} \doi{10.17190/AMF/1246147} \end{APACrefDOI}
\PrintBackRefs{\CurrentBib}

\bibitem [\protect \citeauthoryear {%
Vazquez-Lule%
\ \BBA {} Vargas%
}{%
Vazquez-Lule%
\ \BBA {} Vargas%
}{%
{\protect \APACyear {2020}}%
}]{%
osti_1669695}
\APACinsertmetastar {%
osti_1669695}%
\begin{APACrefauthors}%
Vazquez-Lule, A.%
\BCBT {}\ \BBA {} Vargas, R.%
\end{APACrefauthors}%
\unskip\
\newblock
\APACrefYearMonthDay{2020}{}{}.
\newblock
\APACrefbtitle {FLUXNET-CH4 US-StJ St Jones Reserve.} {Fluxnet-ch4 us-stj st
  jones reserve.}
\newblock
\begin{APACrefDOI} \doi{10.18140/FLX/1669695} \end{APACrefDOI}
\PrintBackRefs{\CurrentBib}

\bibitem [\protect \citeauthoryear {%
Verseghy%
}{%
Verseghy%
}{%
{\protect \APACyear {1991}}%
}]{%
CLASS1}
\APACinsertmetastar {%
CLASS1}%
\begin{APACrefauthors}%
Verseghy, D\BPBI L.%
\end{APACrefauthors}%
\unskip\
\newblock
\APACrefYearMonthDay{1991}{}{}.
\newblock
{\BBOQ}\APACrefatitle {Class—A Canadian land surface scheme for GCMS. I. Soil
  model} {Class—a canadian land surface scheme for gcms. i. soil
  model}.{\BBCQ}
\newblock
\APACjournalVolNumPages{International Journal of Climatology}{11}{2}{111-133}.
\newblock
\begin{APACrefURL}
  \url{https://rmets.onlinelibrary.wiley.com/doi/abs/10.1002/joc.3370110202}
  \end{APACrefURL}
\newblock
\begin{APACrefDOI} \doi{https://doi.org/10.1002/joc.3370110202}
  \end{APACrefDOI}
\PrintBackRefs{\CurrentBib}

\bibitem [\protect \citeauthoryear {%
Verseghy%
}{%
Verseghy%
}{%
{\protect \APACyear {2000}}%
}]{%
verseghy_canadian_2000}
\APACinsertmetastar {%
verseghy_canadian_2000}%
\begin{APACrefauthors}%
Verseghy, D\BPBI L.%
\end{APACrefauthors}%
\unskip\
\newblock
\APACrefYearMonthDay{2000}{}{}.
\newblock
{\BBOQ}\APACrefatitle {The {Canadian} land surface scheme ({CLASS}): {Its}
  history and future} {The {Canadian} land surface scheme ({CLASS}): {Its}
  history and future}.{\BBCQ}
\newblock
\APACjournalVolNumPages{Atmosphere - Ocean}{38}{1}{1--13}.
\newblock
\begin{APACrefURL} \url{https://doi.org/10.1080/07055900.2000.9649637}
  \end{APACrefURL}
\newblock
\begin{APACrefDOI} \doi{10.1080/07055900.2000.9649637} \end{APACrefDOI}
\PrintBackRefs{\CurrentBib}

\bibitem [\protect \citeauthoryear {%
Vivoni%
}{%
Vivoni%
}{%
{\protect \APACyear {2020}}%
}]{%
osti_1660351}
\APACinsertmetastar {%
osti_1660351}%
\begin{APACrefauthors}%
Vivoni, E\BPBI R.%
\end{APACrefauthors}%
\unskip\
\newblock
\APACrefYearMonthDay{2020}{}{}.
\newblock
\APACrefbtitle {AmeriFlux US-SRS Santa Rita Experimental Range Mesquite
  Savanna.} {Ameriflux us-srs santa rita experimental range mesquite savanna.}
\newblock
\begin{APACrefDOI} \doi{10.17190/AMF/1660351} \end{APACrefDOI}
\PrintBackRefs{\CurrentBib}

\bibitem [\protect \citeauthoryear {%
Wagner-Riddle%
}{%
Wagner-Riddle%
}{%
{\protect \APACyear {2019}}%
}]{%
osti_1579541}
\APACinsertmetastar {%
osti_1579541}%
\begin{APACrefauthors}%
Wagner-Riddle, C.%
\end{APACrefauthors}%
\unskip\
\newblock
\APACrefYearMonthDay{2019}{}{}.
\newblock
\APACrefbtitle {AmeriFlux CA-ER1 Elora Research Station.} {Ameriflux ca-er1
  elora research station.}
\newblock
\begin{APACrefURL} \url{https://www.osti.gov/biblio/1579541} \end{APACrefURL}
\newblock
\begin{APACrefDOI} \doi{10.17190/AMF/1579541} \end{APACrefDOI}
\PrintBackRefs{\CurrentBib}

\bibitem [\protect \citeauthoryear {%
Whittaker%
, {Di Luca}%
\BCBL {}\ \BBA {} Roberge%
}{%
Whittaker%
\ \protect \BOthers {.}}{%
{\protect \APACyear {2024}}%
}]{%
SP3/VWMCY0_2024}
\APACinsertmetastar {%
SP3/VWMCY0_2024}%
\begin{APACrefauthors}%
Whittaker, T.%
, {Di Luca}, A.%
\BCBL {}\ \BBA {} Roberge, F.%
\end{APACrefauthors}%
\unskip\
\newblock
\APACrefYearMonthDay{2024}{}{}.
\newblock
\APACrefbtitle {{Surface Atmospheric Variables for the Evaluation of Winds
  (SAVE-Winds) catalog}.} {{Surface Atmospheric Variables for the Evaluation of
  Winds (SAVE-Winds) catalog}.}
\newblock
\APACaddressPublisher{}{Borealis}.
\newblock
\begin{APACrefURL} \url{https://doi.org/10.5683/SP3/VWMCY0} \end{APACrefURL}
\newblock
\begin{APACrefDOI} \doi{10.5683/SP3/VWMCY0} \end{APACrefDOI}
\PrintBackRefs{\CurrentBib}

\end{thebibliography}


\begin{thebibliography}{}

\bibitem [\protect \citeauthoryear {%
Baker%
\ \BBA {} Griffis%
}{%
Baker%
\ \BBA {} Griffis%
}{%
{\protect \APACyear {2018}}%
}]{%
osti_1419507}
\APACinsertmetastar {%
osti_1419507}%
\begin{APACrefauthors}%
Baker, J.%
\BCBT {}\ \BBA {} Griffis, T.%
\end{APACrefauthors}%
\unskip\
\newblock
\APACrefYearMonthDay{2018}{}{}.
\newblock
\APACrefbtitle {AmeriFlux US-Ro4 Rosemount Prairie.} {Ameriflux us-ro4
  rosemount prairie.}
\newblock
\begin{APACrefDOI} \doi{10.17190/AMF/1419507} \end{APACrefDOI}
\PrintBackRefs{\CurrentBib}

\bibitem [\protect \citeauthoryear {%
Baker%
\ \BBA {} Griffis%
}{%
Baker%
\ \BBA {} Griffis%
}{%
{\protect \APACyear {2021}}%
}]{%
osti_1818371}
\APACinsertmetastar {%
osti_1818371}%
\begin{APACrefauthors}%
Baker, J.%
\BCBT {}\ \BBA {} Griffis, T.%
\end{APACrefauthors}%
\unskip\
\newblock
\APACrefYearMonthDay{2021}{}{}.
\newblock
\APACrefbtitle {AmeriFlux FLUXNET-1F US-Ro5 Rosemount I18\_South.} {Ameriflux
  fluxnet-1f us-ro5 rosemount i18\_south.}
\newblock
\begin{APACrefDOI} \doi{10.17190/AMF/1818371} \end{APACrefDOI}
\PrintBackRefs{\CurrentBib}

\bibitem [\protect \citeauthoryear {%
Baker%
\ \BBA {} Griffis%
}{%
Baker%
\ \BBA {} Griffis%
}{%
{\protect \APACyear {2022}}%
}]{%
osti_1881590}
\APACinsertmetastar {%
osti_1881590}%
\begin{APACrefauthors}%
Baker, J.%
\BCBT {}\ \BBA {} Griffis, T.%
\end{APACrefauthors}%
\unskip\
\newblock
\APACrefYearMonthDay{2022}{}{}.
\newblock
\APACrefbtitle {AmeriFlux FLUXNET-1F US-Ro6 Rosemount I18\_North.} {Ameriflux
  fluxnet-1f us-ro6 rosemount i18\_north.}
\newblock
\begin{APACrefDOI} \doi{10.17190/AMF/1881590} \end{APACrefDOI}
\PrintBackRefs{\CurrentBib}

\bibitem [\protect \citeauthoryear {%
Bergamaschi%
\ \BBA {} Windham-Myers%
}{%
Bergamaschi%
\ \BBA {} Windham-Myers%
}{%
{\protect \APACyear {2018}}%
}]{%
osti_1418685}
\APACinsertmetastar {%
osti_1418685}%
\begin{APACrefauthors}%
Bergamaschi, B.%
\BCBT {}\ \BBA {} Windham-Myers, L.%
\end{APACrefauthors}%
\unskip\
\newblock
\APACrefYearMonthDay{2018}{}{}.
\newblock
\APACrefbtitle {AmeriFlux US-Srr Suisun marsh - Rush Ranch.} {Ameriflux us-srr
  suisun marsh - rush ranch.}
\newblock
\begin{APACrefDOI} \doi{10.17190/AMF/1418685} \end{APACrefDOI}
\PrintBackRefs{\CurrentBib}

\bibitem [\protect \citeauthoryear {%
Eichelmann%
\ \protect \BOthers {.}}{%
Eichelmann%
\ \protect \BOthers {.}}{%
{\protect \APACyear {2020}}%
}]{%
osti_1669698}
\APACinsertmetastar {%
osti_1669698}%
\begin{APACrefauthors}%
Eichelmann, E.%
, Knox, S.%
, Sanchez, C\BPBI R.%
, Valach, A.%
, Sturtevant, C.%
, Szutu, D.%
\BDBL {}Baldocchi, D.%
\end{APACrefauthors}%
\unskip\
\newblock
\APACrefYearMonthDay{2020}{}{}.
\newblock
\APACrefbtitle {FLUXNET-CH4 US-Tw4 Twitchell East End Wetland.} {Fluxnet-ch4
  us-tw4 twitchell east end wetland.}
\newblock
\begin{APACrefDOI} \doi{10.18140/FLX/1669698} \end{APACrefDOI}
\PrintBackRefs{\CurrentBib}

\bibitem [\protect \citeauthoryear {%
Flerchinger%
}{%
Flerchinger%
}{%
{\protect \APACyear {2017}}%
{\protect \APACexlab {{\protect \BCnt {1}}}}}]{%
osti_1375202}
\APACinsertmetastar {%
osti_1375202}%
\begin{APACrefauthors}%
Flerchinger, G.%
\end{APACrefauthors}%
\unskip\
\newblock
\APACrefYearMonthDay{2017{\protect \BCnt {1}}}{}{}.
\newblock
\APACrefbtitle {AmeriFlux US-Rms RCEW Mountain Big Sagebrush.} {Ameriflux
  us-rms rcew mountain big sagebrush.}
\newblock
\begin{APACrefDOI} \doi{10.17190/AMF/1375202} \end{APACrefDOI}
\PrintBackRefs{\CurrentBib}

\bibitem [\protect \citeauthoryear {%
Flerchinger%
}{%
Flerchinger%
}{%
{\protect \APACyear {2017}}%
{\protect \APACexlab {{\protect \BCnt {2}}}}}]{%
osti_1375201}
\APACinsertmetastar {%
osti_1375201}%
\begin{APACrefauthors}%
Flerchinger, G.%
\end{APACrefauthors}%
\unskip\
\newblock
\APACrefYearMonthDay{2017{\protect \BCnt {2}}}{}{}.
\newblock
\APACrefbtitle {AmeriFlux US-Rws Reynolds Creek Wyoming big sagebrush.}
  {Ameriflux us-rws reynolds creek wyoming big sagebrush.}
\newblock
\begin{APACrefDOI} \doi{10.17190/AMF/1375201} \end{APACrefDOI}
\PrintBackRefs{\CurrentBib}

\bibitem [\protect \citeauthoryear {%
Flerchinger%
}{%
Flerchinger%
}{%
{\protect \APACyear {2018}}%
}]{%
osti_1418682}
\APACinsertmetastar {%
osti_1418682}%
\begin{APACrefauthors}%
Flerchinger, G.%
\end{APACrefauthors}%
\unskip\
\newblock
\APACrefYearMonthDay{2018}{}{}.
\newblock
\APACrefbtitle {AmeriFlux US-Rls RCEW Low Sagebrush.} {Ameriflux us-rls rcew
  low sagebrush.}
\newblock
\begin{APACrefDOI} \doi{10.17190/AMF/1418682} \end{APACrefDOI}
\PrintBackRefs{\CurrentBib}

\bibitem [\protect \citeauthoryear {%
Flerchinger%
}{%
Flerchinger%
}{%
{\protect \APACyear {2020}}%
}]{%
osti_1617724}
\APACinsertmetastar {%
osti_1617724}%
\begin{APACrefauthors}%
Flerchinger, G.%
\end{APACrefauthors}%
\unskip\
\newblock
\APACrefYearMonthDay{2020}{}{}.
\newblock
\APACrefbtitle {AmeriFlux {US-Rwf} RCEW Upper Sheep Prescibed Fire.} {Ameriflux
  {US-Rwf} rcew upper sheep prescibed fire.}
\newblock
\begin{APACrefDOI} \doi{10.17190/AMF/1617724} \end{APACrefDOI}
\PrintBackRefs{\CurrentBib}

\bibitem [\protect \citeauthoryear {%
Goslee%
}{%
Goslee%
}{%
{\protect \APACyear {2021}}%
}]{%
osti_1811363}
\APACinsertmetastar {%
osti_1811363}%
\begin{APACrefauthors}%
Goslee, S.%
\end{APACrefauthors}%
\unskip\
\newblock
\APACrefYearMonthDay{2021}{}{}.
\newblock
\APACrefbtitle {AmeriFlux US-HWB USDA ARS Pasture Sytems and Watershed
  Management Research Unit- Hawbecker Site.} {Ameriflux us-hwb usda ars pasture
  sytems and watershed management research unit- hawbecker site.}
\newblock
\begin{APACrefURL} \url{https://www.osti.gov/biblio/1811363} \end{APACrefURL}
\newblock
\begin{APACrefDOI} \doi{10.17190/AMF/1811363} \end{APACrefDOI}
\PrintBackRefs{\CurrentBib}

\bibitem [\protect \citeauthoryear {%
Huggins%
}{%
Huggins%
}{%
{\protect \APACyear {2021}}%
}]{%
osti_1832158}
\APACinsertmetastar {%
osti_1832158}%
\begin{APACrefauthors}%
Huggins, D.%
\end{APACrefauthors}%
\unskip\
\newblock
\APACrefYearMonthDay{2021}{}{}.
\newblock
\APACrefbtitle {AmeriFlux FLUXNET-1F US-CF1 CAF-LTAR Cook East.} {Ameriflux
  fluxnet-1f us-cf1 caf-ltar cook east.}
\newblock
\begin{APACrefDOI} \doi{10.17190/AMF/1832158} \end{APACrefDOI}
\PrintBackRefs{\CurrentBib}

\bibitem [\protect \citeauthoryear {%
Huggins%
}{%
Huggins%
}{%
{\protect \APACyear {2022}}%
{\protect \APACexlab {{\protect \BCnt {1}}}}}]{%
osti_1881573}
\APACinsertmetastar {%
osti_1881573}%
\begin{APACrefauthors}%
Huggins, D.%
\end{APACrefauthors}%
\unskip\
\newblock
\APACrefYearMonthDay{2022{\protect \BCnt {1}}}{}{}.
\newblock
\APACrefbtitle {AmeriFlux FLUXNET-1F US-CF2 CAF-LTAR Cook West.} {Ameriflux
  fluxnet-1f us-cf2 caf-ltar cook west.}
\newblock
\begin{APACrefDOI} \doi{10.17190/AMF/1881573} \end{APACrefDOI}
\PrintBackRefs{\CurrentBib}

\bibitem [\protect \citeauthoryear {%
Huggins%
}{%
Huggins%
}{%
{\protect \APACyear {2022}}%
{\protect \APACexlab {{\protect \BCnt {2}}}}}]{%
osti_1881574}
\APACinsertmetastar {%
osti_1881574}%
\begin{APACrefauthors}%
Huggins, D.%
\end{APACrefauthors}%
\unskip\
\newblock
\APACrefYearMonthDay{2022{\protect \BCnt {2}}}{}{}.
\newblock
\APACrefbtitle {AmeriFlux FLUXNET-1F US-CF3 CAF-LTAR Boyd North.} {Ameriflux
  fluxnet-1f us-cf3 caf-ltar boyd north.}
\newblock
\begin{APACrefDOI} \doi{10.17190/AMF/1881574} \end{APACrefDOI}
\PrintBackRefs{\CurrentBib}

\bibitem [\protect \citeauthoryear {%
Huggins%
}{%
Huggins%
}{%
{\protect \APACyear {2022}}%
{\protect \APACexlab {{\protect \BCnt {3}}}}}]{%
osti_1881575}
\APACinsertmetastar {%
osti_1881575}%
\begin{APACrefauthors}%
Huggins, D.%
\end{APACrefauthors}%
\unskip\
\newblock
\APACrefYearMonthDay{2022{\protect \BCnt {3}}}{}{}.
\newblock
\APACrefbtitle {AmeriFlux FLUXNET-1F US-CF4 CAF-LTAR Boyd South.} {Ameriflux
  fluxnet-1f us-cf4 caf-ltar boyd south.}
\newblock
\begin{APACrefDOI} \doi{10.17190/AMF/1881575} \end{APACrefDOI}
\PrintBackRefs{\CurrentBib}

\bibitem [\protect \citeauthoryear {%
Knox%
, Matthes%
, Verfaillie%
\BCBL {}\ \BBA {} Baldocchi%
}{%
Knox%
\ \protect \BOthers {.}}{%
{\protect \APACyear {2016}}%
}]{%
osti_1246140}
\APACinsertmetastar {%
osti_1246140}%
\begin{APACrefauthors}%
Knox, S.%
, Matthes, J\BPBI H.%
, Verfaillie, J.%
\BCBL {}\ \BBA {} Baldocchi, D.%
\end{APACrefauthors}%
\unskip\
\newblock
\APACrefYearMonthDay{2016}{}{}.
\newblock
\APACrefbtitle {AmeriFlux US-Twt Twitchell Island.} {Ameriflux us-twt twitchell
  island.}
\newblock
\begin{APACrefURL} \url{https://www.osti.gov/biblio/1246140} \end{APACrefURL}
\newblock
\begin{APACrefDOI} \doi{10.17190/AMF/1246140} \end{APACrefDOI}
\PrintBackRefs{\CurrentBib}

\bibitem [\protect \citeauthoryear {%
Kusak%
, Sanchez%
, Szutu%
\BCBL {}\ \BBA {} Baldocchi%
}{%
Kusak%
\ \protect \BOthers {.}}{%
{\protect \APACyear {2022}}%
}]{%
osti_1854371}
\APACinsertmetastar {%
osti_1854371}%
\begin{APACrefauthors}%
Kusak, K.%
, Sanchez, C\BPBI R.%
, Szutu, D.%
\BCBL {}\ \BBA {} Baldocchi, D.%
\end{APACrefauthors}%
\unskip\
\newblock
\APACrefYearMonthDay{2022}{}{}.
\newblock
\APACrefbtitle {AmeriFlux FLUXNET-1F US-Snf Sherman Barn.} {Ameriflux
  fluxnet-1f us-snf sherman barn.}
\newblock
\begin{APACrefDOI} \doi{10.17190/AMF/1854371} \end{APACrefDOI}
\PrintBackRefs{\CurrentBib}

\bibitem [\protect \citeauthoryear {%
Ma%
, Xu%
, Verfaillie%
\BCBL {}\ \BBA {} Baldocchi%
}{%
Ma%
\ \protect \BOthers {.}}{%
{\protect \APACyear {2016}}%
}]{%
osti_1245984}
\APACinsertmetastar {%
osti_1245984}%
\begin{APACrefauthors}%
Ma, S.%
, Xu, L.%
, Verfaillie, J.%
\BCBL {}\ \BBA {} Baldocchi, D.%
\end{APACrefauthors}%
\unskip\
\newblock
\APACrefYearMonthDay{2016}{}{}.
\newblock
\APACrefbtitle {AmeriFlux US-Var Vaira Ranch- Ione.} {Ameriflux us-var vaira
  ranch- ione.}
\newblock
\begin{APACrefDOI} \doi{10.17190/AMF/1245984} \end{APACrefDOI}
\PrintBackRefs{\CurrentBib}

\bibitem [\protect \citeauthoryear {%
Matthes%
\ \protect \BOthers {.}}{%
Matthes%
\ \protect \BOthers {.}}{%
{\protect \APACyear {2020}}%
}]{%
osti_1669685}
\APACinsertmetastar {%
osti_1669685}%
\begin{APACrefauthors}%
Matthes, J\BPBI H.%
, Sturtevant, C.%
, Oikawa, P.%
, Chamberlain, S\BPBI D.%
, Szutu, D.%
, Ortiz, A\BPBI A.%
\BDBL {}Baldocchi, D.%
\end{APACrefauthors}%
\unskip\
\newblock
\APACrefYearMonthDay{2020}{}{}.
\newblock
\APACrefbtitle {FLUXNET-CH4 US-Myb Mayberry Wetland.} {Fluxnet-ch4 us-myb
  mayberry wetland.}
\newblock
\begin{APACrefDOI} \doi{10.18140/FLX/1669685} \end{APACrefDOI}
\PrintBackRefs{\CurrentBib}

\bibitem [\protect \citeauthoryear {%
Oikawa%
}{%
Oikawa%
}{%
{\protect \APACyear {2021}}%
}]{%
osti_1832159}
\APACinsertmetastar {%
osti_1832159}%
\begin{APACrefauthors}%
Oikawa, P.%
\end{APACrefauthors}%
\unskip\
\newblock
\APACrefYearMonthDay{2021}{}{}.
\newblock
\APACrefbtitle {AmeriFlux FLUXNET-1F US-EDN Eden Landing Ecological Reserve.}
  {Ameriflux fluxnet-1f us-edn eden landing ecological reserve.}
\newblock
\begin{APACrefDOI} \doi{10.17190/AMF/1832159} \end{APACrefDOI}
\PrintBackRefs{\CurrentBib}

\bibitem [\protect \citeauthoryear {%
Rey-Sanchez%
, Wang%
, Szutu%
, Hemes%
\BCBL {}\ \protect \BOthers {.}}{%
Rey-Sanchez%
, Wang%
, Szutu%
, Hemes%
\BCBL {}\ \protect \BOthers {.}}{%
{\protect \APACyear {2022}}%
}]{%
osti_1871135}
\APACinsertmetastar {%
osti_1871135}%
\begin{APACrefauthors}%
Rey-Sanchez, C.%
, Wang, C\BPBI T.%
, Szutu, D.%
, Hemes, K.%
, Verfaillie, J.%
\BCBL {}\ \BBA {} Baldocchi, D.%
\end{APACrefauthors}%
\unskip\
\newblock
\APACrefYearMonthDay{2022}{}{}.
\newblock
\APACrefbtitle {AmeriFlux FLUXNET-1F US-Bi2 Bouldin Island corn.} {Ameriflux
  fluxnet-1f us-bi2 bouldin island corn.}
\newblock
\begin{APACrefDOI} \doi{10.17190/AMF/1871135} \end{APACrefDOI}
\PrintBackRefs{\CurrentBib}

\bibitem [\protect \citeauthoryear {%
Rey-Sanchez%
, Wang%
, Szutu%
, Shortt%
\BCBL {}\ \protect \BOthers {.}}{%
Rey-Sanchez%
, Wang%
, Szutu%
, Shortt%
\BCBL {}\ \protect \BOthers {.}}{%
{\protect \APACyear {2022}}%
}]{%
osti_1871134}
\APACinsertmetastar {%
osti_1871134}%
\begin{APACrefauthors}%
Rey-Sanchez, C.%
, Wang, C\BPBI T.%
, Szutu, D.%
, Shortt, R.%
, Chamberlain, S\BPBI D.%
, Verfaillie, J.%
\BCBL {}\ \BBA {} Baldocchi, D.%
\end{APACrefauthors}%
\unskip\
\newblock
\APACrefYearMonthDay{2022}{}{}.
\newblock
\APACrefbtitle {AmeriFlux FLUXNET-1F US-Bi1 Bouldin Island Alfalfa.} {Ameriflux
  fluxnet-1f us-bi1 bouldin island alfalfa.}
\newblock
\begin{APACrefDOI} \doi{10.17190/AMF/1871134} \end{APACrefDOI}
\PrintBackRefs{\CurrentBib}

\bibitem [\protect \citeauthoryear {%
Shortt%
, Hemes%
, Szutu%
, Verfaillie%
\BCBL {}\ \BBA {} Baldocchi%
}{%
Shortt%
\ \protect \BOthers {.}}{%
{\protect \APACyear {2022}}%
}]{%
osti_1871144}
\APACinsertmetastar {%
osti_1871144}%
\begin{APACrefauthors}%
Shortt, R.%
, Hemes, K.%
, Szutu, D.%
, Verfaillie, J.%
\BCBL {}\ \BBA {} Baldocchi, D.%
\end{APACrefauthors}%
\unskip\
\newblock
\APACrefYearMonthDay{2022}{}{}.
\newblock
\APACrefbtitle {AmeriFlux FLUXNET-1F US-Sne Sherman Island Restored Wetland.}
  {Ameriflux fluxnet-1f us-sne sherman island restored wetland.}
\newblock
\begin{APACrefDOI} \doi{10.17190/AMF/1871144} \end{APACrefDOI}
\PrintBackRefs{\CurrentBib}

\bibitem [\protect \citeauthoryear {%
Silveira%
}{%
Silveira%
}{%
{\protect \APACyear {2021}}%
}]{%
osti_1832163}
\APACinsertmetastar {%
osti_1832163}%
\begin{APACrefauthors}%
Silveira, M.%
\end{APACrefauthors}%
\unskip\
\newblock
\APACrefYearMonthDay{2021}{}{}.
\newblock
\APACrefbtitle {AmeriFlux FLUXNET-1F US-ONA Florida pine flatwoods.} {Ameriflux
  fluxnet-1f us-ona florida pine flatwoods.}
\newblock
\begin{APACrefDOI} \doi{10.17190/AMF/1832163} \end{APACrefDOI}
\PrintBackRefs{\CurrentBib}

\bibitem [\protect \citeauthoryear {%
Valach%
\ \protect \BOthers {.}}{%
Valach%
\ \protect \BOthers {.}}{%
{\protect \APACyear {2016}}%
}]{%
osti_1246147}
\APACinsertmetastar {%
osti_1246147}%
\begin{APACrefauthors}%
Valach, A.%
, Shortt, R.%
, Szutu, D.%
, Eichelmann, E.%
, Knox, S.%
, Hemes, K.%
\BDBL {}Baldocchi, D.%
\end{APACrefauthors}%
\unskip\
\newblock
\APACrefYearMonthDay{2016}{}{}.
\newblock
\APACrefbtitle {AmeriFlux US-Tw1 Twitchell Wetland West Pond.} {Ameriflux
  us-tw1 twitchell wetland west pond.}
\newblock
\begin{APACrefDOI} \doi{10.17190/AMF/1246147} \end{APACrefDOI}
\PrintBackRefs{\CurrentBib}

\bibitem [\protect \citeauthoryear {%
Vazquez-Lule%
\ \BBA {} Vargas%
}{%
Vazquez-Lule%
\ \BBA {} Vargas%
}{%
{\protect \APACyear {2020}}%
}]{%
osti_1669695}
\APACinsertmetastar {%
osti_1669695}%
\begin{APACrefauthors}%
Vazquez-Lule, A.%
\BCBT {}\ \BBA {} Vargas, R.%
\end{APACrefauthors}%
\unskip\
\newblock
\APACrefYearMonthDay{2020}{}{}.
\newblock
\APACrefbtitle {FLUXNET-CH4 US-StJ St Jones Reserve.} {Fluxnet-ch4 us-stj st
  jones reserve.}
\newblock
\begin{APACrefDOI} \doi{10.18140/FLX/1669695} \end{APACrefDOI}
\PrintBackRefs{\CurrentBib}

\bibitem [\protect \citeauthoryear {%
Vivoni%
}{%
Vivoni%
}{%
{\protect \APACyear {2020}}%
}]{%
osti_1660351}
\APACinsertmetastar {%
osti_1660351}%
\begin{APACrefauthors}%
Vivoni, E\BPBI R.%
\end{APACrefauthors}%
\unskip\
\newblock
\APACrefYearMonthDay{2020}{}{}.
\newblock
\APACrefbtitle {AmeriFlux US-SRS Santa Rita Experimental Range Mesquite
  Savanna.} {Ameriflux us-srs santa rita experimental range mesquite savanna.}
\newblock
\begin{APACrefDOI} \doi{10.17190/AMF/1660351} \end{APACrefDOI}
\PrintBackRefs{\CurrentBib}

\bibitem [\protect \citeauthoryear {%
Wagner-Riddle%
}{%
Wagner-Riddle%
}{%
{\protect \APACyear {2019}}%
}]{%
osti_1579541}
\APACinsertmetastar {%
osti_1579541}%
\begin{APACrefauthors}%
Wagner-Riddle, C.%
\end{APACrefauthors}%
\unskip\
\newblock
\APACrefYearMonthDay{2019}{}{}.
\newblock
\APACrefbtitle {AmeriFlux CA-ER1 Elora Research Station.} {Ameriflux ca-er1
  elora research station.}
\newblock
\begin{APACrefURL} \url{https://www.osti.gov/biblio/1579541} \end{APACrefURL}
\newblock
\begin{APACrefDOI} \doi{10.17190/AMF/1579541} \end{APACrefDOI}
\PrintBackRefs{\CurrentBib}

\end{thebibliography}

\end{document}


\title{Supporting Information for "Evaluating near-surface wind speeds simulated by the CRCM6-GEM5 model using AmeriFlux data over North America"}

\authors{Tim Whittaker\affil{1}, Alejandro Di Luca\affil{1}, Francois Roberge\affil{1}, Katja Winger\affil{1}}

\affiliation{1}{Centre pour l'étude et la simulation du climat à l'échelle régionale (ESCER), D\'epartement des Sciences de la Terre et de l’Atmosph\`ere, Universit\'e du Qu\'ebec \`a Montr\'eal, Montr\'eal, Qu\'ebec, Canada}


\pagebreak 
\section{Stability Functions in the CRCM6-GEM6 model}
\label{APP:StabFunc}
Under stable conditions, the  stability functions used are the following,
\begin{equation*}
	\Psi_m^{stable} = \begin{cases}
		\frac{1}{2}\left[a-\frac{z}{h}-\ln \left(1+\frac{b}{2}z+a\right) - \frac{b}{2\ \sqrt[]{c}}\arcsin\left(\frac{b-2cz}{d}\right) \right] \text{ for DEL}\\
		-(a\zeta + b(\zeta-c/d)e^{-d\zeta}+bc/d) \text{ for BEL}
	\end{cases},
\end{equation*}
with $a = 1, b = 2/3, c = 5, d = 0.35$ and the boundary layer height $h$ for $DEL$ given by $h = max(h_1 , h_2 , h_3 , h_4)$ where
\begin{align*}
	h_1 &= \gamma \cdot (zu + 10 \cdot z_0)\\
	\gamma &= 1.2\\ 
	zu &= \text{height of the model’s lowest prognostic (momentum) level}\\ 
	h_2 &= b_s\ \sqrt[]{\frac{\kappa u_* L}{|f|}}\\ 
	f &= \text{Coriolis parameter}\\
	b_s &= 1\\
	h_3 &= \frac{\gamma}{4 a_s \beta} L \\
	a_s &= 12\\
	\beta &= 1\\
	h_4 &= h_{min} \text{=30m in the operational system}\\
\end{align*}
For unstable conditions, stability functions are given by,
\begin{equation*}
	\Psi_m^{unstable} = \begin{cases}
		 2\ln(w+1)+\frac{1}{2}\ln(w^2-w+1)+\frac{3}{2}\ln(w^2+w+1)-\sqrt[]{3}\arctan\frac{(w^2-1)}{\sqrt[]{3} \ w} \text{ for DEL},\\
		\log z - 2 \log(r + 1) - \log(r^2 + 1) + 2 \arctan(r) \text{ for BEL}
	\end{cases},
\end{equation*}
with, $w(\zeta) = (1-\beta c_i \zeta)^{1/6}$ and $r=(1 - 16\zeta)^{1/4}$
where $\beta$ and $c_i$ are parameters with default values $\beta=1,c_i=40$. 

\pagebreak 
\section{Roughness length calculation in the CRCM6-GEM6 model}
\label{APP:z0}
Over land grid points, the effective roughness length $z_0$ is calculated differently in both land-surface schemes, ISBA and CLASS. In CLASS we have:
\begin{equation}
	z_0 = \text{max}(z_{(0,oro)},z_{(0,veg)}),
	\label{roughGem_C}
\end{equation}
And in ISBA we have:
\begin{equation}
	\frac{1}{\left[\ln{\frac{z_{ref}}{z_{0}}}\right]^2} = \frac{1}{\left[\ln{\frac{z_{ref}}{z_{0,veg}}}\right]^2} +\frac{1}{\left[\ln{\frac{z_{ref}}{z_{0,orog}}}\right]^2}  ,
	\label{roughGem_I}
\end{equation}

\pagebreak 
\section{Extra Figures and Tables}
\begin{figure}[H]
    \centering
    \includegraphics[width=\textwidth]{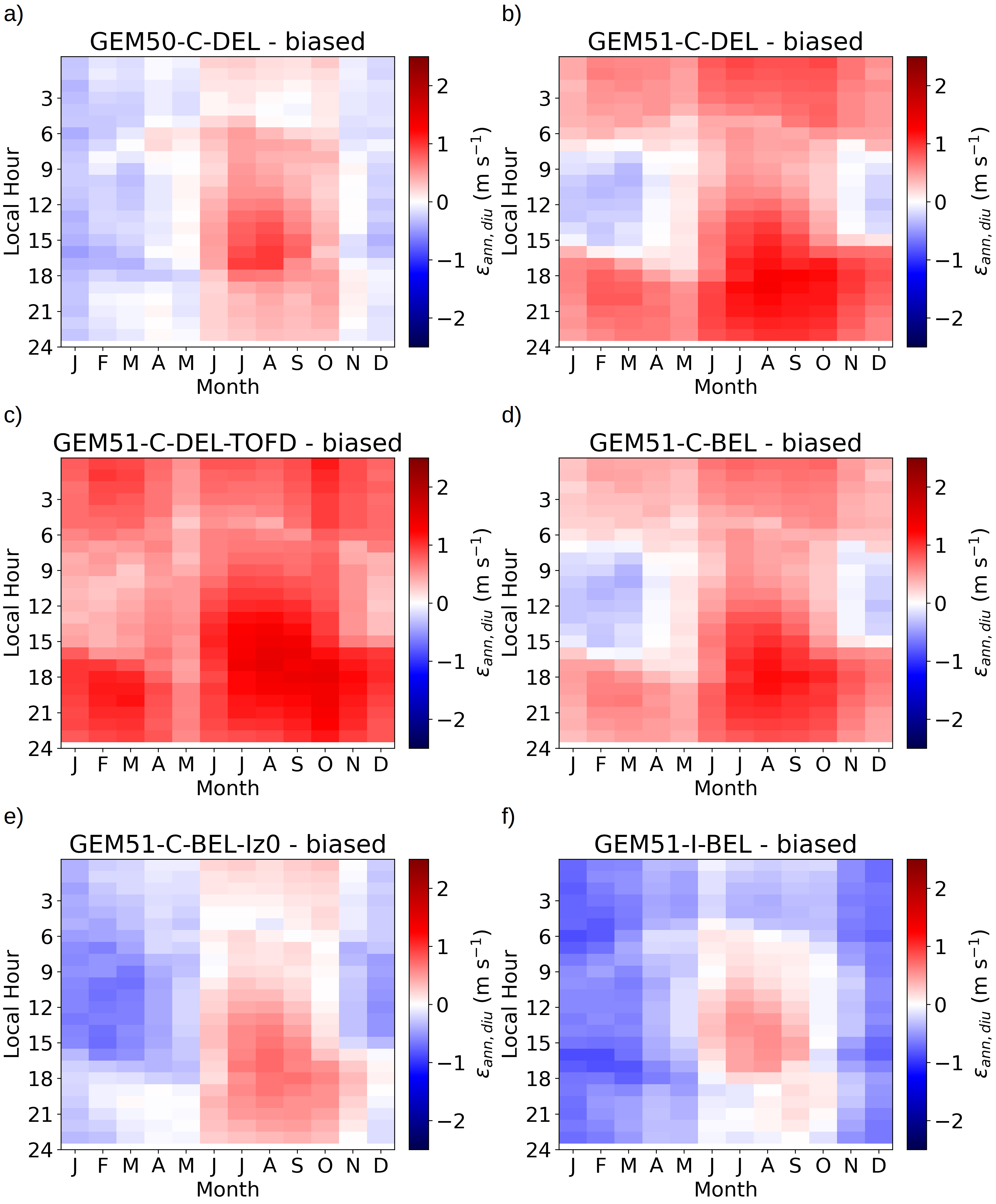}
    \caption{Combined monthly and hourly wind speed cycle errors for all simulations.}
\end{figure}

\begin{figure}[H]
    \centering
    \includegraphics[width=\textwidth]{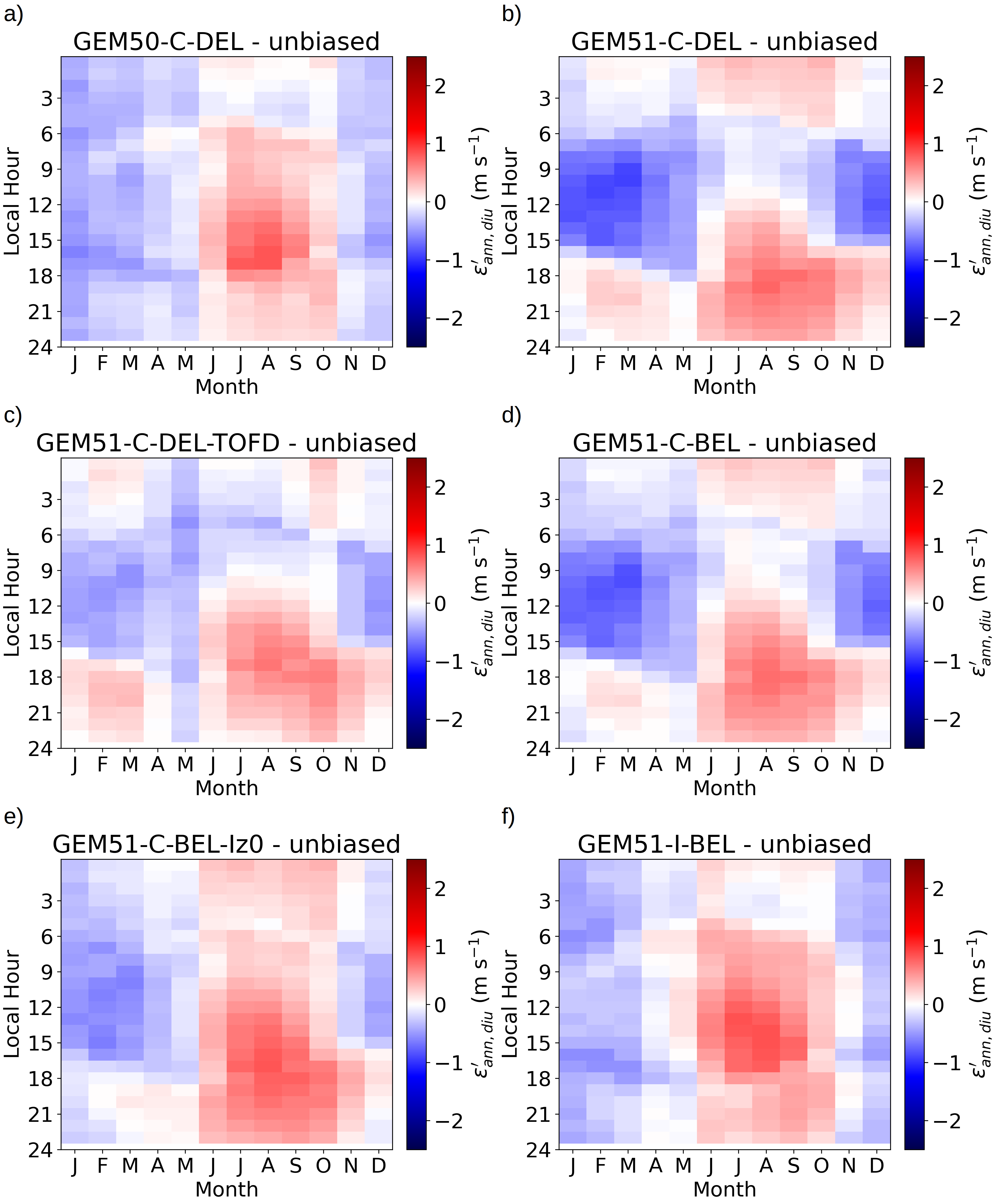}
    \caption{Combined monthly and hourly wind speed cycle errors for all simulations with bias removed.}
\end{figure}

\begin{figure}[H]
    \centering
    \includegraphics[width=\textwidth]{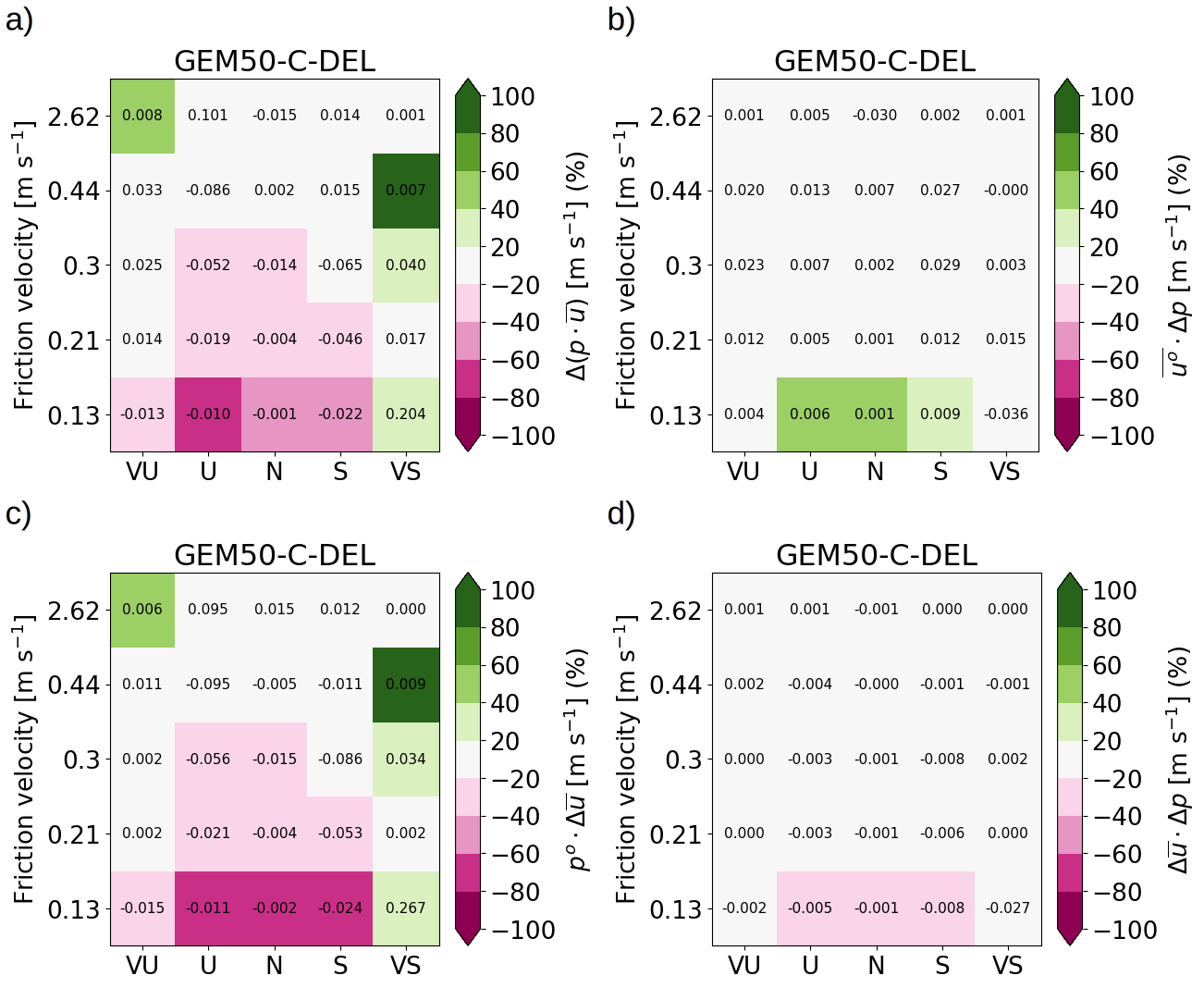}
    \caption{Decomposed percent regime errors for GEM50-C-DEL. Panel a) shows the total binned error, Panel b) shows the binned frequency error, Panel c) shows the binned mean wind speed error, and Panel d) shows the binned residual error. }
\end{figure}

\begin{figure}[H]
    \centering
    \includegraphics[width=\textwidth]{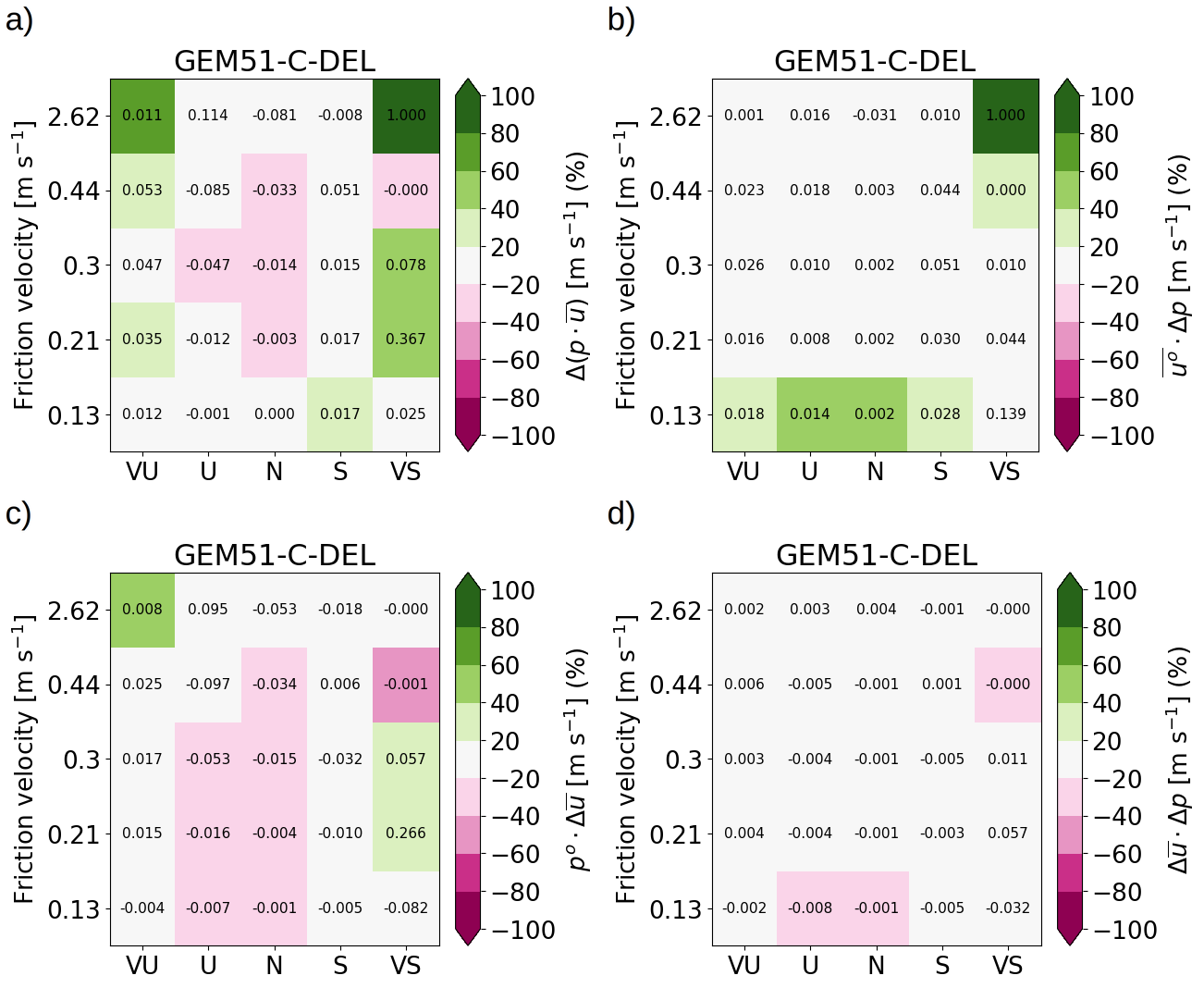}
    \caption{Decomposed percent regime errors for GEM50-C-DEL. Panel a) shows the total binned error, Panel b) shows the binned frequency error, Panel c) shows the binned mean wind speed error, and Panel d) shows the binned residual error. }
\end{figure}

\begin{figure}[H]
    \centering
    \includegraphics[width=\textwidth]{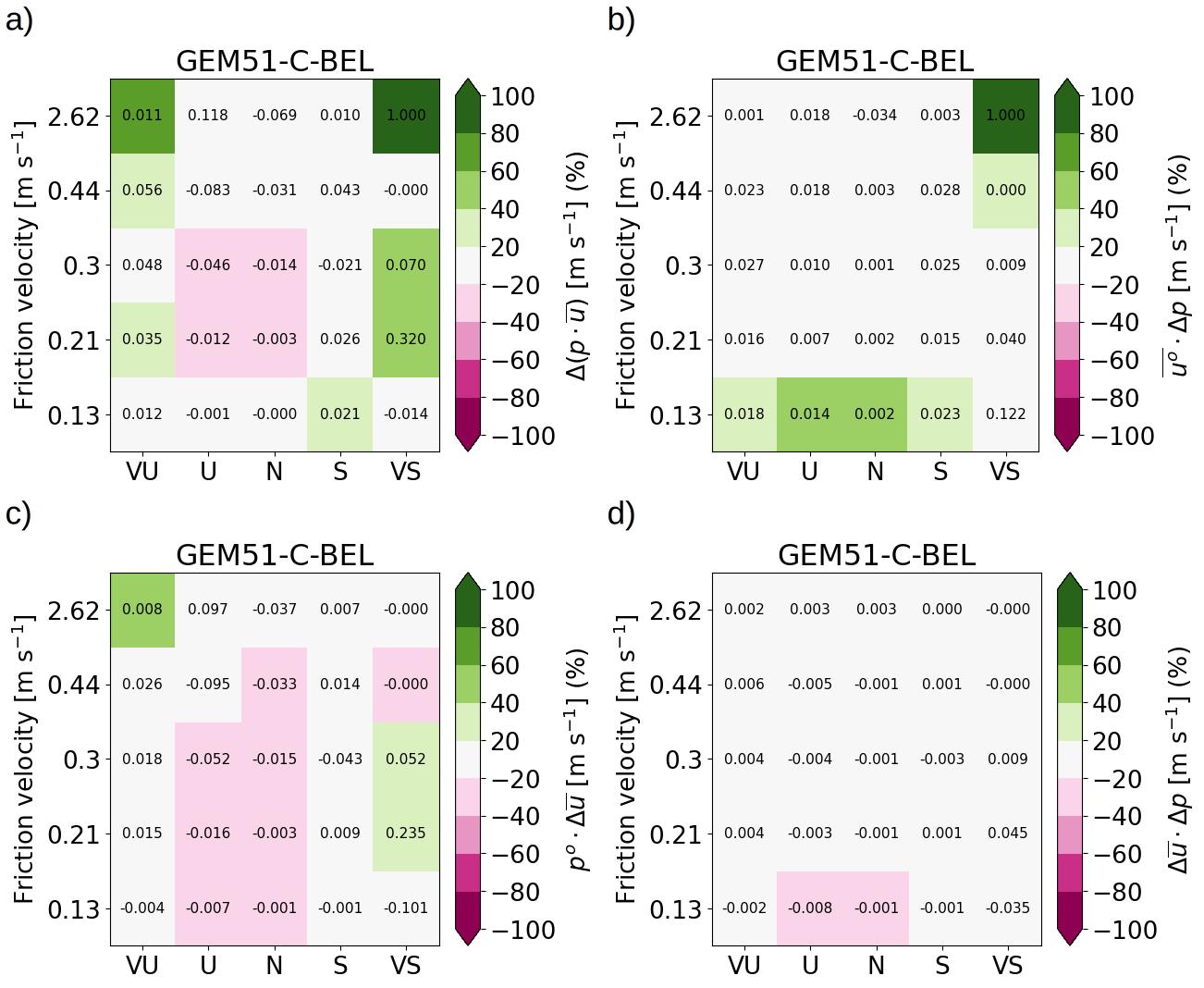}
    \caption{Decomposed percent regime errors for GEM51-C-DEL. Panel a) shows the total binned error, Panel b) shows the binned frequency error, Panel c) shows the binned mean wind speed error, and Panel d) shows the binned residual error. }
\end{figure}

\begin{figure}[H]
    \centering
    \includegraphics[width=\textwidth]{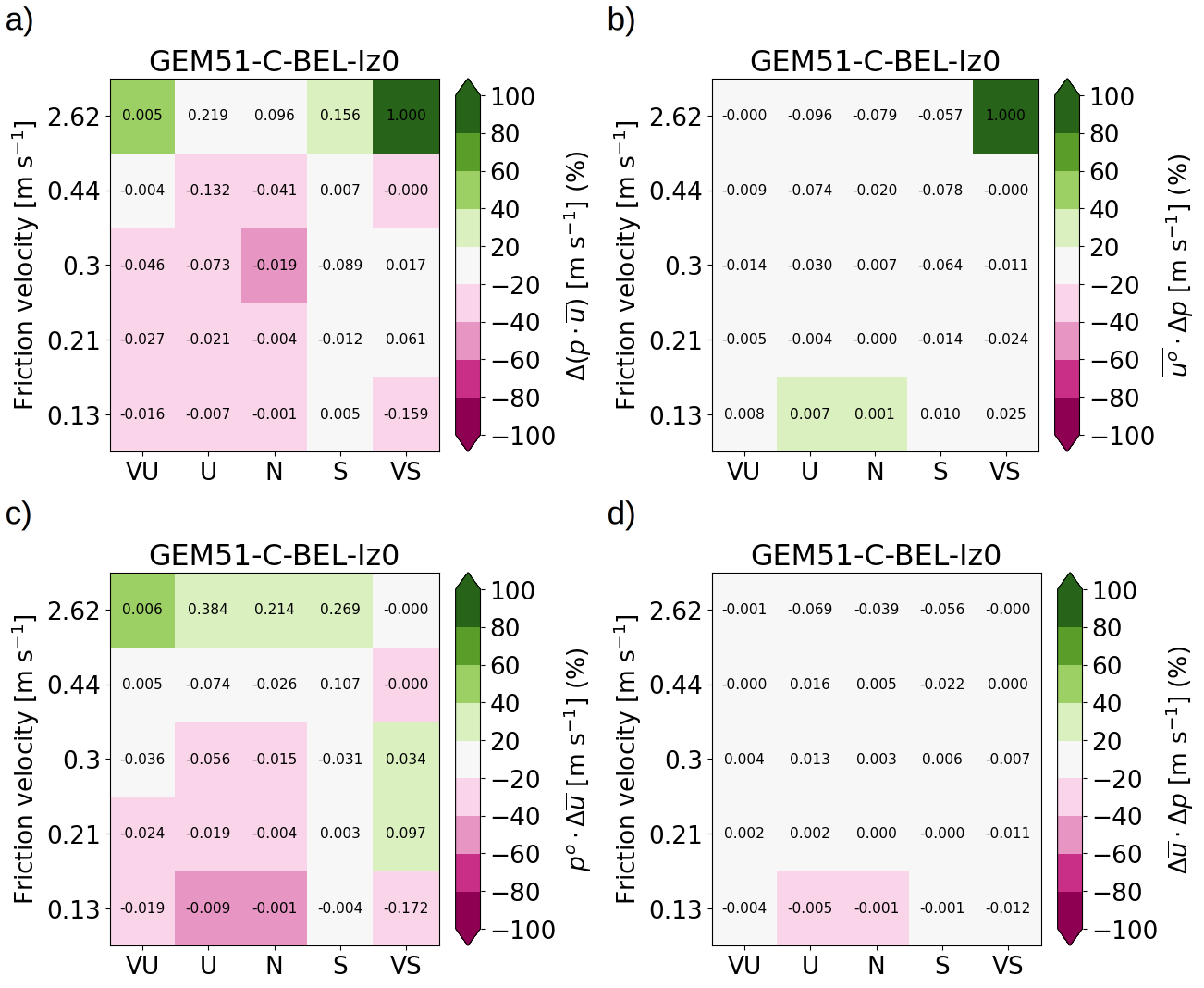}
    \caption{Decomposed percent regime errors for GEM51-C-BEL. Panel a) shows the total binned error, Panel b) shows the binned frequency error, Panel c) shows the binned mean wind speed error, and Panel d) shows the binned residual error. }
\end{figure}

\begin{figure}[H]
    \centering
    \includegraphics[width=\textwidth]{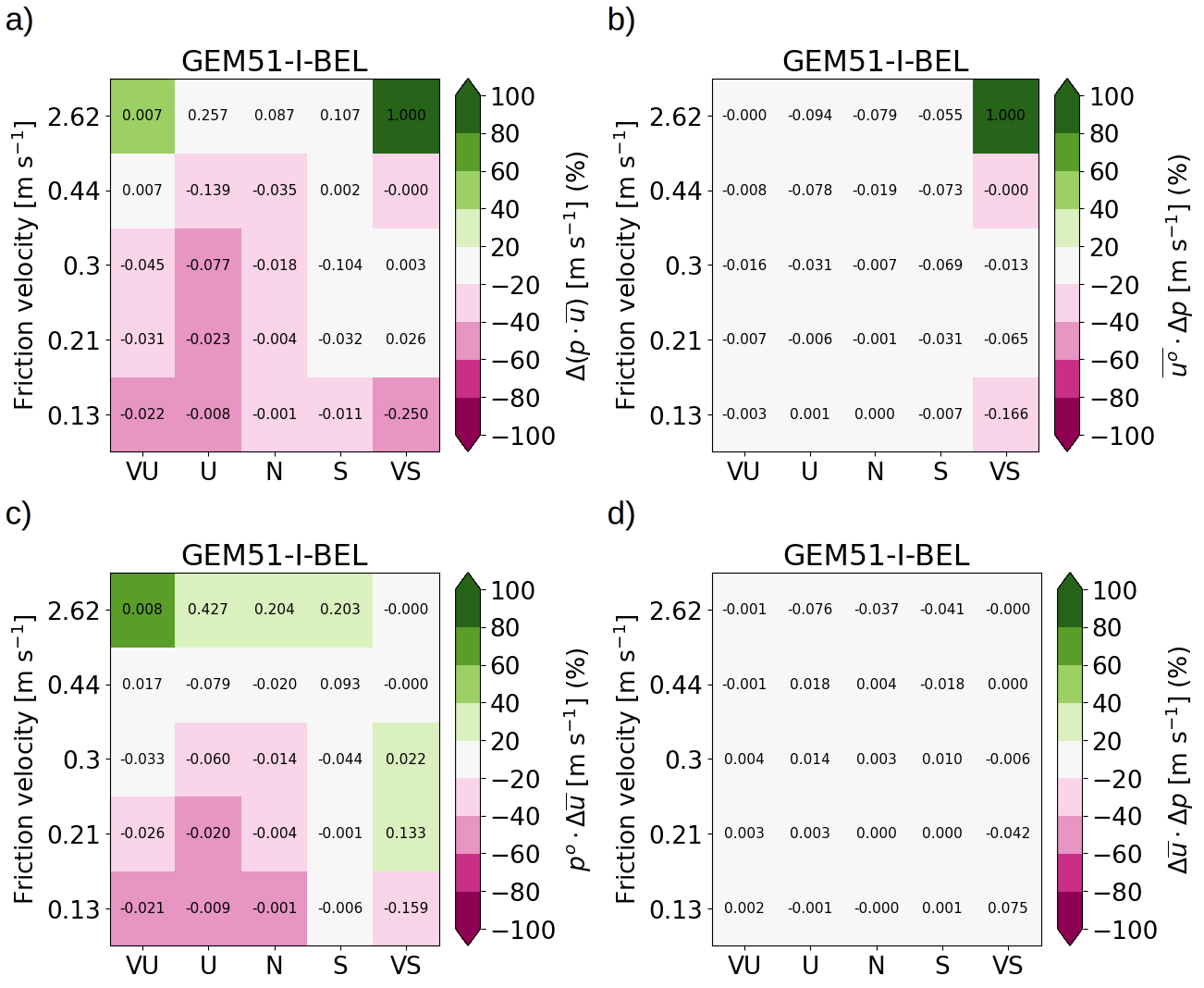}
    \caption{Decomposed percent regime errors for GEM51-C-BEL-Iz0. Panel a) shows the total binned error, Panel b) shows the binned frequency error, Panel c) shows the binned mean wind speed error, and Panel d) shows the binned residual error. }
\end{figure}

\begin{figure}[ht]
	\includegraphics[width=\textwidth]{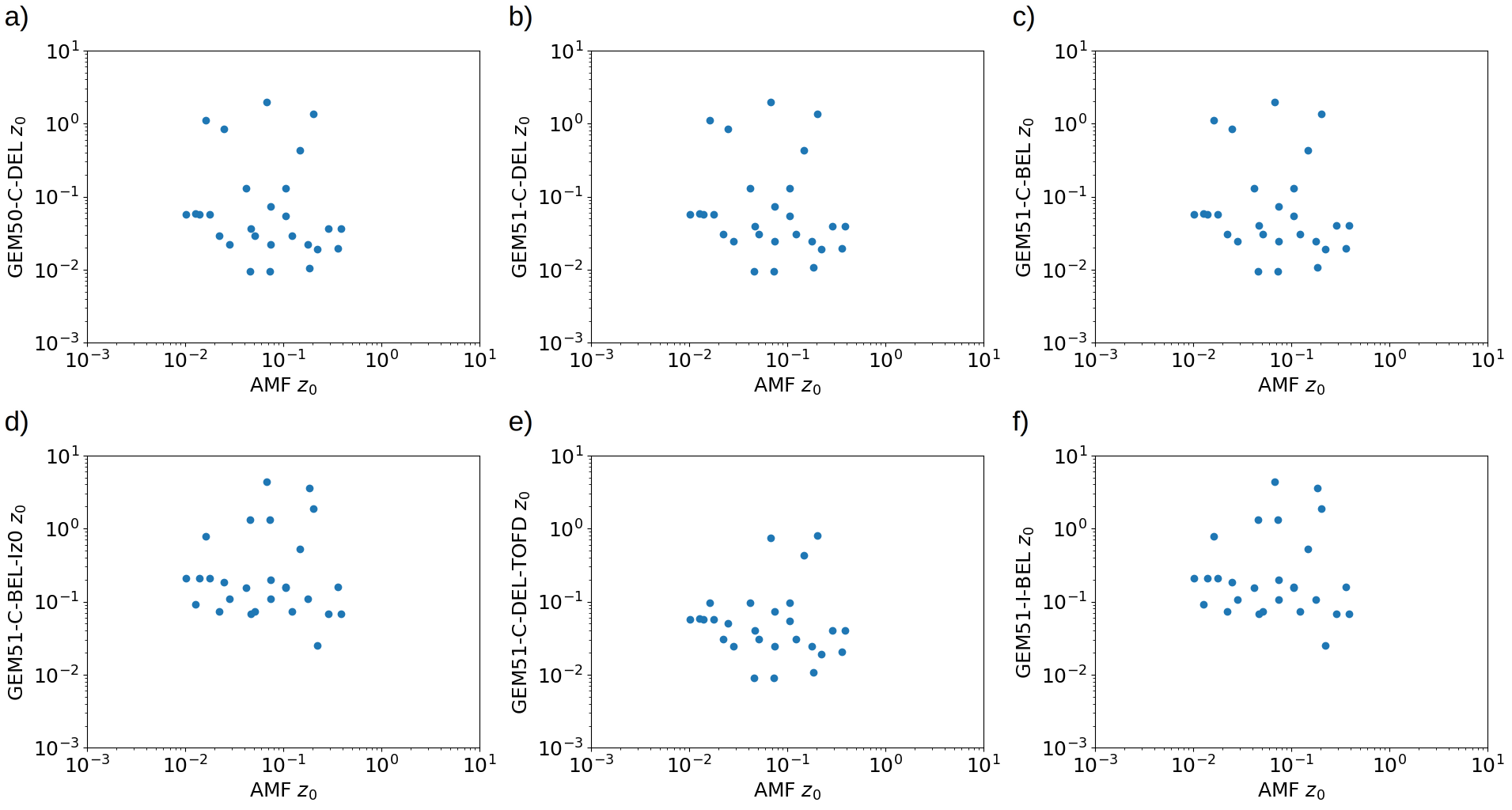}
	\caption{Effective roughness lengths for ISBA (a), CLASS (b), and CLASS using TOFD (c).}
	\label{fig:z0}
\end{figure}

\begin{figure}[ht]
	\includegraphics[width=\textwidth]{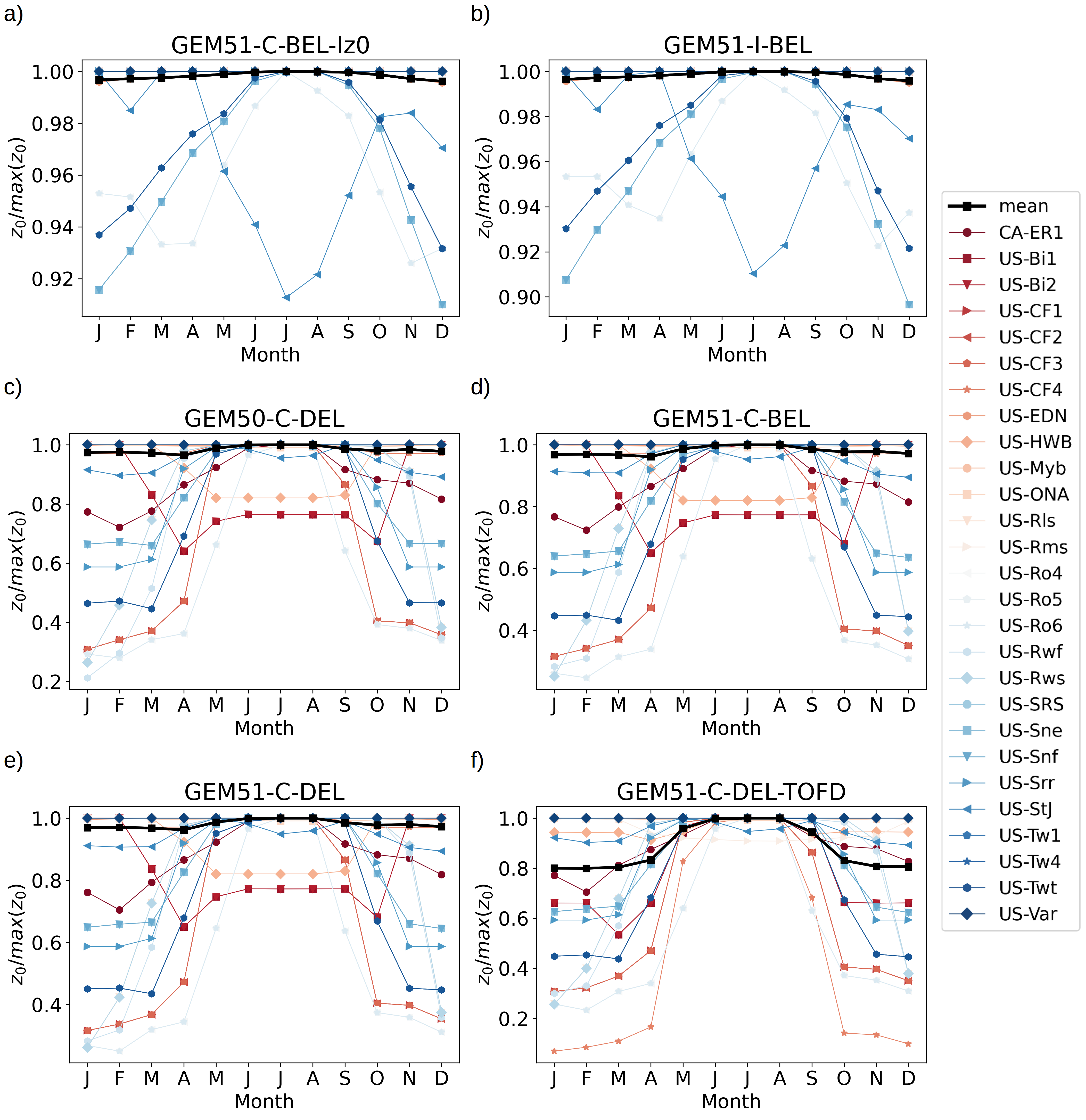}
	\caption{Annual cycle of the effective roughness lengths each station and simulation with the mean cycle given in black.}
	\label{fig:S9}
\end{figure}

\begin{table*}
\begin{tabular}{@{}lccc@{}}
\hline
& $z_0$ mean (m)   & $z_0$ median (m)  &  Corr(AMF,sim) \\ 	
\hline
AMF	& 0.11 &1 \\
GEM50-C-DEL	& 0.25 & 0.037 & -0.09\\
GEM51-C-BEL	& 0.25 & 0.037 & -0.09 \\
GEM51-C-DEL &0.25 & 0.037 & -0.09 \\
GEM51-I-BEL	& 0.60 &	0.158 &	0.01\\
GEM51-C-DEL-TOFD & 0.11	& 0.04 & 0.06\\
GEM51-C-BEL-Iz0 & 0.60 & 0.158 & 0.01\\
\hline
\end{tabular}
\caption{Mean and median and $z_0$ for AMF, ISBA, CLASS and CLASS-TOFD. The third column shows the correlation between AMF and the simulations across all 27 stations.}
\label{z0_tab}
\end{table*}

\begin{table*}[ht]
\center
\resizebox{0.55\textwidth}{!}{
\ra{1.8}
\begin{tabular}{@{}llllll@{}}
\hline
Station Name   & Longitude  &  Latitude &  Altitude (m) & Measurement Height (m)  & Reference  \\ 
\hline
CA-ER1 & 43.64 & -80.41 & 370 & 1.93 & \cite{osti_1579541} \\ 
US-Bi1 & 38.10 & -121.50 & -2.7 & 3.9 & \cite{osti_1871134} \\ 
US-Bi2 & 38.11 & -121.54 & -5 & 5.11 & \cite{osti_1871135} \\ 
US-CF1 & 46.78 & -117.08 & 794 & 4.0 & \cite{osti_1832158} \\ 
US-CF2 & 46.78 & -117.09 & 795 & 4.0 & \cite{osti_1881573} \\ 
US-CF3 & 46.76 & -117.13 & 795 & 2.1 & \cite{osti_1881574} \\ 
US-CF4 &  46.75 & -117.13 & 795 & 2.1 & \cite{osti_1881575} \\ 
US-EDN & 37.62 & -122.11 & - & 4.34 &  \cite{osti_1832159} \\ 
US-HWB & 40.86 & -77.85 & 378 &5  & \cite{osti_1811363} \\ 
US-Myb & 38.04 & -121.76 & 338 & 5.44 &  \cite{osti_1669685} \\ 
US-ONA & 27.38 & -81.95 & 25 & 2.8 & \cite{osti_1832163} \\ 
US-Rls & 43.14 & -116.73 & 1608 & 2.09 & \cite{osti_1418682} \\ 
US-Rms & 43.06 & -116.74 & 2111 & 2.28 & \cite{osti_1375202} \\ 
US-Ro4 & 44.67 & -93.07 & 274 & 2.5 & \cite{osti_1419507} \\ 
US-Ro5 & 44.69 & -93.06 & 283 & 2 & \cite{osti_1818371} \\ 
US-Ro6 & 44.69 & -93.06 & 282 & 2.3 & \cite{osti_1881590} \\ 
US-Rwf & 43.12 & -116.72 & 1878 & 3.5 & \cite{osti_1617724} \\ 
US-Rws & 43.16 & -116.71 & 1425 & 2.05 & \cite{osti_1375201} \\ 
US-SRS & 31.81 & -110.85 & 1169 & 7 & \cite{osti_1660351} \\ 
US-Sne & 38.04 & -121.75 & -5 & 5.43 & \cite{osti_1871144} \\ 
US-Snf & 38.04 & -121.73 & -4 & 3.49 & \cite{osti_1854371} \\ 
US-Srr & 38.20 & -122.02 & 8 & 3.4 & \cite{osti_1418685} \\ 
US-StJ & 39.08 & -75.43 & 6.7 & 3.5 & \cite{osti_1669695} \\ 
US-Tw1 & 38.10 & -121.64 & -5 & 4.64 & \cite{osti_1246147} \\ 
US-Tw4 & 38.10 & -121.64 & -5 & 5.36 & \cite{osti_1669698} \\ 
US-Twt & 38.11 & -121.65 & -7 & 3.18 & \cite{osti_1246140} \\ 
US-Var & 38.41 & -120.95 & 129 & 2 & \cite{osti_1245984} \\ 
\hline
\end{tabular}
}
\caption{AmeriFlux details for the station used in the study.}
\label{amf_details}
\end{table*}

\pagebreak 
\bibliography{agusample}   